\documentclass[aps,prd,amsmath,amssymb,10pt,letterpaper,balancelastpage,showpacs,superscriptaddress,nofootinbib,floatfix,notitlepage,longbibliography]{revtex4-2}

\usepackage{graphicx}
\usepackage{dcolumn}
\usepackage{color}
\usepackage{xcolor}
\usepackage{enumitem}
\usepackage{cancel}
\usepackage{mathtools}
\usepackage{hyperref}
\usepackage{orcidlink}

\graphicspath{{images/}}

\begin{document}

\title{Detecting Nanometer-Scale New Forces with Coherent Neutron Scattering}
\date{\today}

\author{Zachary Bogorad\,\orcidlink{0000-0001-9913-6474}\,}
\email{zbogorad@stanford.edu}
\affiliation{Stanford Institute for Theoretical Physics, Department of Physics, Stanford University, Stanford, CA 94305, USA}

\author{Peter W.~Graham\,\orcidlink{0000-0002-1600-1601}\,}
\email{pwgraham@stanford.edu}
\affiliation{Stanford Institute for Theoretical Physics, Department of Physics, Stanford University, Stanford, CA 94305, USA}
\affiliation{Kavli Institute for Particle Astrophysics \& Cosmology, Department of Physics, Stanford University, Stanford, CA 94305, USA}

\author{Giorgio Gratta}
\email{gratta@stanford.edu}
\affiliation{Department of Physics, Stanford University, Stanford, CA 94305, USA}

\begin{abstract}
Significant effort has been devoted to searching for new fundamental forces of nature.  At short length scales (below approximately 10 nm), the strongest experimental constraints come from neutron scattering from individual nuclei in gases. The leading experiments at longer length scales instead measure forces between macroscopic test masses.
We propose a hybrid of these two approaches: scattering neutrons off of a target that has spatial structure at nanoscopic length scales.
Such structures will give a coherent enhancement to small-angle scattering, where the new force is most significant. 
This can considerably improve the sensitivity of neutron scattering experiments for new forces in the $0.1 - 100$ nm range.
We discuss the backgrounds due to Standard Model interactions and a variety of potential target structures that could be used, estimating the resulting sensitivities. We show that, using only one day of beam time at a modern neutron scattering facility, our proposal has the potential to detect new forces as much as two orders of magnitude beyond current laboratory constraints at the appropriate length scales.
\end{abstract}

\maketitle

\tableofcontents

\section{Introduction}\label{sec:Introduction}

While the Standard Model has been fantastically successful at describing much of the observable universe, several outstanding questions---the nature of dark matter, the Higgs hierarchy problem, and the quantum description of gravity, to name a few---render it necessarily incomplete. Theories that attempt to resolve these problems generally involve the addition of new fields, often leading to a variety of new associated phenomenology. In particular, though the Standard Model includes only four fundamental forces, extensions to it can include a range of additional interactions.

One way that such new forces can arise is via additional gauged $U(1)$ symmetries, such as baryon ($B$) or baryon minus lepton ($B-L$) number \cite{PhysRev.98.1501, Fayet_1996, Fayet_2001, Long_1999}. The resulting gauge bosons will generically mix with the $Z$ boson, leading to a force proportional to some combination of baryon number, lepton number, and hypercharge. Alternatively, new finite-range forces appear in many models with compact extra dimensions \cite{PhysRevD.65.052003, PhysRevD.68.124021, PhysRevD.59.086004}, for example due to messenger fields living in the bulk of such extra dimensions. Other motivations for new forces include proposals to resolve the cosmological constant problem \cite{MotivationCC}, vector models of dark matter \cite{CHOI2020135836, OKADA2020135785} and various new scalar fields \cite{Fujii_1991}. More comprehensive reviews of these various motivations can be found in, for example, \cite{KimReview, AdelbergerReview}.

In this work, we consider new forces independent of the spins of the interacting particles. Such interactions are generally described by a Yukawa potential \cite{MoodyWilczek},
\begin{align}
    V(\mathbf{r}) &= -\frac{g^2Q_1Q_2}{4\pi|\mathbf{r}|}e^{-\mu|\mathbf{r}|} \label{eq:YukawaPotential}
\end{align}
with $Q_{1,2}$ the charges of the two interacting particles, separated by $\mathbf{r}$, with $g$ the coupling to the new force mediator and $\mu$ the mediator's mass.

In most of this work, we will further assume for simplicity that the new force couples to mass, such that the charges of the two particles are simply their respective masses. Since such a new force acts as a short-range modification to gravity, it is conventional to parametrize the new force's strength by its ratio $\alpha$ to that of gravity: $\alpha = g^2m_{\rm Pl}^2/(4\pi)$ with $m_{\rm Pl}$ the Planck mass. Extending our discussion to forces coupled to other charges (e.g. baryon number) is generally simply a matter of rescaling, so long as the interaction remains a Yukawa potential \eqref{eq:YukawaPotential}.

In this work, we will be focused on mediator masses around $10^0-10^4$ eV, corresponding to force ranges $\lambda$ of roughly $10^{-2}-10^2$ nm (we will use $\lambda = 1/\mu$; note that some sources instead define $\lambda = 2\pi/\mu$). This regime is uniquely interesting because it lies around the boundary of two dramatically distinct approaches to new force detection: macroscopic test masses, and neutron scattering. Longer-range interactions can be effectively detected by measuring forces between macroscopic objects \cite{NewForceLimitsBordag, NewForceLimitsKlimchitskaya, NewForceLimitsChen, NewForceLimitsLee, SNOWMASSAdelberger, LevitatedBeadSearch, NewForceLimitsDecca}, with the collection of atoms in one test mass seeing the coherently-summed potential of the other. Conversely, shorter-range interactions are typically probed through the angular distribution of neutrons scattered from a target \cite{XenonPaper1, XenonPaper2, NewForceLimitsHaddock, NewForceLimitsPokotilovski}. These experiments rely on the drastically smaller charge radius of the neutron compared to atomic matter, reducing the backgrounds from electromagnetic interactions and Casimir forces \cite{CasimirForces} which plague force measurements below the $\mu$m scale. An alternative approach employing the Pendell\"{o}sung effect, which is the source of the strongest existing constraints over much of the parameter space we consider, is described in \cite{PendellosungResult}. 

This work's proposal is a combination of these two approaches, using spatial structure such that individual neutrons scatter coherently from collections of many atoms. Such an approach should allow for significantly superior sensitivity to new forces at these length scales. An different technique for detecting new forces with $\lambda \gtrsim 1$ nm has recently been proposed in \cite{MossbauerProposal}.

The remainder of this work is organized as follows: We begin by presenting a summary of our proposal in Section \ref{sec:Overview}. We then specialize to the relatively simple version of our approach that can be performed on targets consisting of only a single element in Section \ref{sec:SingleMaterial}, before addressing targets consisting of two different materials in Section \ref{sec:TwoMaterial}. In both of these sections, we discuss candidate materials and provide projected sensitivities; the former section also includes an explanation of how to separate the effects of a new force from those of structure using X-ray scattering, with the corresponding two-material discussion left for Appendix \ref{app:TwoMaterialSeparation}. Finally, we summarize our results and offer some concluding remarks in Section \ref{sec:Conclusion}.

Because our proposal blends two largely distinct fields---the study of new interactions familiar to particle and nuclear physicists, and scattering techniques used largely for material analysis---we have also included a range of background information, as well as various technical details, in the appendices. We discuss the theory of neutron scattering from single atoms in Appendix \ref{app:NeutronInteractions}. Appendix \ref{app:XRayInteractions} provides an introduction to X-ray scattering from atoms and photoabsorption, as X-ray scattering is necessary in order to normalize the neutron scattering distribution of structured targets. Appendix \ref{app:StructuredScattering} describes how scattering is modified for targets with structure on length scales comparable to the inverse momentum transfer of the scattering process; while our discussion in this appendix is focused on neutron scattering, the ideas are applicable to scattering of any particle. As we noted above, Appendix \ref{app:TwoMaterialSeparation} describes how a new force can be distinguished from a modification of this sort of target structure in two-material targets.

The next several appendices describe a variety of systematic effects that must be controlled in our proposal. Appendix \ref{app:MultipleScattering} describes the impact of multiple scattering events, in which a neutron is scattered multiple times before its detection. Appendix \ref{app:ThermalEffects} considers the effects of finite target temperatures on our proposal. The effects of interactions between atoms within the target are then considered in Appendix \ref{app:AtomicInteractions}.

Some additional information about neutron and X-ray scattering instruments, including the realistically achievable parameters of instruments that are relevant to our proposal, are presented in Appendix \ref{app:InstrumentParameters}. Finally, in Appendix \ref{app:Statistics}, we describe our numerical approach to translating predicted scattering distributions into projections for sensitivity to new forces.

\section{Overview}\label{sec:Overview}

\begin{figure}[b]
    \centering
    \includegraphics[width=0.7\linewidth]{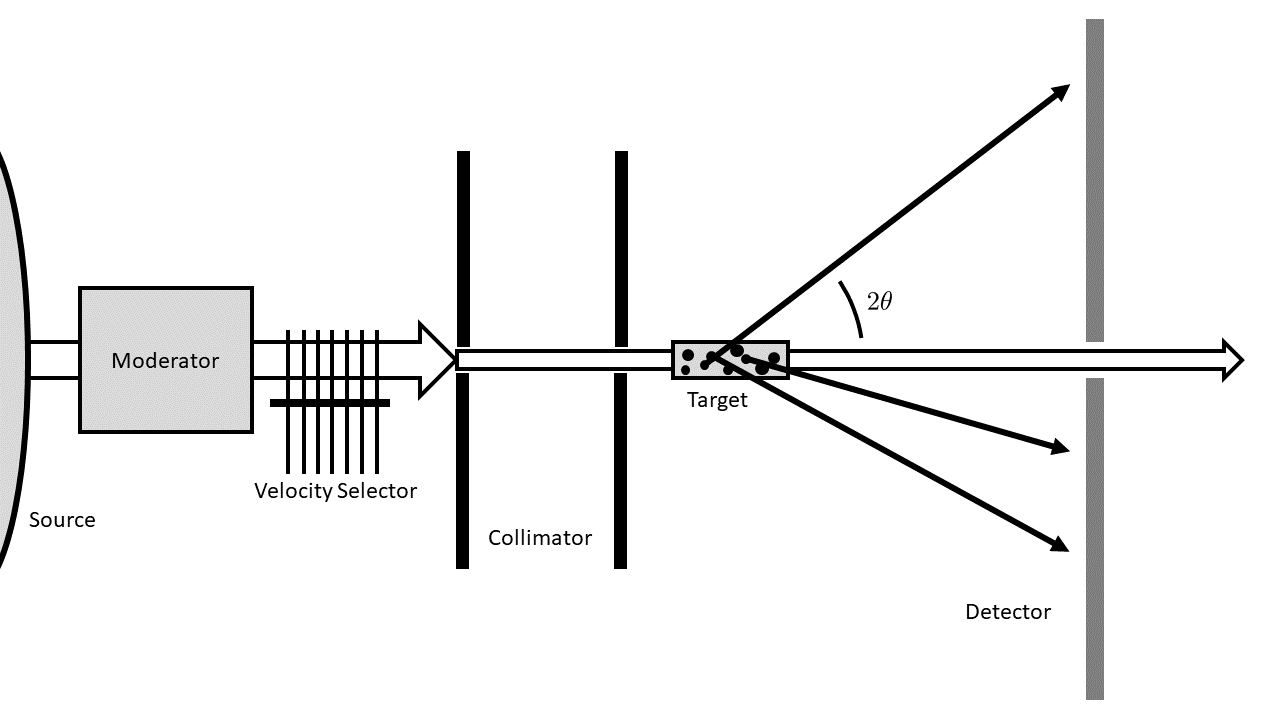}
    \caption{A simplified sketch of a neutron scattering experiment of the type discussed in this work, illustrating the key components of such an experiment; for details, see Appendix \ref{subapp:NeutronBeams}. Neutrons are produced from a reactor \cite{SANSInstrumentNIST, SANSInstrumentORNL, SANSInstrumentBERII, SANSInstrumentJRR3M, SANSInstrumentHANARO1, SANSInstrumentHANARO2} or via spallation \cite{SANSInstrumentSINQ, SANSInstrumentESS}, and then cooled in a moderator. A subpopulation of smaller velocity spread is then selected using either a rotating helical passage \cite{NeutronVelocitySelector} or a series of rotating disks \cite{NeutronVelocitySelectorDisks}, and a collimated beam is formed by passing the neutrons through two or more apertures. This beam is then incident on the target material---which, in this work, will typically have some internal structure---with scattered neutrons detected at some distance beyond the target.}
    \label{fig:ExperimentSummaryDiagram}
\end{figure}

The general principle behind neutron scattering-based searches for new forces is straightforward: a beam of neutrons is scattered off of a target, and the resulting angular distribution of scattered neutrons is measured; any significant deviation from the Standard Model prediction for that distribution is then an indication of new physics. A simplified sketch of such an experiment is shown in Figure \ref{fig:ExperimentSummaryDiagram}.

There are two significant Standard Model sources of neutron scattering: the strong nuclear interaction of neutrons with target nuclei, and the electromagnetic interaction of neutrons with atomic electric and magnetic fields. Nuclear scattering can be treated as hard sphere scattering at the $10^{-2}-10^{2}$ nm length scales that we consider, so accounting for it is a matter of a single, angle-independent fit parameter. Electromagnetic scattering, on the other hand, can be more difficult to model precisely, as it arises from a combination of several different effects and depends sensitively on the target atoms' electronic states. This is frequently circumvented by conducting new force searches using targets composed of noble gases (most often xenon) with zero spin and orbital angular momentum, in which case electromagnetic scattering is far more predictable.

The limiting factors for this procedure are then statistical: though the Standard Model backgrounds are well-understood, the finite number of neutrons scattered from the target sets a minimum strength for a new force that can be detected. This problem is exacerbated by the need to select neutrons scattered with very small momentum transfers in order to detect new forces of interest. Taking the Yukawa force that is the focus of this work as an example, the neutron scattering distribution from a noble gas can be written as (see \eqref{eq:CoherentScatteringLength})
\begin{align}
    \frac{d\sigma}{d\ln\theta} \propto |b_0|^2\left( 1+2\kappa_{\rm EM}f(q_T(\theta))+\frac{2\kappa_{\rm new}}{1+(q_T(\theta)/\mu)^2} \right)\theta\sin2\theta \label{eq:ScatteringDistributionOverview}
\end{align}
where $b_0$ is a characteristic, angle-independent scattering length (primarily due to nuclear scattering, although it receives an electromagnetic correction), $\kappa_{\rm EM}$ and $\kappa_{\rm new}$ are some measures of the relative strength of electromagnetic and new force scattering relative to nuclear scattering (typical values for this work are $\kappa_{\rm EM} \sim 10^{-2}$ and, at our sensitivity goal, $\kappa_{\rm new} \sim 10^{-6}$), $q_T(\theta)$ is the momentum transfer for a scattering angle of $2\theta$ (i.e. $q_T(\theta)=2q_0\sin\theta$ for incident neutrons of momentum $q_0$; see Figure \ref{fig:ExperimentSummaryDiagram}), $f(q)$ is a form factor for the atom, and $\mu$ is the new force mediator's mass. The three terms in this distribution are plotted in Figure \ref{fig:ScatteringContributionPlot} (although note that that figure uses $dp/d\Omega$ rather than $dp/d\ln\theta$). The new force contribution to this distribution is best resolved when $q_T(\theta) \sim \mu$, such that the new force term is not yet heavily suppressed by $(q_T/\mu)^{-2}$ but is no longer an angle-independent offset that cannot be distinguished from the nuclear force, when $q_T/\mu \ll 1$. In terms of the new force's mass coupling $g$, we have
\begin{align}
    \kappa_{\rm new} = \frac{m_n^3 g^2 A}{2\pi \mu^2 b_0} \label{eq:KappaNewDefinitionMT}
\end{align}
(see Appendix \ref{subapp:NewScattering}); $\kappa_{\rm EM}$ is defined by \eqref{eq:KappaEMDefinition}.

The former condition---that the momentum transfer not be too large---can be accomplished through some combination of two approaches: by using colder neutrons, and by considering scattering at small angles. Both methods are statistically costly. ``Cold'' neutrons, with wavelengths conventionally in the $0.4-3$ nm range, are generally produced by thermalizing neutrons in a cryogenic moderator \cite{NISTColdSource}; see Appendix \ref{subapp:NeutronBeams}. Neutrons with longer wavelengths (``ultra-cold'' neutrons, or ``UCNs''), however, are produced via momentum selection of cold neutrons, reducing the available neutron flux. Restricting to small-angle scattering also impacts the statistics, by requiring a more precisely collimated neutron beam. This further reduces the neutron flux, since neutron beams are collimated primarily by rejecting neutrons outside of the chosen phase space. A straightforward optimization shows that, for the application discussed here, looking at small-angle scattering of thermal neutrons is preferable to employing UCNs.

In addition, in experiments done to-date on nuclei of conventional materials, the small-angle scattering is also suppressed by the $\theta\sin2\theta$ term in \eqref{eq:ScatteringDistributionOverview}, corresponding to the limited phase space available for small-angle scattering. In this work, we present a method of circumventing this problem using coherent scattering from structured targets to enhance the fraction of incident neutrons that are scattered at the desired momentum transfers (typically to order unity, in fact). In particular, coherent scattering changes the scattering distribution \eqref{eq:ScatteringDistributionOverview} to (schematically) \cite{ScatteringTheoryBook,NeutronScatteringPolymersHammouda,NeutronScatteringWindsor}
\begin{align}
    \frac{d\sigma}{d\ln\theta} \propto |b_0|^2\left( 1+2\kappa_{\rm EM}f(q_T(\theta))+\frac{2\kappa_{\rm new}}{1+(q_T(\theta)/\mu)^2} \right)S(q_T(\theta)) \, \theta\sin2\theta \label{eq:StructuredScatteringSummary}
\end{align}
where $S(q)$ is the structure factor of the target, which gives the coherent enhancement of scattering at a given momentum transfer. Then, by employing targets such that $S(q_T(\theta)) \, \theta \sin2\theta$ is maximal at $q_T(\theta) \sim \mu$, neutrons can be made to scatter primarily at angles where the new force is most observable, effectively increasing the neutron count available for the measurement. 

The structure factor for a collection of identical target atoms, assuming incident plane wave neutrons, is given by (see \cite{ScatteringTheoryBook,NeutronScatteringPolymersHammouda,NeutronScatteringWindsor}, or the discussion in Appendix \ref{subapp:SimpleStructureFunctions})
\begin{align}
    S(q_T) &= \frac{1}{N}\left|\sum\limits_{j=1}^{N} e^{i\mathbf{q}_T\cdot\mathbf{r}_j}\right|^2, \label{eq:StructureFactorDefinition}
\end{align}
with the sum over the $N$ atoms in the target and $\mathbf{r}_j$ the position of atom $j$. For an ideal gas of spatial extent much larger than $q_T^{-1}$, the positions are effectively uncorrelated, so one expects $N$ atoms to give $S(q_T) \sim 1$. (A similar result arises even for correlated positions if the number of atoms is variable; see Appendix \ref{subapp:SimpleStructureFunctions}.) However, if one considers a cluster of $N$ atoms over some length scale $R$ with $q_TR \ll 1$, one instead expects $S(q_T) \sim N$, giving a factor of $N$ enhancement in the differential scattering cross-section at this momentum transfer. This is the central idea behind this work's proposal: using targets with structures at length scales comparable to $\mu^{-1}$, such that scattering is coherent at small momentum transfers but becomes incoherent at large ones.

As we noted previously, it is generally preferable to perform neutron scattering from noble gases, in order to both reduce and simplify the electromagnetic scattering background. Forming nanometer- to micrometer-scale structures from noble elements alone is likely to be difficult, though perhaps not impossible, as we discuss in Section \ref{sub:SingleMaterialOptions}. A more straightforward option is to employ a combination of two materials: a granular or porous solid and a noble liquid or gas which fills in the gaps in the solid. (In most of this work, we will refer to the noble component of a two-material target as a ``gas,'' although we will ultimately be interested in fluids near liquid density. Distinctions between gases, liquids, and supercritical fluids other than density will generally not be significant for our purposes; see Appendix \ref{subapp:LiquidInteractions}.) We will consider several candidates for such two-material targets in this work, though we will not attempt to catalogue them exhaustively and better options than what we discuss are likely to exist.

Realistic targets' structure factors cannot be predicted \textit{a priori} with sufficient accuracy to remove them from the measured neutron scattering distribution alone. Thus, when using a structured target, a low-angle bump in the scattering distribution cannot be attributed to a new force because it may instead correspond to some additional target structure at that scale. This issue can be circumvented by employing another type of scattering, most probably of X-rays. In the single-material case, the ratio of the neutron to X-ray scattering distributions is then target structure-independent, and remains well predictable within the Standard Model, so a deviation of this ratio from its prediction signals the presence of new physics. Thus, while we will generally focus on neutron scattering---as it is in many ways more technically difficult, and is where many new forces are likely to appear---practical experiments will require both X-ray and neutron scattering and treat them on mostly equal footing.

An additional complication arises in the case of two-material scattering, due to the interference of the solid and noble gas scattering amplitudes. If not for this interference, the solid scattering contribution could simply be measured separately and subtracted out. Dealing with the interference term, however, requires making measurements using at least two, and possibly three, distinct noble elements; we discuss this procedure in Appendix \ref{app:TwoMaterialSeparation}. Nonetheless, while more involved, two-material scattering can still be used to constrain new forces.

\section{Scattering from Single Materials}\label{sec:SingleMaterial}

We begin by considering the more straightforward implementation of our proposal using targets consisting of only a single noble element. Whether such a target could be produced with appropriate structure is unclear: we discuss several potential approaches to doing so below, and there may exist others, but the viability of these target candidates will need to be tested experimentally. Even if none of these approaches can be implemented, however, the single-material version of our proposal is useful as a simple illustration of how neutron and X-ray measurement can be combined to look for new forces, before considering the far more involved analysis required when using two-material targets.

\subsection{Possible Target Materials}\label{sub:SingleMaterialOptions}

The neutron's magnetic moment leads to significant, angle-dependent scattering from atoms with non-zero total orbital angular momenta, total electron spins, or nuclear spins; see Appendix \ref{app:NeutronInteractions}. The uncertainty in the Standard Model predictions for these scattering contributions acts as a background for any neutron scattering search for new forces. Similarly, the electromagnetic interactions between atoms in molecules or solids are likely to induce significant (at the required $\kappa_{\rm new} \sim 10^{-6}$ level) magnetic moments even in atoms that do not otherwise have them, creating an analogous background. Avoiding these two effects makes noble elements particularly attractive target materials \cite{NobleGasChoiceFermi}, as they have no magnetic moments and form nonmolecular gases.

While other elements (e.g. mercury) may in principle make for usable targets, we will focus on noble gases exclusively, considering them alone in this section, and in the presence of a solid in Section \ref{sec:TwoMaterial}. Of the noble elements, xenon is likely the most promising candidate, and has historically been the most used for new force searches, as its large atomic weight enhances the new force scattering contribution of typical models. All of our discussion in this work should hold for any (stable) noble element, however. There should likewise be no qualitative distinction between isotopes of those elements, except through their different nuclear spins, which lead to a small electromagnetic background. (In fact, we will generally focus on isotopes with zero nuclear spin, but this is not critical; see Appendix \ref{app:NeutronInteractions}.) We note, however, that different isotopes of a single element can have wildly different neutron scattering lengths; see, for example, \cite{NeutronScatteringDataENDF, NeutronScatteringDataSears}.

We consider three possible approaches to creating structured targets from a single noble element: noble solids, aerosols, and boiling liquids.

While the solid states of most noble elements are reasonably achievable in laboratory conditions \cite{PhaseDiagramBook}, forming granular structures of such solids may be significantly more difficult. Xenon can form a ``snow-like'' state under appropriate cooling conditions \cite{XenonSnow}. We are not aware of any systematic studies of this state, but it may be possible to create xenon snow with structure on length scales appropriate for our purposes. Similarly, there may or may not be ways to produce snow from other noble elements. Substantial density changes have been discovered when decreasing the temperature of noble solids below a certain critical temperature \cite{NobleSolids}, although this may be due to phase transitions in the solid without changes in homogeneity.

It may also be possible to create granular structures from noble liquids. One way to do this is through aerosolization of a noble liquid. As with the possibility of xenon snow discussed above, we are not aware of any analyses of achievable droplet size distributions for noble elements, but the sub-micrometer sizes we are interested are fairly typical for generic aerosols \cite{AerosolDynamics, AerosolScience}. Since such an aerosol would be unlikely to remain airborne or maintain a constant particle size distribution, this option would require continuous production and extraction of the aerosol in the target chamber. This does not meaningfully change the measurement strategy, however: in the case of a time-varying target, every appearance of the structure factor in the separation of scattering contributions procedure described below can simply be replaced by its average, so variation in the structure factor does not affect final sensitivity. Note that, in this case, the structure factor must remain constant (to a precision of order $\kappa_{\rm new}$) between neutron and X-ray scattering measurements; this should be possible, however, for example by performing these measurements simultaneously \cite{SimultaneousSAXSSANS}.

Finally, scattering could be performed from noble liquids in the process of boiling, with the granular structure formed by the gaseous bubbles that appear during this process. It appears unlikely that the resulting bubbles would be sufficiently small or consistent \cite{BoilingTong, BoilingStephan, XenonMeasurementsLeadbetter,XenonMeasurementsSmith,XenonMeasurementZollweg}, however, so we leave serious consideration of this approach to future work.

For the rough sensitivity projections of this work, we assume that single-material targets consist of isolated granular spheres of approximately equal radii, separated by vacuum; see Appendix \ref{app:StructuredScattering} for a more precise description of our assumptions. This should be a reasonable approximation of aerosol geometry, but may appear less appropriate for snow (which does not consist of spherical grains) or boiling liquids (which have liquid between the grains). However, as Appendix \ref{app:StructuredScattering} further discusses, the general behavior of structure factors is determined solely by the structure's dimensionality and length scale, precluding any large corrections from the differing geometry of snow. Similarly, boiling liquids' structure factors are suppressed by the limited density contrast between the liquid and gaseous states, but are not otherwise affected. All three cases should therefore be approximately described by the same structure factor, to the order-unity precision we desire in this work (see e.g. \cite{NeutronScatteringPolymersHammouda, NeutronScatteringWindsor} and Appendix \ref{app:StructuredScattering}):
\begin{align}
    S(q_T) &\approx  \frac{12\pi}{9+2(q_T\overline{R})^4} n \overline{R}^3 + 1,
\end{align}
where $\overline{R}$ is the typical radius of the grains in the material and $n$ is the number density of the noble atoms. Example structure factors for liquid xenon grains of various size are plotted in Figure \ref{fig:StructureFactorPlot}.

The limiting behaviors of this structure factor are easily understood. For $q_T \overline{R} \ll 1$, $S(q_T) \to (4\pi/3)n\overline{R}^3$: scattering is coherent over individual grains, and thus the scattering distribution is enhanced by the number of atoms per grain. Conversely, for $q_T \overline{R} \gg 1$, $S(q_T) \to 1$, corresponding to fully incoherent scattering, with cross-sections simply summed over all atoms. Accounting for the variation in phase space available at different scattering angles (see \eqref{eq:StructuredScatteringSummary} and the preceding discussion) thus gives a scattering distribution peaked at $q_T \sim R^{-1}$. Granular materials therefore provide a means of increasing scattering probabilities at chosen momentum transfers. In particular, using materials with $R \sim \mu^{-1}$ will allow us to increase experimental sensitivity to a new force of range $\mu^{-1}$.

\subsection{Separating Scattering Contributions}\label{sub:SingleMaterialSeparation}

While it is possible to calculate the structure factors of targets based on their geometric properties (see Appendix \ref{app:StructuredScattering}), such estimates will not be exactly correct for any realistic targets due to variation in their constituent grain size and shape, as well as due to impurities. As a result, it is not sufficient to measure the neutron scattering distribution from a structured target in order to search for a new force, as any bump in low-angle scattering could indicate a bump in the structure factor rather than in the atomic scattering distribution. Circumventing this requires an additional set of measurements to extract the structure factor of the target alone. We now describe how this can be done for a single material, in which case the procedure is relatively straightforward and can be described analytically.

For simplicity, we restrict further to targets consisting of only a single phase or density of the noble element (e.g. xenon snow). Two-phase targets such as a boiling liquids require a slightly modified analysis to account for the density contrast between the two phases, as we discuss for the two-material analysis in Appendix \ref{app:TwoMaterialSeparation}. (Note that this is the only change required, however; the majority of Appendix \ref{app:TwoMaterialSeparation} is devoted to removing uncertain electromagnetic backgrounds, which are not a concern for targets containing only noble atoms.)

The key fact allowing the scattering distributions of individual atoms to be disentangled from the structure factor of the target as a whole is that structure factors are independent of the scattered particle, so long as the scattering lengths of each atom are equal for the scattered particle. For a noble gas, this is likely to hold quite generally, since all of the lowest-order electromagnetic properties (the total electron orbital momentum, the total electron spin, etc.) are zero. Thus, the structure factor can be obtained by performing scattering with X-rays. That structure factor can then be used to extract the neutron scattering distribution from individual atoms.

X-ray scattering is discussed in more detail in Appendix \ref{app:XRayInteractions}; here we will merely cite the corresponding scattering distribution for noble atoms:
\begin{align}
    \frac{d\sigma_X}{d\Omega} &= \left(\frac{Ze^2}{4\pi m_e}\right)^2 \left(f(q_T(\theta)) - \frac{Zm_e}{m_{\rm nuc}}\right)^2 \frac{1+\cos^2 2\theta}{2}, \label{eq:XRayDistributionAveraged}
\end{align}
where $m_{\rm nuc}$ is the mass of the atomic nucleus, $m_e$ is the mass of the electron, $Z$ is the atomic number of the target atoms, and we have averaged over incident polarizations and summed over outgoing ones. Note in particular two features of this distribution: it approaches a constant comparable (or equal) to its maximal value at small angles, and it is fully described by precisely-known parameters except for its dependence on the atomic form factor.

The ratio of the X-ray scattering probability distribution for the structured target to that of a uniform target of the same material is
\begin{align}
    \frac{dp_{\rm X,s}/dq_T}{dp_{\rm X,u}/dq_T} &= S(q_T),
\end{align}
with the `s' and `u' subscripts referring to structured and uniform targets, respectively, and we use $q_T$ in place of $\Omega$ or $\theta$ in order to emphasize that it is the momentum transfer, and not the angle, that should be compared between X-ray and neutron measurements. Here, we have switched to probability rather than cross-section distributions in order to account for normalization (or, equivalently, target thickness): we will generally assume that target thickness is selected so that 10\% of neutrons are scattered above some minimum angle (see Appendix \ref{app:MultipleScattering}), requiring different target thicknesses for different target structures. We will therefore want to compare these normalized scattering probabilities, rather than cross-sections.

The unstructured neutron scattering distribution can then be reconstructed from these two X-ray measurements, combined with a structured neutron measurement:
\begin{align}
    \frac{dp_{\rm n,u}}{dq_T} &= \frac{dp_{\rm n,s}}{dq_T} \left( \frac{dp_{\rm X,u}/dq_T}{dp_{\rm X,s}/dq_T}\right).
\end{align}
Crucially, this combination of measurements can lead to smaller uncertainties at small angles than a single, direct measurement of neutron scattering from a uniform target would, due to the latter's poor statistics at small angles. This requires, up to $\mathcal{O}(1)$ factors,
\begin{subequations}\begin{align}
    N_{n}\frac{dp_{\rm n,s}}{dq_T} &\gtrsim N_{n}\frac{dp_{\rm n,u}}{dq_T} \\
    \frac{N_{X}}{2}\frac{dp_{\rm X,s}}{dq_T} &\gtrsim N_{n}\frac{dp_{\rm n,u}}{dq_T} \\
    \frac{N_{X}}{2}\frac{dp_{\rm X,u}}{dq_T} &\gtrsim N_{n}\frac{dp_{\rm n,u}}{dq_T}
\end{align}\end{subequations}
over the range of momentum transfers useful for detecting the new force ($q_T\sim\mu$), where $N_{n}$ is the total incident neutron count given a fixed available neutron beam time, and $N_{X}$ is the analogous total X-ray count. (The included factors of 2 conservatively account for dividing X-ray beam time equally between the structured and unstructured measurements, although an unequal division may be more efficient.) The first condition holds whenever the structure factor is greater than its average,
\begin{align}
    \overline{S} = \frac{1}{\cos^2\theta_{\rm min}}\int_{\theta_{\rm min}}^{\pi/2} S(q_T(\theta)) \sin(2\theta) d\theta,
\end{align}
where $\theta_{\rm min}$ is the smallest angle observed. Note that this is a stronger condition than merely $S(q_T) > 1$, due to the differing normalization of structured and uniform targets needed to keep the total scattering probability constant. As long as this is the case, the third condition also implies the second. 

In fact, we will be able to satisfy a somewhat stronger condition: that the error on the neutron scattering distribution from the noble element (i.e. from the uniform target) is dominated by the error in the structured neutron scattering distribution rather than by the pair of X-ray measurements. This is the case whenever the number of X-rays scattered in a given angular range from the uniform target is greater than the number of neutrons scattered at those angles from the structured one, i.e. whenever
\begin{align}
    \frac{N_{X}}{2}\frac{dp_{\rm X,u}}{dq_T} \gtrsim N_{n}\frac{dp_{\rm n,s}}{dq_T};
\end{align}
since X-ray scattering is approximately angle-independent at small momentum transfers (see \eqref{eq:XRayDistributionAveraged}), this is approximately equivalent to the requirement that
\begin{align}
    \frac{N_X}{2N_n} \gtrsim \frac{S(q_T)}{\overline{S}} > 1. \label{eq:CombinedMeasurementSuperiorFluxCondition}
\end{align}

The assumption that $S(q_T) > \overline{S}$ will never hold at all momentum transfers: increased scattering at small angles corresponds to decreased scattering at large angles, when total scattering probability is held constant. Depending on the particular measurement, this may be irrelevant (if $\mu$ is small enough that only the enhanced small angles are useful for detecting a new force), or it may indicate that the optimal measurement strategy is to spend some neutron beam time on the uniform target measurement in order to reduce uncertainties at large angles. In this work we will restrict to neutron measurements using only structured targets; we leave a more thorough analysis of optimal measurement strategies to future work, though our results suggest that there is little advantage to neutron scattering from uniform targets (see Figure \ref{fig:SingleMaterialSensitivity}).

As we discuss in Appendix \ref{app:InstrumentParameters}, achievable fluxes for X-ray beams exceed those of neutron beams by a factor of at least $10^6$. Since we will not consider structure factors in excess of approximately $10^5$ (see Figure \ref{fig:StructureFactorPlot}), this is sufficient to ensure that \eqref{eq:CombinedMeasurementSuperiorFluxCondition} should always hold.

It is worth noting, however, that the condition \eqref{eq:CombinedMeasurementSuperiorFluxCondition} is likely too stringent, as it assumes no knowledge of the atomic form factor $f(q)$. In fact, atomic form factors can be calculated numerically from Standard Model parameters (see e.g. \cite{FormFactorsSafari,FormFactorsChantler,FormFactorsHubbell1,FormFactorsHubbell2}), though it is unclear if this can be done with the precision necessary for our purposes. A complete prediction for $f(q)$ is unnecessary, however: since the momentum scale over which $f(q)$ varies ($q_0 \sim 11~Z^{1/3}$ nm$^{-1}$; see Appendix \ref{app:NeutronInteractions}) is known to be much larger than the momentum transfers of interest, the X-ray scattering distribution \eqref{eq:XRayDistributionAveraged} can be accurately described by a combination of known parameters and a series expansion of $f(q_T)$ in powers of $q_T/q_0$. Doing so contributes a few additional degrees of freedom to the fitting procedure, but, crucially, it does not eliminate the signal, as there is no way for such an expansion to replicate the $1/(1+(q_T/\mu)^2)$ behavior of the new force scattering length contribution once $q_T > \mu$. (This is essentially the same reason why electromagnetic effects have little impact on our sensitivity projections, as we discuss in the next subsection.)

Using this separation of scales, only the structured measurements are necessary, and the condition for the neutron measurement to dominate the final uncertainty becomes considerably weaker:
\begin{align}
    \frac{N_X}{N_n} \gg 1. \label{eq:RatioMeasurementSuperiorFluxCondition}
\end{align}
Along with being easily satisfied by a wide range of X-ray instruments, this condition has the added benefit of being intuitively understandable: in this measurement approach, the experiment consists simply of looking at the ratio of the neutron to X-ray scattering distributions of a structured target. The shared dependence of these distributions on the structure factor is eliminated in the ratio, leaving only a measurement of the ratio of scattering distributions of individual atoms; any deviation of this ratio from the Standard Model prediction is then interpreted as a signal of a new force. Since this approach is fully symmetric between the neutron and X-ray measurements, the dominant uncertainty is determined simply by whichever included fewer scattering events.

\subsection{Sensitivity Projections}\label{sub:SingleMaterialProjections}

In the absence of systematics, a new force is detectable if it increases the number of small-angle scattering events by more than the corresponding Poisson error (summed over bins). Accounting for the uncertainties in the nuclear scattering length, electromagnetic scattering length scale, and atomic form factor complicates this criterion: a scattering distribution that includes the new force must be fit with Standard Model parameters, and only new forces for which no combination of these parameters leads to a sufficiently good fit can be detected. A precise statistical description of this criterion is presented in Appendix \ref{app:Statistics}; here we will merely summarize our approach.

Given a pair of new force parameters ($\mu$ and $g$), it is straightforward to generate a predicted total neutron scattering distribution from an assumed target. This distribution can then be fit with Standard Model parameters alone, or with the addition of the two new force parameters. The fits can then be compared using an F-test (see Appendix \ref{app:Statistics}); a significant improvement in the fit when including the new force parameters indicates the presence of such a new force. We assume errors from X-ray scattering (i.e. in our knowledge of the structure factor) to be subdominant. We include three Standard Model parameters in our single-material fits: an overall normalization $\mathcal{N}$ (corresponding to the angle-independent scattering length of the target atom $b_0$), the magnitude of electromagnetic scattering $\kappa_{\rm EM}$, and the momentum scale of electromagnetic scattering $q_0$ (see Appendix \ref{app:NeutronInteractions}). Our two fit functions are therefore given by
\begin{align}
    \frac{d\sigma}{d\ln\theta} = \mathcal{N}^{\rm (fit)}\left( 1 + \frac{2\kappa_{\rm EM}^{\rm (fit)}}{\sqrt{1+\left(q_T(\theta)/q_0^{\rm (fit)}\right)^2}} \left[ + \frac{2\kappa_{\rm new}^{\rm (fit)}}{1+\left(q_T(\theta)/\mu^{\rm (fit)}\right)^2}\right] \right)S(q_T(\theta)) \, \theta\sin2\theta,
\end{align}
with and without the bracketed term; all labeled fit parameters are allowed to vary freely. This ignores any existing constraints on these quantities, but this is unlikely to be particularly conservative: nuclear scattering lengths are generally not known to the required precision, and including the electromagnetic fit parameters had only a small effect on our sensitivity projections. 

While this fitting procedure omits higher-order corrections to the atomic form factor (see Appendix \ref{app:NeutronInteractions}), such terms are unlikely to be significant for $\mu \ll q_0$ given the minimal effect of the leading-order electromagnetic term. This is the result of the same separation of momentum scales discussed in the previous subsection: the large ratio of $q_0/\mu$ precludes a modified electromagnetic term from effectively imitating the new force contribution's momentum dependence. Note as well that the atomic form factor can be empirically determined from X-ray scattering from a uniform target alone, if the approximate analytic form that we employ is insufficient; see Appendix \ref{app:XRayInteractions}. To be conservative, we do not show projections for $\lambda < 10^{-1}$ nm since $\mu$ begins to approach $q_0$ at these length scales; an accurate treatment of this regime is beyond the scope of this work.

Our projections are based on a fiducial beamline with a flux of $10^8$ cm$^{-2}$ s$^{-1}$ neutrons over a target area of $10$ cm$^2$, a typical wavelength of $0.6$ nm, and a minimum resolvable angle of $3$ mrad (i.e. a minimum visible momentum transfer of approximately $(30\text{ nm})^{-1}$). We assume an integration time of 28 hours, giving a total of $10^{13}$ scattered neutrons (i.e. $10^{14}$ incident neutrons; see Appendix \ref{app:MultipleScattering}); sensitivities for other neutron counts are easily estimated from $g^2_{\rm min} \propto N^{-1/2}$. Neutron beam properties relevant to our proposal are discussed further in Appendix \ref{subapp:NeutronBeams}.

\begin{figure}[t]
    \centering
    \includegraphics[width=0.7\linewidth]{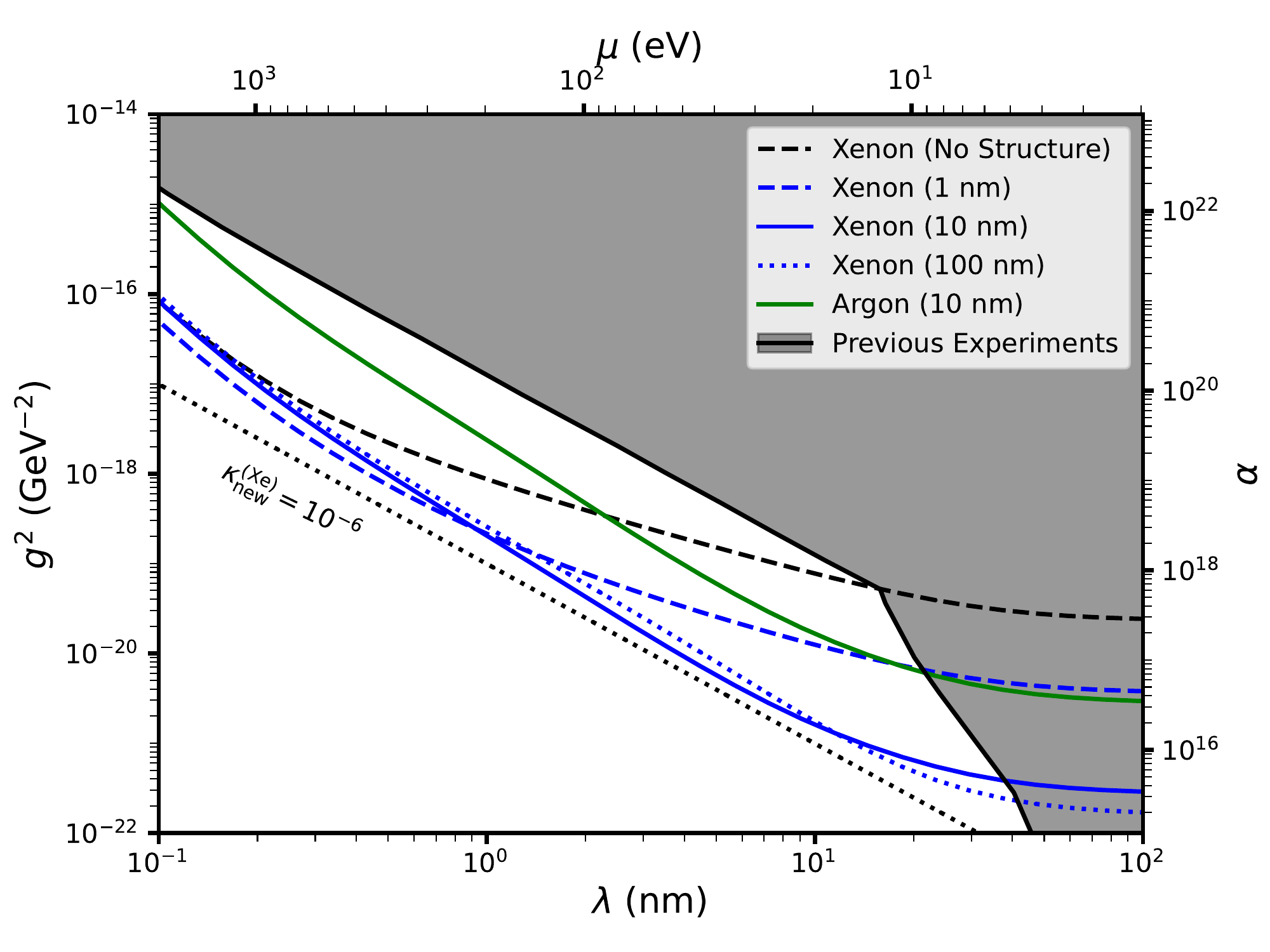}
    \caption{A comparison of projected sensitivities to new mass-coupled Yukawa forces for experiments following the design described in this work, using several different single-material target candidates. Detailed assumptions for each projection are discussed in the main text. Shown is scattering from simple xenon gas (the target of many previous experiments) as well from xenon targets with spherical grains of radius 1, 10 and 100 nm and from argon targets with 10 nm radius grains. (The spacing between these spheres does not affect the sensitivity, so long as their positions remain uncorrelated, as we discuss in the main text.) Also shown are the regions of parameter space already excluded by previous experiments, namely \cite{NewForceLimitsPokotilovski, NewForceLimitsHaddock, XenonPaper1, XenonPaper2, NewForceLimitsBordag, NewForceLimitsDecca, PendellosungResult}, and the line corresponding to $\kappa_{\rm new}=10^{-6}$ for xenon, as an illustration of the target systematic error we use throughout this work. Astrophysical constraints at these masses lie below the bottom edge of the plot, but are somewhat model-dependent; see Section \ref{sec:Conclusion}.}
    \label{fig:SingleMaterialSensitivity}
\end{figure}

We illustrate the projected sensitivities for several single-material targets in Figure \ref{fig:SingleMaterialSensitivity}. We focus primarily on scattering from xenon, as its large atomic weight makes it the optimal target in this case, though we illustrate an achievable sensitivity for argon as well for comparison. Both noble elements are assumed to be at liquid densities, and the number of grains in each target is adjusted to reach a scattering fraction of $0.1$ into angles above the minimum visible angle (see Appendix \ref{app:MultipleScattering}). We assume angular acceptance of $2\theta$ from $3\times10^{-3}$ to $\pi/4$ radians. For xenon, we consider targets with spherical grains of typical radii 1, 10, and 100 nanometers as well as with no structure at all; the spatial density of these grains has no impact on the sensitivity, so long as their positions are uncorrelated (and assuming that the total scattering fraction is held fixed). The no structure curve illustrates the gain in sensitivity over existing experiments that arises simply from assuming more neutrons scattered from the target (as well as from the inevitably somewhat simplified analysis of our estimate compared to actual experiments), as opposed to the benefits of coherent low-angle scattering. As Figure \ref{fig:SingleMaterialSensitivity} illustrates, the use of structures targets can potentially extend a neutron scattering experiment's sensitivity to new forces by nearly two orders of magnitude in $g^2$ at ranges of at least 10 nm, with gains decreasing at shorter ranges until achieving parity at a few angstroms.

It is worth briefly considering some properties of Figure \ref{fig:SingleMaterialSensitivity} in order to confirm that our projections are reasonable. First, the no-structure curve can be reasonably compared to the results of \cite{XenonPaper2}, as the primary difference between the assumptions of our curve and the parameters of that work should be the number of scattered neutrons. We assume approximately $400$ times more scattered neutrons than were used in \cite{XenonPaper2} (noting that the scattering event count reported in that work is restricted to small-angle scattering events), in part due to a higher assumed integrated incident flux and, in part, because the target depth in \cite{XenonPaper2} was insufficient to scatter 10\% of incident neutrons, as we assume. This should lead to a sensitivity improved by a factor of approximately $20$, quite close to our projection at their optimal sensitivity, given the coarseness of this comparison.

Second, it is straightforward to understand the limiting behavior and grain radius dependence of the curves in Figure \ref{fig:SingleMaterialSensitivity}. In the high-$\mu$ limit, all of the sensitivity curves approach $g^2_{\rm min} \propto \mu^4$. This is the result of a combination of two effects: the $\mu^2$ suppression in $\kappa_{\rm new}$ (see \eqref{eq:KappaNewDefinitionMT}), and the fact that the new force scattering distribution is angle-independent up to $(q_{\rm max}/\mu)^2$ corrections; the angle-independent component is therefore absorbed into the nuclear scattering length fit. At small masses $\mu$, the advantages of larger grain radii (10 nm rather than 1 nm) become apparent, as longer-range new forces can be coherent over larger grains. Conversely, increasing the grain radius from 10 to 100 nm is not beneficial with a minimum accepted momentum transfer of $(30\text{ nm})^{-1}$, as the scattering that would be enhanced at these radii is below the minimum visible angle; a sensitivity advantage appears only if one observes scattering at smaller angles, and only at force ranges near the upper limit considered. (Note as well that targets with 100 nm grains may or may not suffer from significant multiple scattering backgrounds that would worsen their sensitivity; see the discussion in Appendix \ref{app:MultipleScattering}.)

The inferior sensitivity of an argon-based experiment is also easily interpretable: the visibility of the new force is given by its relative strength $\kappa_{\rm new} \propto A/b_0$, which is larger for xenon than any other noble element (see Table \ref{tab:MaterialSLDs}). We include argon in Figure \ref{fig:SingleMaterialSensitivity} both to illustrate the large advantage of xenon over other elements and because argon will be a more promising candidate gas in the two-material experiments discussed below.

\section{Scattering from Two Materials}\label{sec:TwoMaterial}

We now turn to scattering from targets consisting of two materials: one structured solid, providing a framework with the required non-uniformity scale, and one noble element filling the spaces within the solid. Crucially, while the neutron scattering length of the noble element is analytically tractable, we do not assume any particular behavior for the scattering length of the solid, whose electronic structure may lead to difficult-to-predict electromagnetic scattering. It is therefore necessary to combine several measurements---including, at minimum, measurements with two different noble gases---from such targets in order to eliminate this background. This procedure lacks a simple analytic description (as we described for the single-material case above), so we limit ourselves to demonstrating that there are sufficient possible measurements in order to constrain the pertinent degrees of freedom; this analysis is presented in Appendix \ref{app:TwoMaterialSeparation}.

As we discuss below, two-material targets have generically inferior sensitivity compared to single-material targets, as a result of systematic effects deriving from the backgrounds due to the solids, as well as of the loss in statistics occurring when a fixed number of neutrons is divided between measurements with different targets. However, unlike the speculative single-material targets of the previous section, the two-material combinations we consider here are readily available. The resulting sensitivity projections are therefore a lower bound on the potential of this proposal.

\subsection{Possible Target Materials}\label{sub:TwoMaterialOptions}

The solid materials that could be appropriate for our purposes can be loosely divided into two categories: porous materials whose pores are filled with a noble gas or liquid, and granular materials whose interstitial volume can be filled with the noble element. We will generally remain agnostic to the particular noble element used in the examples below; as we discuss in Appendix \ref{app:TwoMaterialSeparation}, two-material targets will generally require two or even three noble elements in order to separate the different scattering contributions, but the procedure does not otherwise depend on the specific element.

Different solid materials may, however, be more or less compatible with particular noble elements. One reason for this is the enhancement of the scattering probability with increasing difference in scattering length density (SLD, i.e. the product of atomic number density and scattering length, summed over constituent atoms) between the solid and the gas (see Appendix \ref{app:StructuredScattering}): solid materials whose SLD is much higher than that of the gas will scatter too much, potentially forcing an experiment to use impractically thin targets in order to avoid excessive multiple scattering (see Appendix \ref{app:MultipleScattering}). Note that, since scattering lengths of different isotopes of a single element can differ by large factors (see \cite{NeutronScatteringDataENDF, NeutronScatteringDataSears}), we will focus on particular isotopes of noble gases in this section; see Table \ref{tab:MaterialSLDs}. We assume natural abundances for all solids.

Even if thickness is not a concern, however, large solid SLDs lead to a loss of sensitivity because they reduce the fraction of scattering events that come from the noble gas, which are the only events for which a new force can be resolved from the electromagnetic background.

For both of these reasons, solids whose SLD is not much greater than the maximum achievable value for a given noble element---essentially the SLD for its liquid form, given the incompressibility of liquids---are preferable. Table \ref{tab:MaterialSLDs} lists approximate SLDs for all of the stable noble liquids and several example solid materials. As this list illustrates, finding solids with sufficiently small SLDs (and the necessary granular structure) may be challenging, motivating much of our discussion below. 

Note that we do not include electromagnetic scattering lengths in Table \ref{tab:MaterialSLDs} or in our calculations, as these depend on the detailed electronic structure of solids and are thus difficult to predict. While electromagnetic scattering lengths of atoms with non-zero total spins or orbital angular momenta can be comparable to their nuclear scattering lengths, they sum incoherently (see Appendix \ref{app:StructuredScattering}), and thus should have little impact on the sensitivity projections, as the expected scattering distributions are dominated by small-angle coherent scattering.

\begin{table}[b]
\centering
\begin{tabular}{| c | c c c |} 
    \hline
    Material & $b_c$ (fm) & $n_{\rm liquid}$ (nm$^{-3}$) & SLD$_{\rm liquid}$ (fm nm$^{-3}$) \\ 
    \hline
    He-4 & 3.3 & 22 & 72 \\
    Ne-20 & 4.6 & 37 & 170 \\
    Ar-36 & 25 & 21 & 530 \\
    Kr-86 & 8.1 & 18 & 140 \\
    Xe-136 & 9.0 & 14 & 120 \\
    \hline
    Material & $b_c^{\rm unit}$ (fm) & $n^{\rm unit}$ (nm$^{-3}$) & SLD (fm nm$^{-3}$) \\ 
    \hline
    SiO$_2$ & 16 & 27 & 420 \\
    Al$_2$O$_3$ & 24 & 24 & 580 \\
    Al$_2$Ti$_3$O$_9$ & 49 & 5.6 & 275 \\
    BaTiO$_3$ & 19 & 15 & 290 \\
    CeO$_2$ & 16 & 25 & 410 \\
    CNTs & 6.7 & 100 & 670 \\
    \hline
\end{tabular}
\caption{The coherent scattering lengths, densities, and scattering length densities of all of the stable noble elements and of some candidate solid materials: alumina, silica, and carbon nanotubes (``CNTs''). Noble element densities are given for their liquid state, as an upper bound; the optimal isotope was chosen for each. For the solid ceramics, the scattering lengths and densities are for the full ``molecule,'' e.g. they treat all five atoms of Al$_2$O$_3$ as one unit; note that this is an inaccurate measure of coherent scattering once the inverse momentum transfer is comparable to the interatomic spacing. The CNT number density assumes a skeletal mass density of 2 g/cm$^3$, though estimates for the true value vary; see, for example, \cite{CNTReview, CNTDensityMeasurement, MWCNTDensityDiscussionEsconjauregui, MWCNTDensityDiscussionLaurent}. All of the solid results assume a natural mixture of isotopes. Nuclear incoherent scattering and absorption should not significantly affect measurements using any of these materials, so we do not include them here. Values for scattering lengths are from \cite{NeutronScatteringDataENDF, NeutronScatteringDataSears}, while those for densities are from \cite{CRCHandbook, PubChemAl2Ti3O9}.}
\label{tab:MaterialSLDs}
\end{table}

Several example candidate solid materials are presented in this section, but this is likely not a complete list and better materials may result from a more thorough search. The first is silica, SiO$_2$, which we use as our primary benchmark for sensitivity projections. Silica may be a candidate material in either of two forms: as a porous ceramic, or as a collection of spherical grains. Silica gels (in particular aerogels and xerogels) can have very high porosities, and can be produced with a variety of pore sizes, including some appropriate for our purposes (for a review, see, for example, \cite{SilicaReview}). Silica can also be manufactured in the form of small spheres, typically via the St\"{o}ber process (see e.g. \cite{StoberProcess, SilicaMicrospheres}). Silica has a reasonably low SLD of $420\text{~fm~nm}^{-3}$ (see Table \ref{tab:MaterialSLDs}): lower than the maximum of argon, though not of any other noble element.

Cerium oxide (CeO$_2$) has a very similar SLD of $410\text{~fm~nm}^{-3}$, and is widely available in powdered form due to its application in surface polishing \cite{CeO2Review}. Note, however, that the typical grain size of common cerium oxide powders is too large for our purposes.

Alumina, Al$_2$O$_3$ can be produced as a porous ceramic with pores of variable sizes \cite{AluminaCompany}, as it is used in filtration and various industrial processes; it cannot, to the authors' knowledge, be produced in the form of appropriately-sized beads. However, its SLD is $580\text{~fm~nm}^{-3}$, somewhat higher than that of silica.

Carbon nanotubes (CNTs) can provide appropriate granularities but, unlike the preceding materials, in a non-isotropic geometry. The structure factors of forests of CNTs, with incident neutrons parallel to the average nanotube direction, are estimated in Appendix \ref{subapp:CNTStructureFunction}. While not as sharply peaked as the spherical grain structure factor, the CNT result nonetheless accommodates the same low-angle scattering enhancement. Forests of multi-walled CNTs with appropriate radii ($\mathcal{O}$(10 nm)) have been produced \cite{CNTDiametersMamalis, CNTDiametersHinds, CNTDiametersIbrahim, CNTDiametersTruong}, though it is unclear whether they can be produced with appropriate thickness, density, substrate, etc. for this work's procedure.

Finally, we broadly consider the potential of some alloys to be effective targets. Alloys can be attractive solid targets because they can be chosen to have small or even vanishing SLD: since some elements (e.g. hydrogen, lithium, titanium \cite{NeutronScatteringDataENDF, NeutronScatteringDataSears}) have negative coherent scattering lengths, it is possible for the total coherent scattering length to be suppressed at length scales greater than the inverse interatomic spacing. (A degree of this effect can be achieved even with non-metals, as the two titanium-containing entries in Table \ref{tab:MaterialSLDs} illustrate; the advantage of alloys is the greater freedom to adjust their atomic compositions.) This can potentially reduce the solid-background issues discussed above. We note that we are not aware of any materials with this characteristic that can be produced with the necessary granular structure, but the attractiveness of such a target means that this may be a direction worth exploring in future work.

Our projections for the two-material case assume the same geometry as the single-material projections---a collection of uncorrelated spherical grains of gas---but now with the space between those grains filled with a solid material rather than vacuum. As we noted in Section \ref{sub:SingleMaterialOptions}, the exact shape of the grains should only modify our projections by order-unity factors, leaving us to compute only a single structure factor. The structure factor in this case is defined analogously to the single-material definition \eqref{eq:StructureFactorDefinition}, but now including the contributions of both materials:
\begin{align}
    S(q_T) &= \frac{1}{\sum_j N_j|b_j(\mathbf{q}_T)|^2}\left|\sum\limits_{j}\sum\limits_{k=1}^{N_j}b_j(\mathbf{q}_T)e^{i\mathbf{q}_T\cdot\mathbf{r}_{j,k}}\right|^2,
\end{align}
where the target contains $N_j$ atoms of the j'th element, each with scattering length $b_j(q_T)$. As we show in Appendix \ref{subapp:SimpleStructureFunctions}, this is well-approximated for our geometry by
\begin{align}
    S(q_T) &\approx  \frac{12\pi\overline{R}^3}{9+2(q_T\overline{R})^4} 
    \left( \frac{f\left|\Delta\mathcal{S}\right|^2}{fn_g|b_g(\mathbf{q}_T)|^2 + (1-f)\sum_j n_{s,j}|b_{s,j}(\mathbf{q}_T)|^2} \right) + 1, \label{eq:TwoMaterialStructureFactorMainText}
\end{align}
where $f$ is the volume fraction of the target occupied by the gas, $n_{g}$ ($n_{s,j}$) is the number density of the gas (the j'th element in the solid), $b_{g}(\mathbf{q}_T)$ ($b_{s,j}(\mathbf{q}_T)$) is the neutron scattering length of the gas (the j'th solid element), and $\Delta\mathcal{S}$ is the difference in SLDs between the two materials. The key difference from the single-material case is the dependence on this difference: coherent scattering vanishes in the limit of equal SLD, as the two materials become essentially equivalent for it.

\subsection{Sensitivity Projections}\label{sub:TwoMaterialProjections}

\begin{figure}[t]
    \centering
    \includegraphics[width=0.7\linewidth]{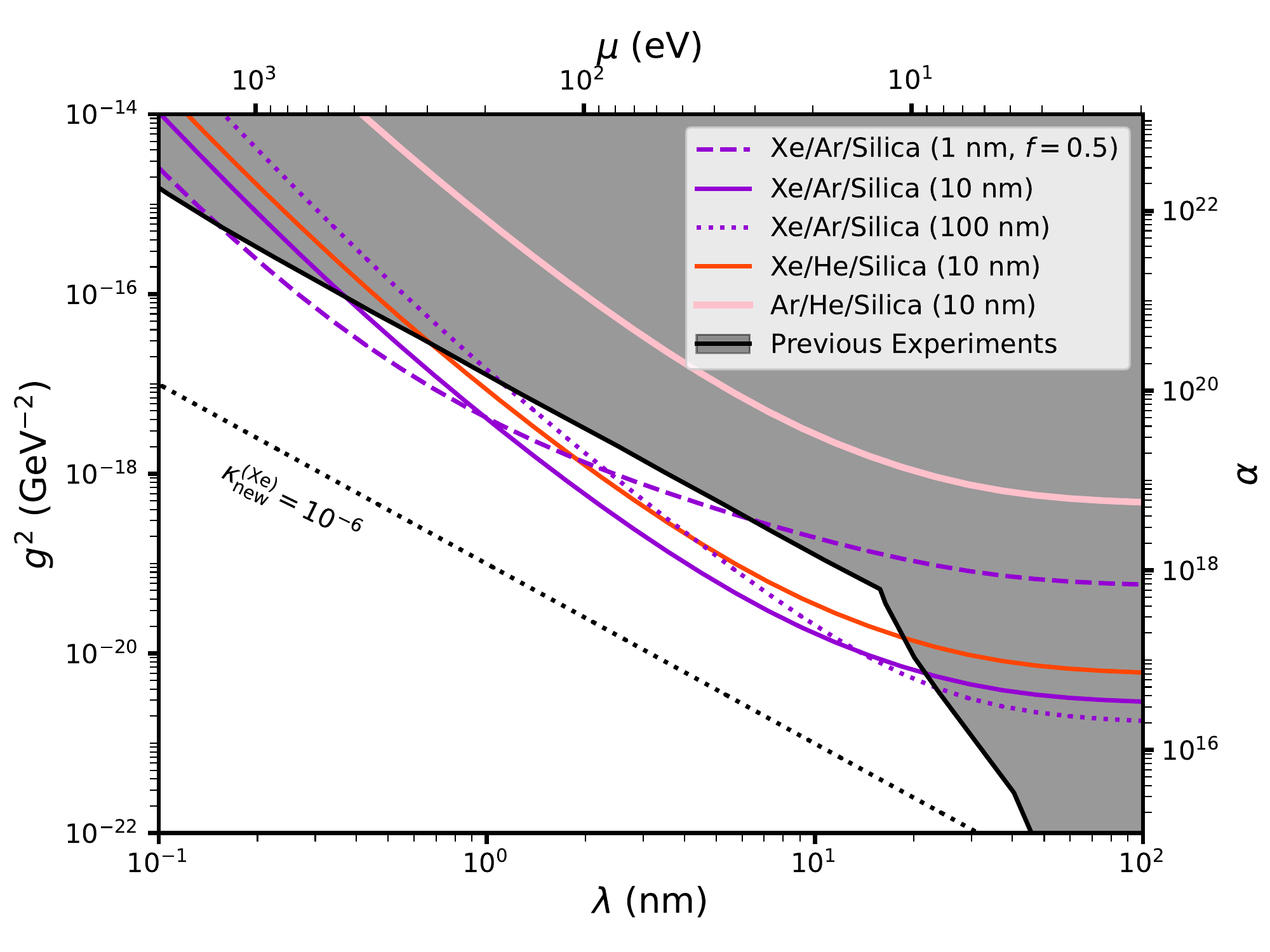}
    \caption{A comparison of projected sensitivities to new mass-coupled Yukawa forces for experiments following the design described in this work, using several different two-material target candidates. As discussed in Appendix \ref{app:TwoMaterialSeparation}, the two-material measurement requires at least two different noble elements, so each projection is labeled by two noble elements and one porous solid. Detailed assumptions for each projection are discussed in the main text. Shown is scattering from xenon and argon within silica with spherical pores of radius 1, 10 and 100 nm, as well as from the 10 nm case with helium replacing either of the gases. Also shown are the regions of parameter space already excluded by previous experiments, namely \cite{NewForceLimitsPokotilovski, NewForceLimitsHaddock, XenonPaper1, XenonPaper2, NewForceLimitsBordag, NewForceLimitsDecca, PendellosungResult}, and the line corresponding to $\kappa_{\rm new}=10^{-6}$ for xenon, as an illustration of the target systematic error we use throughout this work. Astrophysical constraints at these masses lie below the bottom edge of the plot, but are somewhat model-dependent; see Section \ref{sec:Conclusion}.}
    \label{fig:TwoMaterialSensitivity}
\end{figure}

Sensitivity projection is considerably more involved for two-material targets; we describe our approximation of it in Appendix \ref{subapp:TwoMaterialStatistics}. We assume the same neutron beam parameters as for the single-material case; see Section \ref{sub:SingleMaterialProjections} and Appendix \ref{subapp:NeutronBeams}. The two-material analysis requires several different neutron scattering measurements, so, in this case, we assume that the $10^{13}$ total scattered neutrons are divided evenly between them. We choose the porosity of each composite target such that 10\% of neutrons are scattered by the minimum accepted angle ($3\times10^{-3}$ radians) or more, when both solid and gas are present, in a thickness of 0.1 cm. (The one exception to this is xenon and argon in 1 nm radius-grain silica, for which 0.1 cm thickness is never sufficient; in that case we assume a porosity of $0.5$ and increase the thickness to reach 10\% scattering.) The resulting sensitivity projections are shown in Figure \ref{fig:TwoMaterialSensitivity}. 

We focus on the promising of the readily-available target candidates: granular silica, with measurements taken from both xenon and argon within it, showing the resulting sensitivities for several grain radii. Sensitivities for the 10 nm grain case with helium in place of xenon and in place of argon are also shown for reference; notably, the latter performs comparably to the argon option. This is the result of two competing effects with opposite influences on the resulting sensitivity: helium's low scattering length density makes it difficult to resolve above the solid background, but its small atomic weight means that comparison to it removes less of the new force contribution of xenon than comparison to argon does; see Appendix \ref{subapp:TwoMaterialStatistics}.

Figure \ref{fig:TwoMaterialSensitivity} illustrates many of the same features discussed in the single-material case in Section \ref{sub:SingleMaterialProjections}. However, the projected sensitivities are notably worse in the two-material case, due to the loss in statistics from dividing neutron flux between measurements and from eliminating the solid background. Nonetheless, scattering from silica with 10 nm radius grains surrounded by xenon and argon has the potential to surpass the reach of traditional, xenon-only scattering experiments by a factor of several for forces with ranges at or slightly above 10 nm; see Figure \ref{fig:SensitivitySummary}.

Sensitivity projections for solid materials other than silica were calculated but are not shown as they differed from the silica results only by order-unity factors. This is consistent with the comparable SLDs of all solids we considered (see Table \ref{tab:MaterialSLDs}). We emphasize, however, that our projections in Figure \ref{fig:TwoMaterialSensitivity} should be interpreted as conservative estimates, as there may exist better solid material candidates (i.e. granular solids with smaller SLDs) than we have considered. Thus the true potential reach of two-material neutron scattering likely lies somewhere between our one- and two-material projections.

\section{Conclusion}\label{sec:Conclusion}

We have presented an improved approach to constraining short-range forces using neutron scattering. By taking advantage of the enhancement of scattering at length scales comparable to target structures, neutrons can be made to preferentially scatter at small angles, where a new force may be most visible. The effects of such substructure can then be separated from those of a new force by combining measurements with different targets and by using X-ray scattering. The technique we describe could be implemented using a variety of different targets, including both single-element targets---which offer superior sensitivity but may be significantly more difficult to produce---and two-material targets.

\begin{figure}[t]
    \centering
    \includegraphics[width=0.7\linewidth]{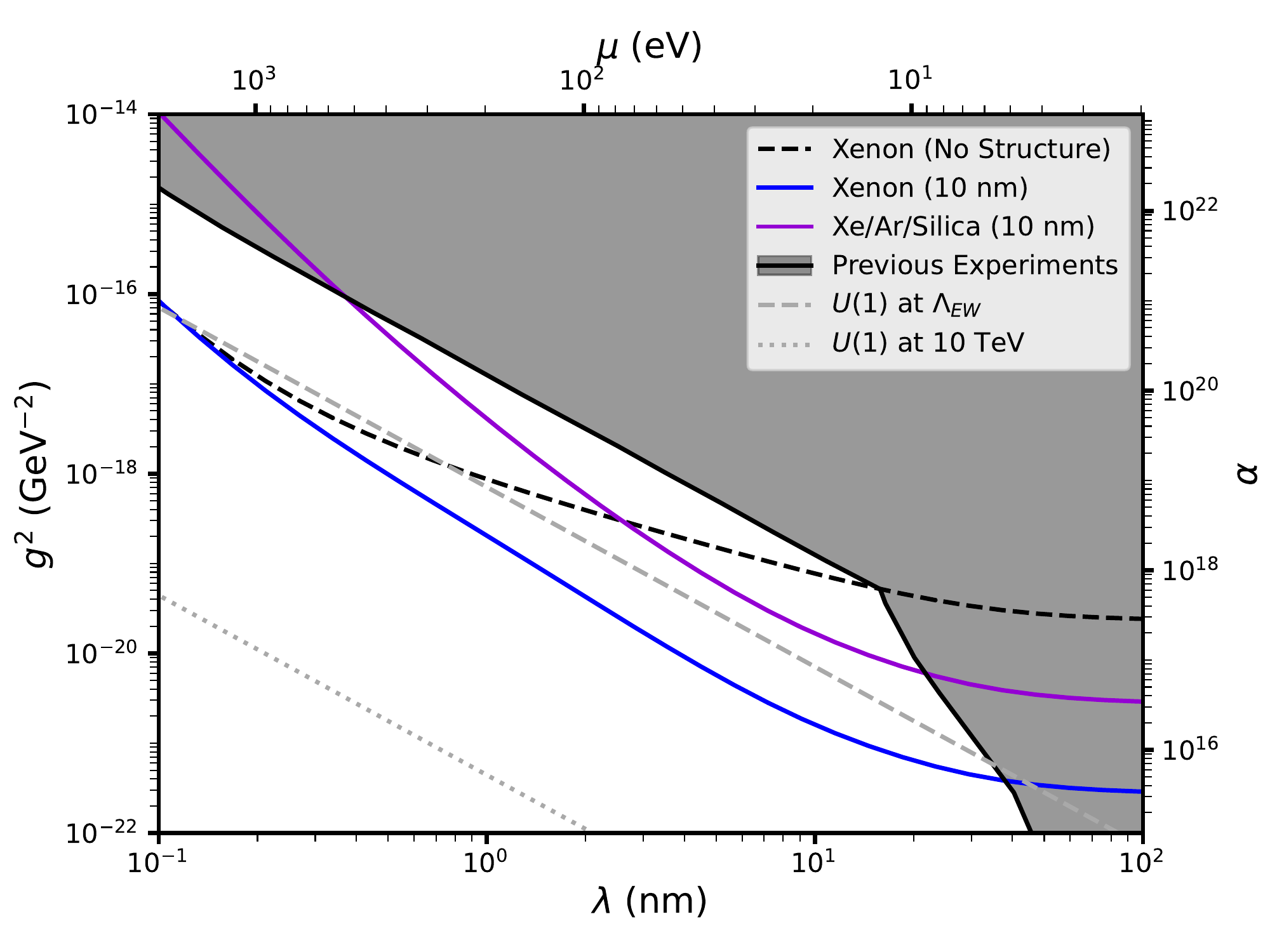}
    \caption{A summary of the sensitivity to new mass-coupled Yukawa forces for the most promising single- and two-material targets considered in this work (see Figures \ref{fig:SingleMaterialSensitivity} and \ref{fig:TwoMaterialSensitivity}). Here, the solid black region indicates the portion of parameter space excluded by previous experiments \cite{NewForceLimitsPokotilovski, NewForceLimitsHaddock, XenonPaper1, XenonPaper2, NewForceLimitsBordag, NewForceLimitsDecca, PendellosungResult} (exclusion from astrophysical observations lies below the bottom edge of the plot; see text), the two gray lines show the mass-coupling relations expected for new forces arising from $U(1)$ gauge symmetries broken at the electroweak scale or at 10 TeV \cite{Fayet_1996, Fayet_2001}, while the dotted black line corresponds to an estimate of the sensitivity that could be obtained using the conventional, uniform targets of previous experiments but assuming the $10^{13}$ scattered neutrons that we use for our projections. Any parameter space below this dotted line reachable using structured targets then corresponds to the benefits of coherently enhanced low-angle scattering. Note that, while the additional achievable parameter space for the silica-based target is fairly small, this projection is quite conservative: significantly better sensitivity may be achievable using porous or granular solids with smaller SLDs; see Section \ref{sub:TwoMaterialOptions}.}
    \label{fig:SensitivitySummary}
\end{figure}

Our estimates for spin-independent forces proportional to mass, assuming approximately one day of neutron beam time, are summarized in Figure \ref{fig:SensitivitySummary}; these projections can be generalized to other couplings (e.g. couplings to baryon number or baryon minus lepton number) simply by rescaling. We do not consider parametrically different forces (e.g. spin-dependent interactions) in this work. Such forces can be detected in scattering measurements using polarized neutrons, but we leave consideration of such approaches to future work.

Figure \ref{fig:SensitivitySummary}, like the two preceding figures, also shows the parameter space excluded by previous experiments, including both other neutron-based experiments \cite{NewForceLimitsHaddock, XenonPaper1, XenonPaper2, PendellosungResult} and searches for Casimir forces between many-atom test masses \cite{NewForceLimitsBordag, NewForceLimitsDecca, NewForceLimitsPokotilovski}. A variety of other experiments have obtained limits that are now subdominant to those plotted for the force ranges we consider; see e.g. \cite{NeutronScatteringNesvizhevsky, NewForceLimitsHaddock, NewForceLimitsPokotilovski, NeutronLimitsKamyshkov, NeutronOpticsLimits, AtomicTestsMurata, AtomicTestsTanaka}. Other proposals to explore similar parameter space include \cite{MossbauerProposal, CannexProposal}.

Our focus in this work has been on experimental searches for new forces, but we note that typical sources of new interactions are strongly constrained by astrophysical observations. In particular, measurements of stellar cooling generally restrict $g^2 \lesssim 10^{-24}$ GeV$^{-2}$ for all mediator masses we consider \cite{AstroConstraintsLasenby, WDLuminosity}, excluding most models to which our proposal is sensitive. Note, however, that astrophysical bounds are generically quite model-dependent: for example, stellar cooling constraints on forces coupled to $B-L$ are several orders of magnitude stronger, due to their interactions with electrons \cite{AstroConstraintsLasenby}; conversely, modifications to gravity due to extra dimensions will generally evade cooling constraints entirely. For a variety of particle physics models that can avoid standard astrophysical bounds, see \cite{EvasionBrax, EvasionDeRocco, EvasionJaeckel, EvasionJain, EvasionMasso1, EvasionMasso2, EvasionMohapatra}; although none of these models are immediately applicable to the new forces we consider, they illustrate the general model-dependence of such constraints. See also \cite{AstroConstraintsRaffelt, AxionForceBounds} for a discussion of a broader class of generally weaker astrophysical bounds on new forces.

Notably, the techniques discussed here are expected to achieve significantly improved sensitivity in the $1-100$ nm regime even using only existing facilities and materials. Substantial additional sensitivity improvement over the entire $0.1-100$ nm range should be possible through the development of appropriate granular materials.

\acknowledgments

The authors would like to thank Daniel Hussey, Young Lee, and Thomas Weiss for helpful discussions about SANS and SAXS techniques and instruments. 
This work was supported by the Simons Investigator Award No.~824870, NSF Grant No.~PHY-2014215, DOE HEP
QuantISED Award No.~100495, the Gordon and Betty Moore Foundation Grant No.~GBMF7946, and the U.S.~Department of Energy (DOE), Office of Science, National Quantum Information Science Research Centers, Superconducting Quantum Materials and Systems Center (SQMS) under contract No.~DE-AC02-07CH11359. 
ZB is supported by the National
Science Foundation Graduate Research Fellowship under Grant No. DGE-1656518 and by
the Robert and Marvel Kirby Fellowship and the Dr. HaiPing and Jianmei Jin Fellowship
from the Stanford Graduate Fellowship Program. GG is supported, in part, by DoE grant DE-SC0017970.

\appendix
\section{Neutron Scattering from Atoms}\label{app:NeutronInteractions}

\begin{figure}[b]
    \centering
    \includegraphics[width=0.7\linewidth]{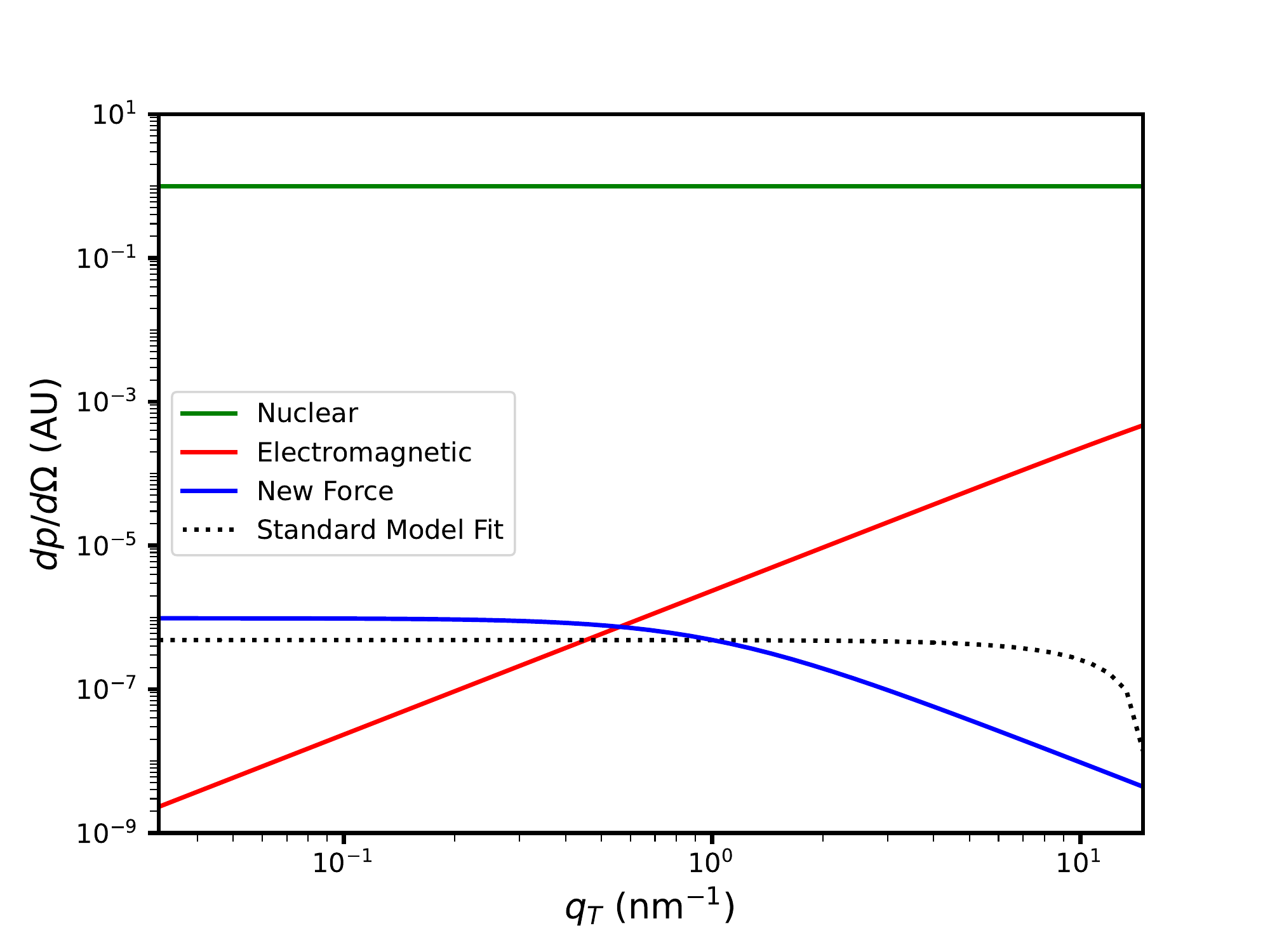}
    \caption{An illustration of the relative sizes of the three scattering probability contributions in \eqref{eq:ScatteringDistributionOverview}---nuclear, electromagnetic, and new force---for xenon gas, assuming a new force with $\mu^{-1} = 1$ nm and coupling $g^2 = 10^{-19}$, near our projected sensitivity at this range. Also shown is a linear combination of the nuclear and electromagnetic contributions that attempts to reproduce the new force's behavior, illustrating the inability of the new force contribution to be absorbed into the Standard Model fit parameters. Since the total scattering distribution is proportional to the target density and depth, the three solid lines shown here have been normalized so that $dp/d\Omega = 1$ for the nuclear contribution (at all angles, since it is angle-independent). Note that the electromagnetic and new force contributions shown here correspond to the interference terms between those forces' contributions and the nuclear contribution, since this is their dominant effect.}
    \label{fig:ScatteringContributionPlot}
\end{figure}

In this section, we tabulate leading contributions to the neutron-atom scattering length, including both Standard Model backgrounds and the potential fifth force contribution. As discussed in the main text, we break these contributions into three categories: nuclear scattering, which arises due to quantum chromodynamics (QCD) effects; electromagnetic scattering, due to quantum electrodynamics (QED) effects; and new force scattering, due to an assumed new spin-independent force coupling neutrons to nucleons. A plot of these three scattering contribution is shown in Figure \ref{fig:ScatteringContributionPlot}. Scattering due to weak interactions is negligible, so we will not discuss it here. More detailed reviews of neutron-atom interactions can be found in  \cite{SearsNeutronAtomInteractions, NeutronReviewAbele}.

Since we will be interested in the interference of scattering contributions (both of individual and distinct atoms), it is most convenient to work in terms of scattering lengths $b(\mathbf{q}_T)$, where $\mathbf{q}_T$ is the momentum transfer and the differential cross-section is
\begin{align}
    \left.\frac{d\sigma}{d\Omega}\right|_{\mathbf{q}_T(\theta)=\mathbf{q}} &= |b(\mathbf{q})|^2.
\end{align}
This will be appropriate when incident neutrons are accurately described as plane waves; we discuss the alternative in Appendix \ref{app:StructuredScattering}.

\subsection{Nuclear Scattering}\label{subapp:NuclearScattering}

Nuclear scattering of neutrons from atoms arises due to the strong force, which has a range of only $\mathcal{O}(1-10~\text{fm})$. Since the maximum momentum transfers that we consider in this work are $\mathcal{O}(10~\text{nm}^{-1})$, we can model nuclear scattering as scattering from a delta function potential (the ``Fermi pseudopotential'') up to corrections of order $\mathcal{O}((|\mathbf{q}_T| b_{\rm nuc})^2) \lesssim \mathcal{O}(10^{-8})$ for momentum transfer $\mathbf{q}_T$ and nuclear scattering length scale $b_{\rm nuc}$ \cite{FermiPseudopotentialOriginal, NeutronScatteringKoester, NeutronScatteringSearsAppendix, NeutronEffectiveRangeBlatt, NEScatteringKoester,  NeutronEffectiveRangeHackenburg}. This is sufficiently small that we will not be concerned with corrections to the delta function form in this work. Subject to this approximation, the nuclear scattering length is therefore angle-independent.

This scattering length can, however, depend on the neutron's spin with respect to the neutron's spin. In particular, the most general expression we can write for it is
\begin{align}
    b_{\rm nuc}(\mathbf{q}_T) = b_{\rm nuc,c} + \sqrt{I(I+1)} b_{\rm nuc,i}\boldsymbol{\sigma}\cdot\mathbf{I}
\end{align}
with $\boldsymbol{\sigma}$ the neutron's spin, $\mathbf{I}$ the nuclear spin, and $b_{\rm nuc,c}$ and $b_{\rm nuc,i}$ constants that must be determined empirically. In this work, we focus in large part on noble element isotopes with zero nuclear spin, in which case the second term vanishes and nuclear scattering is fully isotropic.

Here, the subscripts $c$ and $i$ in the two components of the nuclear scattering length mark these as being coherent and incoherent contributions, respectively. As the name suggests, incoherent scattering contributions do not generally combine coherently in bulk targets, since different atoms' nuclear spins should not be significantly correlated in any systems we consider. This is discussed in more detail in Appendix \ref{app:StructuredScattering}.

\subsection{Electromagnetic Scattering}\label{subapp:EMScattering}

Electromagnetic scattering of neutrons is significantly more complex than nuclear scattering. Here, we will not attempt to provide a complete description of the electromagnetic contributions to neutron-atom scattering, but will simply summarize the results. For a more detailed discussion, see, for example, \cite{SearsNeutronAtomInteractions,NEScatteringKoester,NEScatteringKopecky}.

The largest source of electromagnetic neutron-atom scattering is the interaction of the neutron with atoms' magnetic dipole moments. Including the contributions from electron spin, electron orbital angular momentum, and nuclear spin, this gives a total scattering length of 
\begin{align}
    b_{\rm dipole}(\mathbf{q}_T) &= \frac{g_n e^2}{8\pi m_e} \boldsymbol{\sigma} \cdot (\mathbf{1}-\hat{\mathbf{q}}_T\hat{\mathbf{q}}_T) \cdot \left( g_e f_S(\mathbf{q}_T)\mathbf{S} + f_L(\mathbf{q}_T)\mathbf{L} + \frac{m_e}{m_n}g_I \mathbf{I} \right)
\end{align}
where $g_n$ is the neutron g-factor ($g_n \approx -3.8$ \cite{PDG2022}), $\boldsymbol{\sigma}$ is the neutron's spin (with magnitude $1/2$), $\mathbf{q}_T$ is the momentum transfer, $\hat{\mathbf{q}}_T$ is a unit vector along the direction of the momentum transfer, $g_e$ is the electron g-factor ($g_e \approx 2$), $g_I$ is the nuclear g-factor of the target atom, $\mathbf{S}$, $\mathbf{L}$ and $\mathbf{I}$ are the atom's total electron spin, electron orbital angular momentum, and nuclear spin, respectively, and the functions $f_{S,L}(\mathbf{q})$ are form factors, defined below. (Note that we work in units where $\varepsilon_0 = 1$, which leads to a factor of $4\pi$ difference relative to, for example, \cite{SearsNeutronAtomInteractions}.)

The form factors $f_{S,L}(\mathbf{q})$ account for the spatial correlations of electron spins and angular momenta, which affect the relative phases of scattering contributions from different electrons. (Form factors are entirely analogous to structure factors, discussed in the main text and in Appendix \ref{app:StructuredScattering}.) Thus, while scattering from all of the electrons will add coherently in low momentum transfer scattering ($q_Tr_{\rm atom} \ll 1$), this will not be the case at momentum transfers comparable to or larger than the inverse atomic radius. For spin, this form factor is defined by
\begin{align}
    f_S(\mathbf{q})\mathbf{S} &= \left\langle \sum\limits_{j} \mathbf{s}_je^{i\mathbf{q}\cdot\mathbf{r}_j}\right\rangle
\end{align}
where the expectation value is over one atom, the sum is over the electrons, and the $j$'th electron has spin $\mathbf{s}_j$ and position $\mathbf{r}_j$. The definition of $f_L$ is somewhat more involved \cite{AngularMomentumFormFactorBook} and we will not be concerned with its form in this work. Analogous form factors for the nucleus are not relevant for our purposes, as the nuclear radius is far smaller than the inverse momentum transfers we consider (see the discussion of nuclear scattering above).

Since the neutron is moving with respective to the atom, the magnetic field it sees also acquires a contribution from the Lorentz transformation of the atom's electric field. This leads to an additional scattering contribution known as the ``Schwinger term,'' given by \cite{SchwingerTerm}
\begin{align}
    b_{\rm Schwinger}(\mathbf{q}_T) = -i\frac{g_n Ze^2}{8\pi m_n}(1-f(\mathbf{q}_T))\boldsymbol{\sigma}\cdot\hat{\mathbf{n}}\cot\theta
\end{align}
where $Z$ is the atomic number of the atom, $\hat{\mathbf{n}}$ is a unit vector along the cross product of the incident and outgoing neutron momenta, and $f(\mathbf{q})$ is the atomic form factor,
\begin{align}
    f(\mathbf{q})Z &= \left\langle \sum\limits_{j} e^{i\mathbf{q}\cdot\mathbf{r}_j}\right\rangle. \label{eq:AtomicFormFactorDef}
\end{align}
This atomic form factor is reasonably well-approximated by
\begin{align}
    f(q) \approx \frac{1}{\sqrt{1+(q/q_0)^2}}
\end{align}
where $q_0 \approx 11~Z^{1/3}\text{ nm}^{-1}$. Note that, while $\cot\theta$ diverges at small scattering angles, $1-f(\mathbf{q}_T(\theta))$ approaches 0 as $\theta$ goes to zero quickly enough ($\propto \theta^2$) for the Schwinger term to also go to zero at small angles \cite{SearsNeutronAtomInteractions}.

Neutrons can also scatter from purely electric fields, due to their internal charge distribution. Though the neutron is charge-neutral to extremely high precision and has a vanishing or negligible electric dipole moment \cite{PDG2022}, positive and negative charge densities within it may still be physically separated. Radial dependence of the charge density then leads to a potential depending on the Laplacian of the electric potential, and thus to a scattering contribution of the form \cite{SearsNeutronAtomInteractions,NEScatteringKopecky}
\begin{align}
    b_{E}(\mathbf{q}_T) = -\frac{m_nZ}{3a_0m_e}\left\langle r_n^2 \right\rangle (1-f(\mathbf{q}_T)) \label{eq:bChargeRadius}
\end{align}
where $a_0$ is the Bohr radius and $\left\langle r_n^2 \right\rangle \sim 10^{-1}$ fm$^2$ \cite{PDG2022} is the neutron mean-square charge radius, defined as
\begin{align}
    \left\langle r_n^2 \right\rangle = \int r^2 \rho(\mathbf{r})\ d^3\mathbf{r}
\end{align}
with $\rho(\mathbf{r})$ the charge density within the neutron.

We note that, historically, it was common to include one additional source of neutron scattering of the same form, known as the ``Foldy term'' (see e.g. \cite{SearsNeutronAtomInteractions, FoldyTermOriginal, NEScatteringKoester}), though this is now understood to be incorrect \cite{FoldyTermDiracEquation, FoldyTermFormFactor}. This change in understanding is of no phenomenological importance, however, as the Foldy term took a form identical to that of \eqref{eq:bChargeRadius}, and could thus be absorbed into the value of the (empirically determined) neutron charge radius.

Finally, there is one more noteworthy contribution to neutron-atom scattering, which arises due to the electric polarizability of the neutron. Though the neutron's electric dipole moment is known to be extremely small (or zero) in the absence of external electric fields \cite{PDG2022}, it may acquire one in their presence. This leads to additional scattering of neutrons from electric fields, which can be shown to take the form \cite{SearsNeutronAtomInteractions}
\begin{align}
    b_{P}(\mathbf{q}_T) = -\sqrt{\frac{3}{\pi}}\frac{m\alpha_n(Ze)^2}{4 \pi \sqrt{\left\langle r_n^2 \right\rangle}} \left(1 + \mathcal{O}\left(\frac{\sqrt{\left\langle r_n^2 \right\rangle}}{r_A}\right) + \mathcal{O}\left(q_T\sqrt{\left\langle r_n^2 \right\rangle}\right) \right)
\end{align}
where $\alpha_n$ is the neutron's electric polarizability ($\sim10^{-3}$ fm$^{3}$ \cite{PDG2022}) and $r_A$ is the atomic radius. Note that both $\sqrt{\left\langle r_n^2 \right\rangle}/r_A$ and $q_T\sqrt{\left\langle r_n^2 \right\rangle}$ are $\lesssim 10^{-5}$, while the overall magnitude of $b_P$ is around $10^{-4}~b_{\rm nuc}$. We can therefore ignore both of these terms at our systematic error target of $10^{-6}$ (see Appendix \ref{sub:SystematicErrors}), in which case the polarizability scattering term is angle-independent and is indistinguishable from a change in the nuclear scattering length.

Throughout this discussion, we have omitted terms at higher order in $m_e/m_n$. Since $(m_e/m_n)^2 \sim 3\times10^{-7}$, such terms should generally be too small to matter for our purposes. However, the large value of $Z$ for many targets of interest could potentially result in some such terms becoming significant. We leave the calculation of such higher-order terms to future work, noting that these are purely electromagnetic effects and should therefore be precisely calculable if necessary.

\subsection{New Force Scattering}\label{subapp:NewScattering}

The scalar-scalar fifth forces that we consider in this work correspond to a Yukawa potential of \cite{MoodyWilczek}
\begin{align}
    V(r) = -\frac{g^2M_1M_2}{r}e^{-\mu r}
\end{align}
with $g$ the new force coupling, $M_{1,2}$ the masses of the interacting particles, and $\mu$ the mediator mass. The resulting neutron scattering length of an atom of atomic weight $A$ (i.e. mass $Am_n$) is (see e.g. \cite{XenonPaper1})
\begin{align}
    b_{\rm new}(\mathbf{q}_T) = \frac{m_n^3g^2A}{2\pi}\frac{1}{\mu^2+q_T^2}.
\end{align}

It is convenient to define a relative strength 
\begin{align}
    \kappa_{\rm new} = \frac{m_n^3 g^2 A}{2\pi \mu^2 b_0} \label{eq:KappaNewDefinition}
\end{align}
of the new force scattering length $b_{\rm new}$ to the sum of angle-independent Standard Model scattering lengths $b_0$, where the latter is dominated by nuclear scattering but also receives contributions from electromagnetic scattering. (There is, of course, some ambiguity in this definition, since we can always subtract a constant from the angle-dependent scattering length and add it to $b_0$. We define $b_0$ explicitly after our discussion of how scattering is simplified for noble gases in Appendix \ref{app:StructuredScattering}; see \eqref{eq:b0Definition}.) In terms of $\kappa_{\rm new}$, we have
\begin{align}
    b_{\rm new}(\mathbf{q}_T) = \frac{\kappa_{\rm new}b_0}{1+(q_T/\mu)^2}.
\end{align}

\section{X-Ray Scattering from Atoms}\label{app:XRayInteractions}

The elastic X-ray scattering distribution from the electrons of a single atom, ignoring near-resonance effects, is given by \cite{XRayScatteringBlume, XRayScatteringGibbs, XRayScatteringStirling}
\begin{align}
    \left.\frac{d\sigma}{d\Omega}\right|_X &= \left(\frac{e^2}{4\pi m_e}\right)^2 
    \left| \left\langle f\left| \sum_j e^{i\mathbf{q}_T\cdot\mathbf{r}_j} \right| i\right\rangle \hat{\varepsilon}\cdot\hat{\varepsilon}' 
    - \frac{iE}{m_e} \left\langle f\left| \sum_j e^{i\mathbf{q}_T\cdot\mathbf{r}_j} \left(\frac{i\mathbf{q_T}\times\mathbf{p}_j}{E^2}\cdot\mathbf{A} + \mathbf{s}_j\cdot\mathbf{B} \right) \right| i\right\rangle \right|^2
\end{align}
where $E$ is the photon energy, $\varepsilon$ and $\varepsilon'$ are the incident and outgoing photon polarizations, $\mathbf{r}_j$, $\mathbf{p}_j$ and $\mathbf{s}_j$ are the position, momentum, and spin of the $j$'th charge, $i$ ($f$) is the initial (final) atomic state, and $A$ and $B$ are matrices depending on $\varepsilon$ and $\varepsilon'$ whose exact forms will not matter for our purposes. Scattering from the nucleus is described by analogous terms \cite{PhysRev.84.523}. As in the case neutron scattering, these expressions simplify considerably for noble atoms, leaving only
\begin{align}
    \left.\frac{d\sigma}{d\Omega}\right|_X &= \left(\frac{e^2}{4\pi m_e}Zf(q_T) - \frac{Z^2e^2}{4\pi m_{\rm nuc}}\right)^2 \left( \hat{\varepsilon}\cdot\hat{\varepsilon}' \right)^2
\end{align}
where $f(q_T)$ is the usual atomic form factor (see Appendix \ref{app:NeutronInteractions}), $m_{\rm nuc}$ is the mass of the nucleus, and we are ignoring the finite size of the nucleus. If we assume unpolarized incident X-rays and sum over outgoing polarizations, the observed scattering distribution becomes
\begin{align}
    \left.\frac{d\sigma}{d\Omega}\right|_X &= \left(\frac{e^2}{4\pi m_e}Zf(q_T) - \frac{Z^2e^2}{4\pi m_{\rm nuc}}\right)^2 \frac{1+\cos^2\theta}{2}. \label{eq:XRayScatteringDistributionAtomAverage}
\end{align}

X-ray scattering in practice is significantly complicated by photoabsorption, with typical attenuation lengths of 10-1000 $\mu$m for 10 keV X-rays; higher-weight elements typically lead to stronger absorption \cite{XRayAttenuation}. While attenuation lengths at the higher end of this range are unlikely to be a problem for our proposal, the much shorter attenuation length of, for example, xenon would lead to essentially complete absorption of 10 keV X-rays for the targets we consider. X-ray attenuation lengths generically increase rapidly with photon energy, however, so this issue can be largely circumvented using higher-energy X-ray sources, at the cost of requiring measurements at smaller scattering angles in order to study the same momentum transfers. At 40 keV, X-rays have an attenuation length of several hundred micrometers in liquid xenon \cite{XRayAttenuation}; using X-rays at or above this energy should therefore be sufficient for our purposes.

\section{Scattering from Structured Materials}\label{app:StructuredScattering}

In this appendix, we present a largely pedagogical introduction to structure factors; other sources on this topic include \cite{ScatteringTheoryBook,NeutronScatteringPolymersHammouda,NeutronScatteringWindsor}. We begin by summarizing the structure factors of several target material configurations, beginning with relatively simple targets before considering the structured targets that are the focus of this work. In the first subsection of this appendix, we ignore the distinction between the coherent and total scattering lengths of an atom (i.e. the incoherent scattering length) for simplicity; while this is a reasonable approximation for noble gases, it is not for many other elements, so we return to the more general case in the second subsection. Finally, we present an estimate of the structure factor of a carbon nanotube forest: this serves both as an illustrative example of the behavior of structure factors, and provides an alternative (albeit likely inferior) structure to the grain-based targets that are the focus of this work.

We emphasize that the structure factors computed below are intended only as rough predictions in order to estimate the projected sensitivities of our proposal. The structure factors of actual targets will need to be measured in order to separate their effects from those of a new force, as discussed in Section \ref{sub:SingleMaterialSeparation} and Appendix \ref{app:TwoMaterialSeparation}.

\subsection{Structure Factors of Simple Geometries}\label{subapp:SimpleStructureFunctions}

Our focus in this work is scattering from noble gases, which have zero total electron spin and angular momentum. We further assume zero nuclear spin; while this is not true of all noble element isotopes, it holds for all of the isotopes we consider. Moreover, at the target temperatures relevant to our proposal, there are no significant excited state populations for any atoms we consider. Thus, in the absence of any internal state variation, scattering from individual noble atoms depends exclusively on the momentum transferred during the scattering process. We therefore begin by considering scattering from targets where every atom has the same scattering length, before returning to the more general case in the next subsection of this appendix.

Scattering lengths of different atoms within the target sum, though the differing path lengths corresponding to scattering from different target atoms lead to relative phase factors. In the limit of large distances to the neutron source and detector, the resulting total scattering length is then
\begin{align}
    b_{\rm tot}(\mathbf{q}_T) &= \sum\limits_j b(\mathbf{q}_T)e^{i\mathbf{q}_T\cdot\mathbf{r}_j}, \label{eq:TotalScatteringLengthGeneric}
\end{align}
where the sum if over atoms at positions $\mathbf{r}_j$.

\begin{figure}[t]
    \centering
    \includegraphics[width=0.7\linewidth]{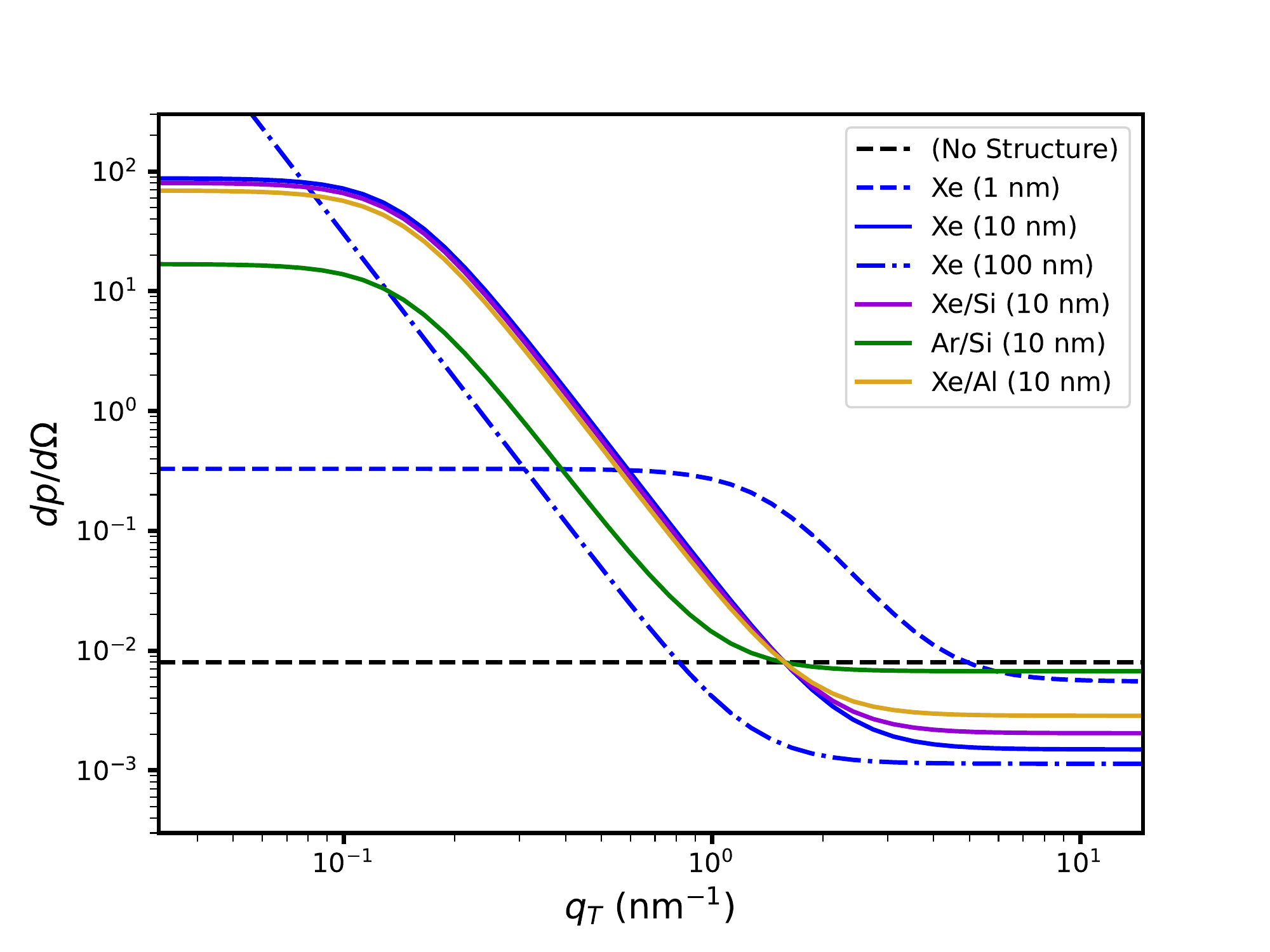}
    \caption{A comparison of the neutron scattering distributions of different targets, including a target with no structure, three targets consisting of spherical grains of xenon of radius 1, 10 or 100 nm, and three targets consisting of xenon or argon within spherical 10 nm-radius pores in silica or alumina. All targets are assumed to have depths that lead to 10\% scattering at or above $3\times10^{-3}$ radians for a neutron wavelength of 0.6 nm, with the two-material targets' porosities set to make that depth 0.1 cm, consistent with our assumptions in Section \ref{sec:TwoMaterial}. The left and right limits of the plot correspond to scattering angles ($2\theta$) of $3\times10^{-3}$ and $\pi$ radians, respectively.}
    \label{fig:StructureFactorPlot}
\end{figure}

The simplest relevant geometry for which we can evaluate the result of \eqref{eq:TotalScatteringLengthGeneric} is a large volume of ideal gas. Here, the ``largeness'' requirement is satisfied if all length scales of the target volume are much larger than the inverse momentum transfers $q_T^{-1}$ considered. If this is the case, the phase factors associated with scattering from each atom are independent and uniformly distributed, such that the expected total scattering length of the target is $0$. The expected total cross-section, however, is given by the variance of this distribution,
\begin{align}
    \left.\frac{d\sigma}{d\Omega}\right|_{\rm tot} &= N|b(\mathbf{q}_T)|^2, \label{eq:ScatteringSumIncoherent}
\end{align}
with $N$ the total number of atoms. The structure factor for this case is thus $S(q_T) = 1$. This is the usual incoherent sum of scattering cross-sections, and is plotted in Figure \ref{fig:StructureFactorPlot} as ``(No Structure).'' 

Note that this result depended only on two conditions: the absence of a preferred phase (such that the expected total scattering length was zero), and the variance in the total of the phase factors $e^{i\mathbf{q}_T\cdot\mathbf{r}_j}$ being equal to the number of atoms. The second condition, in particular, can arise in various ways: while the independent random phases present in this example are one option, variation in the number of atoms is another. Thus, even in the limit of $\mathbf{q}_T \to 0$, incoherent cross-sections can arise if the number of atoms within a target structure is a random variable, with the factor of $N$ in \eqref{eq:ScatteringSumIncoherent} replaced by the variance in the number of atoms (e.g. the expectation value of $N$, in the case of Poisson statistics).

Up to this point, we have implicitly assumed that the incident and outgoing neutrons are plane waves, such that the momentum transfer and scattering cross-section are well-defined. In practice, the finite momentum spread (or, equivalently, finite spatial extent) of neutrons complicates this result. Consider a neutron propagating along $\hat{z}$ with transverse wavefunction
\begin{align}
    \psi_\perp(\mathbf{r}_\perp) &= \frac{1}{\sqrt{2\pi\Delta r_\perp^2}}e^{-\mathbf{r}_\perp^2/(4\Delta r_\perp^2)};
\end{align}
we will ignore any $z$-dependence of the neutron wavefunction throughout this discussion since it does not affect the result in the limit of small momentum transfers, where $\mathbf{q}_T$ is orthogonal to $\hat{z}$; generalizing to larger angles is straightforward. The scattering probability is then
\begin{align}
    \left.\frac{dp}{d\Omega}\right|_{\rm tot} &= \left| \sum\limits_j \psi_\perp(\mathbf{r}_{j,\perp})b(\mathbf{q}_T)e^{i\mathbf{q}_T\cdot\mathbf{r}_j} \right|^2.
\end{align}
To calculate the coherent prediction for this value, we take the standard continuum limit, such that
\begin{align}\begin{split}
    \left.\frac{dp}{d\Omega}\right|_{\rm coh} &= \left| \int \psi_\perp(\mathbf{r}_{j,\perp})b(\mathbf{q}_T)e^{i\mathbf{q}_T\cdot\mathbf{r}_j} dN \right|^2 \\
    &= 8\pi|b(\mathbf{q}_T)|^2\left(\Delta r_\perp nL\right)^2 e^{-2q_T^2\Delta r_\perp^2},
\end{split}\end{align}
with $n$ the atomic number density and $L$ the depth of the target. Accounting for incoherent scattering, subject to our usual assumption that the resulting phase factors are entirely random, gives the usual contribution of $nL|b(\mathbf{q}_T)|^2$, for a total scattering distribution of
\begin{align}
    \left.\frac{dp}{d\Omega}\right|_{\rm tot} &= nL|b(\mathbf{q}_T)|^2 \left(1 + 8\pi(nL\Delta r_\perp^2)e^{-2q_T^2\Delta r_\perp^2} \right). \label{eq:FullyCoherentDistribution}
\end{align}

In particular, while the scattering distribution of Gaussian neutrons is unchanged when $q_T\Delta r_\perp \gg 1$, it is enhanced by an additional factor of $8\pi nL\Delta r_\perp^2$ (effectively the number of atoms seen by a single neutron) for $q_T\Delta r_\perp \ll 1$: scattering is ``fully coherent'' in this regime, even for targets with no structure whatsoever. We will assume that $q_T\Delta r_\perp \gg 1$ throughout most of this work, but we note that this requirement may (or may not) constrain realistic implementations of our proposal due to the impact of multiple scattering events; this is discussed in Appendix \ref{app:MultipleScattering}.

The more typical source of non-trivial structure factors is the arrangement of the atoms in the target. Most of the geometries of interest in this work will be characterized by regions with similar length scales in all directions (the sole exception, carbon nanotubes, is discussed later in this appendix). Such isotropic geometries can be qualitatively understood by considering scattering from a sphere; the exact structure factors of generic geometries must be computed numerically.

The structure factor for a sphere of monatomic ideal gas with radius $R$ is (see, for example, \cite{NeutronScatteringPolymersHammouda, NeutronScatteringWindsor}; we also derive this result, including corrections from incoherent scattering, in the next subsection)
\begin{align}
    S(q_T) &= \left( \frac{3(\sin(q_TR)-q_TR\cos(q_TR))}{(q_TR)^3} \right)^2 N + 1, \label{eq:StructureFactorSingleSphereExact}
\end{align}
with $N$ the total number of atoms in the sphere. Here, the first term corresponds to coherent scattering from the sphere as a whole, while the (frequently omitted) second term accounts for incoherent scattering due to random variation in atom positions as discussed in the large-volume case above.

All targets that we consider will consist of many distinct grains. We will generally assume that these grains' positions are essentially uncorrelated on the scale of $q_T^{-1}$, such that we can treat their sum in the same way as an ideal gas; we consider deviations from this assumption below. Thus, the structure factor of many, randomly-positioned spheres is still \eqref{eq:StructureFactorSingleSphereExact}, with $N$ understood to be the number of atoms per sphere rather than the total number of atoms in the target.

In practice, the grains of realistic targets are unlikely to have identical radii (or even, in most cases, identical shapes). It will therefore be convenient for our purposes to work with a version of \eqref{eq:StructureFactorSingleSphereExact} averaged over a small spread of radii:
\begin{align}
    S(q_T) &\approx  \frac{12\pi}{9+2(q_T\overline{R})^4} n \overline{R}^3 + 1, \label{eq:CloudScatteringDistributionAveraged}
\end{align}
with $n$ the number density of atoms within each grain, is accurate to within a few percent for relevant $q_T$ and $R$, assuming 10\% variation in $R$ around its average $\overline{R}$. This simplified form will be preferable for the approximate sensitivity projections we make in this paper; more precise calculations, using measured distributions of $R$, can be performed numerically.

Since creating isolated grains of noble atoms is potentially quite difficult (though perhaps not impossible; see Section \ref{sub:SingleMaterialOptions}), many of the targets we consider in this work consist of two separate materials: a solid that creates the granular structure, and a noble gas filling in the unoccupied space within that solid. There are two general categories of solids that could be used for this: porous materials, consisting of a single solid block with holes that are filled by the noble gas, and piles of grains, where the noble gas fills the spaces between the grains. Notably, these are largely equivalent from the perspective of coherent scattering: since coherent scattering from a uniform target is negligible for $q_T\Delta r_\perp \gg 1$ (see \eqref{eq:FullyCoherentDistribution}), the coherent scattering cross-section of a collection of grains filled with an ideal gas is equal to the coherent scattering cross-section from a target where the same ideal gas occupies all of space except those grains.

Predicting scattering distributions from solid targets is significantly more challenging, due to the non-trivial correlations between atomic positions. In the limiting case of scattering plane wave neutrons from a perfect crystal lattice, scattering is infinitely peaked at momentum transfers that are integer multiples of the inverse lattice spacing, generally less than 1 nm. Since our interest in this work is in maximizing scattering at much smaller momentum transfers, we will instead focus on amorphous solids, whose lack of regular structure should dramatically reduce this source of coherence. Nonetheless, some degree of short-range order, and consequently some amount of coherent enhancement, is likely to persist. We will ignore this effect below, as it should not have a significant impact on our sensitivity projections so long as it is smaller than the coherent enhancement caused by the target's granular structure; while this appears likely, it should be checked empirically for any particular target solid candidate.

For targets consisting of multiple elements, the definition of the structure factor \eqref{eq:StructureFactorDefinition} must be generalized to
\begin{align}
    S(q_T) &= \frac{1}{\sum_j N_j|b_j(\mathbf{q}_T)|^2}\left|\sum\limits_{j}\sum\limits_{k=1}^{N_j}b_j(\mathbf{q}_T)e^{i\mathbf{q}_T\cdot\mathbf{r}_{j,k}}\right|^2 \label{eq:StructureFactorDefinitionGeneral}
\end{align}
where the target contains $N_j$ atoms of the j'th element, each with scattering length $b_j(q_T)$. Ignoring the effects of regular solid structure discussed above, this can be rewritten as (see e.g. \cite{NeutronScatteringPolymersHammouda, NeutronScatteringWindsor}, or, again, the derivation in the next subsection)
\begin{align}
    S(q_T) &= \frac{\left| \int\limits_{R_{\rm gas}} \left(\mathcal{S}_{\rm gas}(\mathbf{q}_T) - \mathcal{S}_{\rm solid}(\mathbf{q}_T)\right) e^{i\mathbf{q}_T\cdot\mathbf{r}}  d^3\mathbf{r} \right|^2}{\sum_j N_j|b_j(\mathbf{q}_T)|^2} + 1, \label{eq:TwoMaterialStructureFactorIntegrated}
\end{align}
where $R_{\rm gas}$ is the region filled with gas and $\mathcal{S}_{\rm solid}$ ($\mathcal{S}_{\rm gas}$) is the scattering length density or ``SLD'' of the solid (gas), defined as $\mathcal{S}(q_T) = \sum_j n_j b_j(q_T)$ with the sum over the distinct elements making up the solid (gas). In particular, coherent scattering from two-material targets is dependent on the difference between the SLDs of the two materials, and vanishes when they are equal. Taking as an example our usual geometry of isolated spherical grains, but now with a different material between the grains, we have
\begin{align}
    S(q_T) &\approx  \frac{12\pi\overline{R}^3}{9+2(q_T\overline{R})^4} \frac{f\left|\Delta\mathcal{S}\right|^2}{fn_g|b_g(\mathbf{q}_T)|^2 + (1-f)\sum_j n_{s,j}|b_{s,j}(\mathbf{q}_T)|^2} + 1, \label{eq:TwoMaterialStructureFunctionUnimproved}
\end{align}
where $\Delta\mathcal{S}$ is the difference of the two materials' SLDs, $f$ is the fraction of the total volume taken up by the gas, and the remaining sum is over the distinct elements making up the solid. (The apparent asymmetry in $f\leftrightarrow 1-f$ here is a consequence of our assumption that the grain locations are uncorrelated, which can only hold exactly in the $f \ll 1$ limit. Realistic geometries will have $f \sim 1-f$, giving order-one corrections to this result.) Approximate scattering distributions for a selection of two-material targets of interest in this work are illustrated in Figure \ref{fig:StructureFactorPlot}.

\subsection{Coherent and Incoherent Scattering} \label{subapp:CoherenceCategories}

The discussion of scattering above assumed that the scattering length of every atom in the sphere was equal. At minimum, this requires all of the atoms in the target to be the same isotope. Even for a single isotope, however, electromagnetic scattering depends on the electronic (and nuclear spin) state of the atom, which will vary from atom to atom in every system we consider; see Appendix \ref{subapp:EMScattering}. It is easy to see that a uniform mixture of isotopes and states can be handled simply by replacing $b$ and $|b|^2$ in \eqref{eq:StructureFactorDefinitionGeneral} by their average values; we consider effects that might lead to spatially correlated states in later appendices.

Throughout this section, we will restrict to scattering of unpolarized neutrons, for which the neutron spin $\boldsymbol{\sigma}$ is uniformly distributed. Averaging over spins will therefore allow us to simplify our scattering distributions significantly.

For an isotropic medium, the expectation values of $\mathbf{L}$, $\mathbf{S}$ and $\mathbf{I}$ are all zero. The spin average of $\boldsymbol{\sigma}\cdot\hat{\mathbf{n}}\cot\theta$ is also zero for every scattering direction (but note that it is generally non-zero for any particular neutron, so Schwinger scattering is enhanced by structure factors; averaging over neutrons merely eliminates any interference between it and coherent scattering). Then the expectation value of the scattering length is given by the remaining terms from Appendix \ref{app:NeutronInteractions}:
\begin{align}\begin{split}
    b_c(\mathbf{q}_T) &= \left\langle b_n(\mathbf{q}_T) \right\rangle = b_{\rm nuc,c} + b_E(\mathbf{q}_T) + b_P + b_{\rm new}(\mathbf{q}_T) \\
    &= \left(b_{\rm nuc,c} - \sqrt{\frac{3}{\pi}}\frac{m\alpha_n(Ze)^2}{4 \pi \sqrt{\left\langle r_n^2 \right\rangle}} - \frac{m_n Z}{3a_0m_e}\left\langle r_n^2 \right\rangle \right) + \left( \frac{m_n Z}{3a_0 m_e}\left\langle r_n^2 \right\rangle\right)f(q_T) + \frac{m_n^3 g^2A}{2\pi \mu^2}\frac{1}{1+(q_T/\mu)^2} \\
    &= b_{0}\left(1 + \kappa_{\rm EM}f(q_T) + \frac{\kappa_{\rm new}}{1+(q_T/\mu)^2}\right) \label{eq:CoherentScatteringLength}
\end{split}\end{align}
where we have split the expectation value of the scattering length (i.e. the coherent scattering length) into an angle-independent contribution
\begin{align}
    b_0 &= b_{\rm nuc,c} - \sqrt{\frac{3}{\pi}}\frac{m\alpha_n(Ze)^2}{4 \pi \sqrt{\left\langle r_n^2 \right\rangle}} - \frac{m_n Z}{3a_0m_e}\left\langle r_n^2 \right\rangle \label{eq:b0Definition}
\end{align}
and two angle-dependent components: one from electromagnetic interactions, and one from any new force. Here,
\begin{align}
    \kappa_{\rm EM} = \frac{m_n Z}{3a_0 m_e b_0}\left\langle r_n^2 \right\rangle \label{eq:KappaEMDefinition}
\end{align}
parametrizes the relative strength of electromagnetic scattering compared to nuclear scattering (the noble elements we consider typically have $\kappa_{\rm EM} \sim 10^{-2}$), and $\kappa_{\rm new}$ (defined by \eqref{eq:KappaNewDefinition}) does the same for the new force. The contributions of the three terms in \eqref{eq:CoherentScatteringLength} are plotted in Figure \ref{fig:ScatteringContributionPlot}, although we do not absorb the angle-independent component of electromagnetic scattering into $b_0$ in that figure (i.e. we plot the physical contribution $b_0\kappa_{\rm EM}(1-f(q_T))$ of electromagnetism, rather than the $b_0\kappa_{\rm EM}f(q_T)$ that we use in most of the text for convenience).

For an anisotropic medium, the coherent scattering length will pick up additional terms proportional to the expectation values of $\mathbf{L}$, $\mathbf{S}$ and $\mathbf{I}$. We will never need these terms, however: as we discuss in the main text, we do not expect scattering from solids to be sufficiently predictable anyway, so we will always use combinations of measurements in which solid scattering cancels out. This leaves only scattering from the gas (or perhaps liquid) component, which should be sufficiently isotropic on its own. We demonstrate that solids will not lead to significant anisotropy in the gas or liquid near their surface in Appendix \ref{app:AtomicInteractions}.

Polarized neutrons are less inherently problematic, but require somewhat more tedious calculations: preferential polarization along some direction leads to Schwinger scattering that depends on the angle of scattering around the beam axis. Since we focus on unpolarized neutron beams, however, we omit this contribution.

Incoherent scattering is significantly more complicated, since the various electromagnetic terms generally do not average to zero. It will be helpful to organize the scattering length contributions as follows:
\begin{align}
    b_n(\mathbf{q}_T) = b_c(\mathbf{q}_T) + b_{\rm Sch}(\mathbf{q}_T)\boldsymbol{\sigma}\cdot\hat{\mathbf{n}} + b_{i,S}(\mathbf{q}_T)\boldsymbol{\sigma}\cdot(\mathbf{1}-\hat{\mathbf{q}}_T\hat{\mathbf{q}}_T)\cdot\mathbf{S} + b_{i,L}(\mathbf{q}_T)\boldsymbol{\sigma}\cdot(\mathbf{1}-\hat{\mathbf{q}}_T\hat{\mathbf{q}}_T)\cdot\mathbf{L} + b_{i,I}\boldsymbol{\sigma}\cdot(a\mathbf{1}-\hat{\mathbf{q}}_T\hat{\mathbf{q}}_T)\cdot\mathbf{I}
\end{align}
where
\begin{subequations}\begin{align}
    b_{\rm Sch}(\mathbf{q}_T) &= -i\frac{g_n Z e^2}{8\pi m_n}(1-f(\mathbf{q}_T))\cot\theta \\
    b_{i,S}(\mathbf{q}_T) &= \frac{g_n g_e e^2}{8\pi m_e} f_S(\mathbf{q}_T) \\
    b_{i,L}(\mathbf{q}_T) &= \frac{g_n e^2}{8\pi m_e} f_L(\mathbf{q}_T) \\
    b_{i,I} &= \frac{g_n g_I e^2}{8\pi m_n} \\
    a &= 1 + \sqrt{I(I+1)} \frac{ b_{{\rm nuc},i} }{ b_{i,I} }
\end{align}\end{subequations}
with the various constants and functions defined in Appendix \ref{subapp:EMScattering}. Note that $a$ can be much larger than 1 for some atoms, but is exactly equal to one for atoms with zero nuclear spin, which including all of the most promising target gases considered in this work.

The expectation value for the norm squared of the scattering length receives contributions from the square of each of these, as well as from any cross-terms that are non-zero. Fortunately, many of these terms turn out to be negligible for the situations we consider. Isotropic targets and unpolarized neutrons make all of the cross-terms including $b_c$ zero. (This is precisely what we assumed when discussing the coherent scattering length above.) The same holds for cross-terms involving $b_{\rm Sch}(\mathbf{q}_T)$. This leaves only cross-terms from terms that depend on the target atoms' electronic and nuclear spin states.

The largest of these cross-terms is
\begin{align}
    2b_{i,S}(\mathbf{q}_T)b_{i,L}(\mathbf{q}_T)\left\langle \left( \boldsymbol{\sigma}\cdot(\mathbf{1}-\hat{\mathbf{q}}_T\hat{\mathbf{q}}_T)\cdot\mathbf{S} \right) \left( \boldsymbol{\sigma}\cdot(\mathbf{1}-\hat{\mathbf{q}}_T\hat{\mathbf{q}}_T)\cdot\mathbf{L} \right) \right\rangle = \frac{1}{9} b_{i,S}(\mathbf{q}_T) b_{i,L}(\mathbf{q}_T) \left\langle \mathbf{S}\cdot\mathbf{L} \right\rangle
\end{align}
where the right hand side has been averaged over neutron spin and assumes an isotropic target, leaving only the expectation value over that target. (Recall that $|\boldsymbol{\sigma}|=1/2$ in our conventions.) There are analogous terms for the cross-terms including $b_{i,I}$, but in practice the expectation values of $\left\langle \mathbf{S}\cdot\mathbf{I} \right\rangle$ and $\left\langle \mathbf{L}\cdot\mathbf{I} \right\rangle$ are suppressed by the small value of the hyperfine coupling compared to the target temperature, which renders these cross-terms negligible for the targets we consider.

With these simplifications, and performing the rest of the spin averages, we find, for a single atom,
\begin{align}\begin{split}
    \left|b_i(\mathbf{q}_T)\right|^2 = \left\langle |b_n(\mathbf{q}_T)|^2 \right\rangle - \left|b_c(\mathbf{q}_T)\right|^2 - \frac{1}{12}\left|b_{\rm Sch}(\mathbf{q}_T)\right|^2 =&\  \frac{1}{18}\sqrt{S(S+1)}b_{i,S}(\mathbf{q}_T)^2 + \frac{1}{18}\sqrt{L(L+1)}b_{i,L}(\mathbf{q}_T)^2 \\ 
    &+ \frac{3a^2-2a+1}{36}\sqrt{I(I+1)}b_{i,I}^2 + \frac{1}{9} b_{i,S}(\mathbf{q}_T) b_{i,L}(\mathbf{q}_T) \left\langle \mathbf{S}\cdot\mathbf{L} \right\rangle.
\end{split}\end{align}
Note that the $b_{\rm Sch}(\mathbf{q}_T)^2$ term is suppressed by $(Zm_e/m_N)^2$ relative to the leading-order electromagnetic terms. While this makes it a small correction to the total scattering length, the large values of $Z$ for target atoms we consider, and the fact that it is coherently enhanced in structured materials, make it potentially non-negligible.

In much of this work, we restrict to scattering from noble gases with zero nuclear spin. In this case, the incoherent scattering length is simply zero, which is why we generally ignore the distinction between coherent and total scattering length. Note, however, that the scattering lengths throughout this text should be understood to include the small Schwinger correction,
\begin{align}
    \Delta b(\mathbf{q}_T) = \frac{1}{\sqrt{12}}b_{\rm Sch}(\mathbf{q}_T) = -i\frac{g_n Z e^2}{16\sqrt{3}\pi m_n}(1-f(\mathbf{q}_T))\cot\theta,
\end{align}
which behaves like a coherent scattering length but must always be added to scattering probabilities in quadrature after averaging over neutron polarizations (as well as due to its relative factor of $i$).

We now return to scattering from a collection of spherical grains. Our estimate of the resulting scattering distribution above \eqref{eq:CloudScatteringDistributionAveraged} did not distinguish between coherent and incoherent scattering lengths. We can correct this by more carefully evaluating the total scattering length of a sphere of ideal gas:
\begin{align}
    \left\langle |b_{\rm sphere}(\mathbf{q}_T)|^2 \right\rangle &= \left|\left\langle b_{\rm sphere}(\mathbf{q}_T)\right\rangle\right|^2 + \text{Var}\left(b_{\rm sphere}(\mathbf{q}_T)\right).
\end{align}
As discussed above, the expectation value of the total scattering length of one sphere can be evaluated in terms of the integrated SLD:
\begin{align}\begin{split}
    \left\langle b_{\rm sphere}(\mathbf{q}_T)\right\rangle &= \int \mathcal{S}(\mathbf{q}_T)e^{i\mathbf{q}_T\cdot\mathbf{r}}d^3\mathbf{r} \\
    &= \mathcal{S}(\mathbf{q}_T) \iint 2\pi r^2 \sin\theta ~ e^{i\mathbf{q}_T\cdot\mathbf{r}} ~ dr ~ d\theta \\
    &= \frac{4\pi(\sin(q_TR)-q_TR\cos(q_TR))}{q_T^3}\mathcal{S}(\mathbf{q}_T).
\end{split}\end{align}
Note that, since this corresponds to coherent scattering, only the coherent scattering length should be included in the SLD, i.e. $\mathcal{S}(\mathbf{q}_T) = nb_c(\mathbf{q}_T)$.

The variance of the total scattering length of the sphere conversely corresponds to incoherent scattering and, as discussed previously, should simply be given by the sum of the total scattering length's norm squared over all of the atoms within it. Thus
\begin{align}
    \left\langle |b_{\rm sphere}(\mathbf{q}_T)|^2 \right\rangle &= \left(\frac{4\pi(\sin(q_TR)-q_TR\cos(q_TR))}{q_T^3}n\right)^2 |b_c(\mathbf{q}_T)|^2 + \frac{4\pi}{3}nR^3 |b_n(\mathbf{q}_T)|^2.
\end{align}
We are interested in the structure factor of a target containing many such spheres; assuming that these sum incoherently, the total scattering lengths of each sphere should add in quadrature, which does not affect the structure factor of the target. Armed with this result, and performing the same radius-averaging as above, correcting \eqref{eq:CloudScatteringDistributionAveraged} is straightforward:
\begin{align}
    S(q_T) &\approx  \frac{12\pi}{9+2(q_T\overline{R})^4} n \overline{R}^3 \frac{|b_c(\mathbf{q}_T)|^2}{|b_n(\mathbf{q}_T)|^2} + 1. \label{eq:CloudScatteringDistributionImproved}
\end{align}

When grains are closely spaced (e.g. in highly porous materials) and for small $q_T$ (e.g. $q_T^{-1} \sim 30$ nm, near the limit of what we consider), there may be some degree of coherence even between the grains of the target, as their positions become correlated on scales of $q_T^{-1}$. Similarly to the effect of short-range correlations in amorphous solids discussed previously, this is likely to have some effect on the total scattering distribution but should not significantly change our conclusions so long as it does not prevent an order-unity fraction of neutrons from scattering at small angles.

We can now extend this approach to two-material targets in which one of the materials is arranged in spherical grains or pores; above, we merely cited the result \eqref{eq:TwoMaterialStructureFactorIntegrated}. The expectation value of the total scattering length is now
\begin{align}\begin{split}
    \left\langle b_{\rm tot}(\mathbf{q}_T) \right\rangle &= \int\limits_{R_{\rm gas}} \mathcal{S}_{\rm gas}(\mathbf{q}_T) e^{i\mathbf{q}_T\cdot\mathbf{r}} d^3\mathbf{r} + \int\limits_{R_{\rm solid}} \mathcal{S}_{\rm solid}(\mathbf{q}_T) e^{i\mathbf{q}_T\cdot\mathbf{r}} d^3\mathbf{r} \\
    &= \int\limits_{R_{\rm gas}} (\mathcal{S}_{\rm gas}(\mathbf{q}_T) - \mathcal{S}_{\rm solid}(\mathbf{q}_T)) e^{i\mathbf{q}_T\cdot\mathbf{r}} d^3\mathbf{r} + \int\limits_{R_{\rm total}} \mathcal{S}_{\rm solid}(\mathbf{q}_T) e^{i\mathbf{q}_T\cdot\mathbf{r}} d^3\mathbf{r}.
\end{split}\end{align}
As discussed above, the last term is suppressed as $\exp(-q_T^2\Delta r_\perp^2/2)$ for neutrons with gaussian transverse profile of width $\Delta r_\perp$. We assume throughout this work that this is renders it irrelevant for all momentum transfers we wish to observe; the potentially significant effects of this term through multiple scattering events are the subject of Appendix \ref{app:MultipleScattering}. Then
\begin{align}
    \left\langle b_{\rm tot}(\mathbf{q}_T) \right\rangle &= \int\limits_{R_{\rm gas}} (\mathcal{S}_{\rm gas}(\mathbf{q}_T) - \mathcal{S}_{\rm solid}(\mathbf{q}_T)) e^{i\mathbf{q}_T\cdot\mathbf{r}} d^3\mathbf{r}.
\end{align}

The variance in the total scattering length can be broken into three contributions: the variance of the total scattering length of the gas, the variance of the total scattering length of the solid, and the covariance of those two scattering lengths. We assume, as usual, that the first two are given simply by the sums of the squares of the atomic scattering lengths; the third is discussed in Appendix \ref{app:AtomicInteractions}, where we show that it should be negligible. We can then immediately write down the structure factor of the two-material target:
\begin{align}
    S(q_T) &\approx  \frac{12\pi\overline{R}^3}{9+2(q_T\overline{R})^4} \left( \frac{f\left|\Delta\mathcal{S}\right|^2}{fn_g|b_g(\mathbf{q}_T)|^2 + (1-f)\sum_j n_{s,j}|b_{s,j}(\mathbf{q}_T)|^2} \right) + 1. \label{eq:TwoMaterialStructureFactorImproved}
\end{align}
This looks identical to the form we obtained previously \eqref{eq:TwoMaterialStructureFunctionUnimproved}, but it should now be understood that the SLDs in the numerator include only the coherent scattering lengths, while the scattering lengths in the denominator are the total scattering lengths (which may be quite different for non-noble elements).

\subsection{Structure Factors of Nanotube Forests}\label{subapp:CNTStructureFunction}

There is one other geometry considered in this work: a collection of tubes, such as that of a carbon nanotube forest. We will therefore be interested in calculating the structure factor for this geometry as well. We will calculate it for the single-material case; generalizing to the (more applicable) two-material case can be straightforwardly done as above.

Consider an array of long, thin tubes approximately aligned with the neutron beam axis $z$. Let one tube have length $L$ and a circular cross-section of radius $R$, and let its center line be described by the function $x(z)\hat{\mathbf{x}}+y(z)\hat{\mathbf{y}}$. If we assume that the tubes are, on average, radially symmetric about $z$, we can take $x$ to be momentum transfer direction without loss of generality, in which case the total coherent scattering length from one tube is approximately
\begin{align}
    \left\langle b_{\rm tube}(\mathbf{q}_T) \right\rangle &= \iiint dz\ r dr\ d\varphi\ ne^{iq_T(x(z)+r\sin\varphi)}b_c(\mathbf{q}_T).
\end{align}
Note that the effects of the tube's transverse size are completely separable from the effects of the variation in the centerline position:
\begin{align}\begin{split}
    \left\langle b_{\rm tube}(\mathbf{q}_T) \right\rangle &= n b_c(\mathbf{q}_T) \iint r dr\ d\varphi\ e^{iq_T r\sin\varphi} \int dz\ e^{iq_Tx(z)}.
\end{split}\end{align}
The first factor, from the tube cross-section, is analytically expressible in terms of a Bessel function:
\begin{align}
    \iint r dr\ d\varphi\ e^{iq_T r\sin\varphi} &= \frac{2\pi R}{q_T} J_1\left(q_TR\right).
\end{align}
The second factor, however, is strongly dependent on the details of the tube shape. In general, this must be calculated numerically for a chosen target geometry. For the purpose of obtaining an estimate of the sensitivities we can achieve, however, it is helpful to compute an approximate result for it.

If the variation in $x(z)$ is large compared to the inverse momentum transfers considered, we can approximate the center line integral using the stationary phase approximation. In this case,
\begin{align}
    \int dz\ e^{iq_Tx(z)} \approx \sum\limits_{z_0\in\Sigma} e^{iq_Tx(z_0)}e^{i\frac{\pi}{4}\text{sign}(x''(z_0))}\sqrt{\frac{2\pi}{q_Tx''(z_0)}}
\end{align}
where $\Sigma$ is the set of points $z_0$ satisfying $x'(z_0)$. Intuitively, this corresponds to assuming that the rapidly varying phases along slanted portions of the tube average to zero, leaving only contributions from around the points where the tube is temporarily parallel to the $z$ axis.

At this point, it is helpful to switch to evaluating the norm squared of the coherent scattering length, $\left|\left\langle b_{\rm tube}(q_T)\right\rangle\right|^2$, which allows us to simplify these expressions further. Averaging the first factor squared over some variation in $R$ (just as we did for spheres) allows us to make the approximate replacement
\begin{align}
    \left| \iint r dr\ d\varphi\ e^{iq_T r\sin\varphi} \right|^2 &\approx \frac{(\pi R^2)^2}{1+0.7(q_TR)^3}.
\end{align}
While not exact, this expression is convenient and will be good enough for order-of-magnitude sensitivity estimate we wish to obtain.

Since the $x$ coordinates of the points $z_0$ are essentially random on the scale of $q_T^{-1}$, the contributions from each point add in quadrature (i.e. incoherently) even for the coherent scattering contribution. (This is analogous to our assumption of spheres summing incoherently above, with similar caveats about small $q_T$.) The second factor then simplifies to 
\begin{align}
    \left|\int dz\ e^{iq_Tx(z)}\right|^2 \approx \sum\limits_{z_0\in\Sigma} \frac{2\pi}{q_Tx''(z_0)}.
\end{align}
This is still a somewhat awkward expression, so it is helpful to rewrite it in terms of more intuitive variables. If we let $\lambda$ and $A$ be the typical wavelength and amplitude of the tube's center line undulation, the integral can be approximated, very roughly, by
\begin{align}
    \left|\int dz\ e^{iq_Tx(z)}\right|^2 \sim \frac{L\lambda}{q_TA}.
\end{align}
Combining these two approximations, we have a total coherent scattering length per tube of
\begin{align}
    \left|\left\langle b_{\rm tube}(q_T)\right\rangle\right|^2 \sim \frac{(\pi R^2)^2}{1+0.7(q_TR)^3}\frac{L\lambda}{q_TA}(nb_c(q_T))^2.
\end{align}

If carbon nanotubes' atomic structure were highly irregular, the incoherent scattering contribution would be given by a sum of $|b_n(\mathbf{q}_T)|^2$ over all of the atoms in the tube. For a forest of tubes, we assume, just as for the sphere case, that the individual tubes' phases are uncorrelated. Then the structure factor of the full forest would be the same as that of a single tube,
\begin{align}
    S(\mathbf{q}_T) &\sim \frac{\pi n R^2\lambda}{1+0.7(q_TR)^3} \frac{1}{q_TA}\frac{|b_c(\mathbf{q}_T)|^2}{|b_n(\mathbf{q}_T)|^2} + 1.
\end{align}
In fact, carbon nanotubes are likely to have highly regular atomic structures, which may significantly modify this result. Accounting for this is beyond the scope of this work (and is likely to require measurements of particular nanotube forests), but we note that, as usual, this should not meaningfully affect our conclusions unless it changes the dominant angular scale of scattering from the forest.

\section{Separating Scattering Contributions with Two-Material Targets}\label{app:TwoMaterialSeparation}

Many of the targets that we consider in this work combine a solid responsible for the target's granular structure with a noble liquid or gas within that structure. This significantly complicates the task of separating features of the scattering distribution that are the result of the target structure from any that are the result of a new force. We will not attempt to illustrate this process explicitly (though a crude approximation is presented in Appendix \ref{subapp:TwoMaterialStatistics}), as we did for the single-material case above; in practice this will need to be done numerically. In this section, we demonstrate merely that this is possible: that is, that enough parameters can be measured to constrain every pertinent degree of freedom, including the two (coupling and mass) for the new force.

The generic neutron scattering distribution from such a solid-gas combination is (see Appendix \ref{app:NeutronInteractions})
\begin{align}\begin{split}
    \frac{d\sigma_{n,2}}{d\theta} = &\left( \vphantom{\left|\frac{1}{(1)^1}\right|^1} \left|B_0(q_T) + b_{0}\left(1+\kappa_{\rm EM}f(q_T)+\kappa_{\rm new}\frac{1}{1+(q_T/\mu)^2}\right)W(q_T)\right|^2 \right. \\ 
    &+ \left. \vphantom{\left|\frac{1}{(1)^1}\right|^1} \left|\left(\mathbf{B}_S(q_T) - ib_{S,0}(1-f(q_T))\cot\theta\  W(q_T)\hat{\mathbf{n}}\right)\cdot\boldsymbol{\sigma}\right|^2\right) \sin\theta \label{eq:TwoMaterialScatteringDistribution}
\end{split}\end{align}
where 
\begin{align}
    W(q_T) &= \sum\limits_{j=1}^{N_{\rm gas}} e^{i\mathbf{q}_T\cdot\mathbf{r}_j}
\end{align}
is the sum of the scattering phases from the atoms in the gas (such that, if the gas remained in place but there was no solid, the structure factor would be $S'(q_T) = |W(q_T)|^2/N_{\rm gas}$),
\begin{align}
    b_{S,0} = \frac{g_n Z e^2}{8\pi m_n}
\end{align}
is the magnitude of the gas's Schwinger scattering length, $B_0(q_T)$ is an unknown total spin-independent scattering length for the solid (already summed over all the atoms) and $\mathbf{B}_S(q_T)\cdot\boldsymbol{\sigma}$ is the analogous spin-dependent total scattering length, which sums coherently with the spin-dependent Schwinger scattering length of the noble gas. We do not assume that either $B_0$ or $B_S$ takes any particular form or value, as they are likely to both depend heavily on the complicated interactions within the solid. We do, however, assume that the electronic structures of the noble gas atoms are not significantly affected by the presence of the solid (or by the likely high pressure of the gas); this is confirmed in Appendix \ref{app:AtomicInteractions}.

For simplicity, we begin by assuming that the spin-dependent terms are negligible; we reintroduce them below. In this simplified case, we have
\begin{align}\begin{split}
    \frac{d\sigma_{n,2}}{d\theta} = &\left( \vphantom{\left|\frac{1}{(1)^1}\right|^1} \left|B_{0}(q_T)\right|^2 + b_{0}\left(1+\kappa_{\rm EM}f(q_T)+\kappa_{\rm new}\frac{1}{1+(q_T/\mu)^2}\right)\left|W(q_T)\right|^2 \right. \\ 
    &+ \left. \vphantom{\left|\frac{1}{(1)^1}\right|^1} 2\text{Re}\left( B_{0}(q_T) W(q_T)^*\right) b_{0}\left(1+\kappa_{\rm EM}f(q_T)+\kappa_{\rm new}\frac{1}{1+(q_T/\mu)^2}\right) \right) \sin\theta.
\end{split}\end{align}
We can separate the various parameters in this expression by taking a series of measurements. In particular, consider measurements performed using $P$ different particles (neutrons, X-rays, electrons, etc.; we will generally restrict to $P=2$ in this work, assuming neutron and X-ray scattering) and $Q$ different noble gases, with a shared solid. Then we can measure the following set of scattering distributions:
\begin{itemize}
    \item $PQ$ distributions for each of the $P$ particles scattered from the solid filled with each of the $Q$ noble elements;
    \item $P$ distributions scattering each particle from the solid alone; and
    \item $(P-1)Q$ distributions for each of the $P$ particles, except neutrons, scattered from the $Q$ noble elements alone.
\end{itemize}
We do not include neutron scattering from the gas alone here since, as we discussed for the single-material case in Section \ref{sub:SingleMaterialSeparation}, it suffers from poor statistics at small angles; it is, however, still potentially useful at large angles, as we discuss below. We can thus take a total of
\begin{align}
    N_{\rm measure} = 2PQ + P - Q
\end{align}
independent measurements.

This should be compare to the number of degrees of freedom in the scattering distributions above:
\begin{itemize}
    \item One sum of phase factors $W(q_T)$. This sum is complex, but there is an arbitrary overall phase corresponding to the choice of origin (i.e. only the relative phases between $W$ and each $B$ appears in scattering distributions) so we can take $W(q_T)$ to be real without loss of generality.
    \item $Q$ real atomic form factors $f(q_T)$ for the noble elements.
    \item $Q$ real neutron nuclear scattering lengths for the noble elements.
    \item $P$ complex solid scattering lengths ($B_{n,0}(q_T)$, $B_{X,0}(q_T)$, etc.).
    \item One real, angle-independent electromagnetic scattering length scale for neutrons (i.e. $b_{0}\kappa_{EM}$).
    \item $P-1$ angle-independent scattering length scales for each of the other particles. We will assume that each of these particles' scattering distributions is fully described by some combination of this length scale, the atomic form factor, and the scattering angle, as is the case for X-ray scattering:
    \begin{align}
        \left.\frac{d\sigma}{d\Omega}\right|_X &= \left(\frac{Ze^2}{4\pi m_e}\right)^2 \left(f(q_T) - \frac{Zm_e}{m_{\rm nuc}}\right)^2 \frac{1+\cos^2 2\theta}{2}.
    \end{align}
    Generalizing this discussion to more complicated scattering lengths is straightforward, but we will find below that neutron and X-ray scattering alone should be sufficient for our purposes.
    \item Two angle-independent parameters describing the new force: the coupling $g$ and the mass $\mu$.
\end{itemize}

We thus find a total of
\begin{align}
    N_{\rm{dof,}\theta} = 2P + Q + 1
\end{align}
degrees of freedom per angular bin, plus the $P+Q+2$ angle-independent parameters. Of these, the $Q$ neutron scattering lengths can be measured using the otherwise-unhelpful gas-only neutron scattering measurements discussed above, as they are both inherently angle-independent and not suppressed by other effects at large angles.

Assuming we have at least $P+2$ bins per measurement, we can conservatively treat the remaining angle-independent parameter as one additional angle-dependent degree of freedom. Note that at least 2 of these bins must be at small angles, since the new force parameters are suppressed at larger momentum transfers; this minimum may be increased if any of the scattered particles have distributions that are similarly peaked. We will assume that there are sufficient bins for this to be the case; see Appendix \ref{subapp:TargetGeometry}.

In order to detect a new force, we need $N_{\rm measure} \geq N_{\rm{dof,}\theta} + 1$, with the $1$ accounting for these angle-independent parameters. This can be achieved with at least 2 particles and at least 2 noble gases. The most promising candidates are likely neutron and X-ray scattering from xenon and argon (see Section \ref{sub:TwoMaterialOptions}), though there may be circumstances in which other options are preferable.

We now return to the full scattering distribution \eqref{eq:TwoMaterialScatteringDistribution}, including spin-dependent scattering. This leads to three new terms, which can be handled in different ways:
\begin{itemize}
    \item After averaging over neutron polarizations, $\left|\mathbf{B}_S(q_T)\cdot\boldsymbol{\sigma}\right|^2$ acts simply as a correction to the pure-solid $|B_0(q_T)|^2$ term; it can therefore be absorbed into the spin-independent description by a replacement of $B_0$ with $B_0'$ such that $|B_0'(q_T)|^2 = |B_0(q_T)|^2 + \overline{\left|\mathbf{B}_S(q_T)\cdot\boldsymbol{\sigma}\right|^2}$ while $2\text{Re}\left( B_{0}(q_T) W(q_T)^*\right) = 2\text{Re}\left( B_{0}'(q_T) W(q_T)^*\right)$. This term therefore does not require any modification to the analysis above.
    \item  $\left|ib_{S,0}(1-f(q_T))\cot\theta~ W(q_T)\hat{\mathbf{n}}\cdot\boldsymbol{\sigma}\right|^2$ is the corresponding gas-only term; it is suppressed relative to spin-independent scattering both by two powers of the small Schwinger scattering length $b_{S,0}$, typically of order $10^{-3}$ times the nuclear scattering length $b_{0}$, generally rendering it irrelevant at our systematic error target. Even if this is insufficient, note that, at small angles, it is additionally suppressed by $(1-f(q_T))^2\cot^2\theta \propto \theta^2$.
    \item This leaves $2\text{Re}(i\mathbf{B}_S(q_T))\cdot\boldsymbol{\sigma}b_{S,0}(1-f(q_T))\cot\theta~ W(q_T)\hat{\mathbf{n}}\cdot\boldsymbol{\sigma}$, the cross-term from these two effects. This is suppressed at small angles by a factor of $\theta b_{S,0}/b_{0}$, or approximately $10^{-5}$ for the smallest angles we consider in this work. There may be additional suppression relative to the spin-independent scattering contributions from the magnitude of $\mathbf{B}_S(q_T)$, though this is uncertain: the magnitude of spin-dependent scattering from individual, non-noble atoms is generally comparable to that of spin-independent scattering (see Appendix \ref{app:NeutronInteractions}), but the spin dependence may limit coherence at small angles. 
    
    One additional source of suppression is created by the target structure. If the atoms in the solid had random spins and angular momenta, their spin-dependent contributions would never sum coherently; in practice, there will be some correlations between nearby atoms due to interactions within the solid, leading to coherent scattering up to the correlation length scale $\xi$. Nonetheless, if $\xi < R$, coherent spin-dependent scattering will still be suppressed relative to spin-independent scattering when $q_T\xi \lesssim 1 \lesssim q_TR$, which may be sufficient to make this cross-term insignificant at small angles. Otherwise, this term produces one additional degree of freedom per angular bin.
\end{itemize}

If we assume, optimistically, that the cross-term is insignificant at small angles, none of the spin-independent discussion above needs to be modified and we can still separate all scattering contributions with 2 scattered particles and 2 noble gases.

If not, however, we now have one additional degree of freedom; requiring $N_{\rm measure} \geq N'_{\rm{dof,}\theta} + 1$ then requires either a third scattered particle (e.g. electrons) or, likely more simply, a third noble gas. Thus, even in this pessimistic scenario, new force scattering can be separated from generic solid backgrounds, though doing so requires a combination of several measurements.

\section{Multiple Scattering Events}\label{app:MultipleScattering}

Throughout most of this work, we have worked in terms of scattering probabilities. While this is a good approximation when scattering events are rare, it is not sufficient at the level of precision we require. In fact, the various scattering probabilities estimated throughout this work should be interpreted as expected numbers of scatterings per neutron, with the scattering count per neutron Poisson distributed. This has an unfortunate consequence: a neutron that is observed to have scattered with some momentum transfer $\mathbf{q}_T$ may in fact have scattered two or more times with momentum transfers that summed to $\mathbf{q}_T$. This becomes especially problematic when low-angle scattering is coherently enhanced, as this coherent enhancement will extend not only to the small angles we want to measure, but also to even smaller angles. This results in the observed small-angle scattering distribution being enhanced more than expected, due to combinations of even-smaller angle scatterings, which could simulate a new force signal. In this appendix, we estimate the magnitude of this effect and discuss how one can account for it.

\begin{figure}[t]
    \centering
    \includegraphics[width=0.7\linewidth]{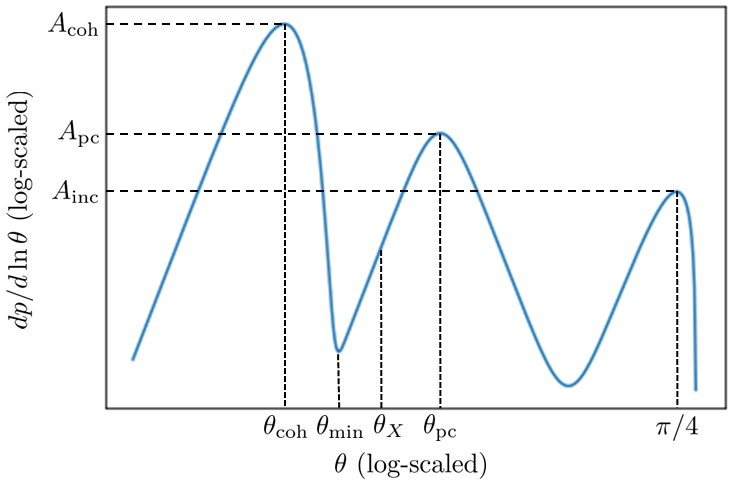}
    \caption{A schematic illustration of the scattering distribution \eqref{eq:ScatteringDistributionSchematic}. The locations and values of local extrema indicated on the plot are approximate; we have omitted various $\mathcal{O}(1)$ factors here for simplicity. Also marked is the minimum angle $\theta_{X}$ accessible via X-ray scattering (conservatively assumed to be larger than $\theta_{\rm min}$).}
    \label{fig:MultipleScatteringExample}
\end{figure}

For concreteness, we continue to use the collection of spherical grains model that we use throughout much of this work; we begin with the single-material case for simplicity. It is helpful to rewrite the resulting scattering distribution (given by \eqref{eq:CloudScatteringDistributionAveraged}, plus the fully-coherent contribution analogous to \eqref{eq:FullyCoherentDistribution}) in terms of the logarithm of the scattering angle:
\begin{align}
    \frac{dp}{d\ln\theta} &= \left( fnL \left(\frac{12\pi}{9+2(q_TR)^4} n R^3 + 1\right) + 8\pi\left( fnL\Delta r_\perp\right)^2 e^{-\mathbf{q}_T^2\Delta r_\perp^2/2} \right) \left|b(q_T)\right|^2 2\pi\theta\sin2\theta
\end{align}
where $f$ is the fraction of the target volume occupied by the grains, each with atomic number density $n$ and average radius $R$, and $L$ is the thickness of the target. We can rewrite this as the sum of three terms, which dominate at different angles:
\begin{align}
    \frac{dp}{d\ln\theta} &= A_{\rm inc}\theta\sin2\theta + A_{\rm pc}\frac{(\theta/\theta_{\rm pc})^2}{1+(\theta/\theta_{\rm pc})^4} + A_{\rm coh}\left(\frac{\theta}{\theta_{\rm coh}}\right)^2e^{-(\theta/\theta_{\rm coh})^2} \label{eq:ScatteringDistributionSchematic}
\end{align}
where $A_{\rm inc}$, $A_{\rm pc}$ and $A_{\rm coh}$ are constant prefactors for incoherent, partially-coherent (i.e. coherent over grains but not over the full target), and fully-coherent scattering, respectively, and $\theta_{\rm pc}$ and $\theta_{\rm coh}$ are the angular scales of partial and full coherence (i.e. the angle above which scattering from individual spheres becomes incoherent and the angle above which the neutron size no longer matters). Here we use $q_T \propto \theta$ rather than the exact form $q_T\propto\sqrt{2-2\cos(2\theta)} = 2\sin(\theta)$ for simplicity, since we are interested only in small-angle effects in this section. The resulting scattering distribution is illustrated schematically in Figure \ref{fig:MultipleScatteringExample}.

In a typical experiment, $\theta_{\rm pc}$ is likely to be comparable to the smallest scattering angle at which neutrons can be observed, since the entire purpose of employing structured targets is to maximize small-angle scattering. The question, then, is whether neutrons observed to scatter by an angle around $\theta_{\rm pc}$ are sourced by a single significant scattering of order that angle, or by a series of smaller-angle scatterings that summed to it. In particular, any experiment must ensure that the observed scattering distribution is not significantly affected by scatterings at angles too small to study using X-ray scattering, since there is no way to determine the structure factor at such small momentum transfers.

To establish conditions for this to be the case, divide angles below $\theta_{\rm pc}$ into two regimes: those large enough to be measured using X-ray scattering ($\theta>\theta_{X}$), and those below that threshold. At least some of the former will certainly affect the final observed scattering distribution, but their effects can be numerically predicted once the structure factor at these angles is determined with X-ray scattering. The effects of angles below $\theta_X$, on the other hand, should be insignificant as long as $\theta_X/\theta_{\rm pc}$ is sufficiently small, as we show below.

Now, further subdivide the angles below $\theta_X$ further into two ranges: those above the local minimum in the scattering distribution at $\theta_{\rm min}$ (see Figure \ref{fig:MultipleScatteringExample}), and those below it. We begin with the former. The number of scatterings per neutron from the $\theta_{\rm min}<\theta<\theta_X$ range is upper bounded by a Poisson distribution with expected value $\alpha = dp/d\ln\theta|_{\theta=\theta_X}\ln(\theta_X/\theta_{\rm min})$, with each scattering by an angle upper bounded by $\theta_X$ by assumption; we will conservatively use both of these upper bounds.

Under these assumptions, the probability of $N$ scatterings from this range is
\begin{align}
    p(N) &= \frac{e^{-\alpha}\alpha^N}{N!}.
\end{align}
The minimum number of scattering events from this angular range needed to create an observed scattering by at least $\theta_{\rm pc}$ is $N_{\rm min} = \lceil \theta_{\rm pc}/\theta_X \rceil$, though contributions from larger numbers of scatterings may dominate when $N_{\rm min} \gg 1$ or $N_{\rm min} - \theta_{\rm pc}/\theta_X \ll 1$ as it is unlikely for all of the small-angle scatterings to be in the same direction. Even this weak lower bound is likely sufficient for our purposes, however: the probability of at least $N_{\rm min}$ scatterings is at most \cite{PoissonTailBound}
\begin{align}
    p(N \geq N_{\rm min}) \leq \frac{(e\alpha)^{N_{\rm min}}e^{-\alpha}}{(N_{\rm min})^{N_{\rm min}}}
\end{align}
and we have
\begin{align}
    \frac{dp}{d\ln\theta} \propto \theta^2
\end{align}
in this angle range, so this bound becomes
\begin{align}
    p(N \geq N_{\rm min}) \leq \frac{\left(eA_{\rm pc}\frac{\theta_{X}^2}{\theta_{\rm pc}^2}\right)^{N_{\rm min}}e^{-A_{\rm pc}\theta_X^2/\theta_{\rm pc}^2}}{(N_{\rm min})^{N_{\rm min}}}.
\end{align}
Taking a typical value of $A_{\rm pc} \sim 0.1$ to balance maximizing statistics with not being swamped by multiple scatterings at angles around the peak, and assuming $\ln(\theta_X/\theta_{\rm min}) \sim 1$, this gives
\begin{align}
    p(N \geq 3) \lesssim 10^{-6}
\end{align}
so we should only need to perform X-ray scattering down to an angle of around one third (perhaps one fourth) of the minimal neutron scattering angle in order to be able to accurately numerically predict the effects of multiple scatterings in the $\theta_{\rm min}<\theta<\theta_X$ range. As we discuss in Appendix \ref{subapp:XRayBeams}, this should not be particularly difficult.

We can similarly bound the effects of scattering at angles below $\theta_{\rm min}$, which are enhanced by total coherence over the neutron's transverse extent. In this case, the minimum number of scatterings to reach the first angular bin outside of the beam is $\lceil\theta_{\rm pc}/\theta_{\rm coh}\rceil$, which should be at least of order 10 for realistic experiments. The fully-coherent scattering peak height $A_{\rm coh}$ can be estimated using \eqref{eq:FullyCoherentDistribution}; relative to the partially-coherent peak, it is
\begin{align}
    \frac{A_{\rm coh}}{A_{\rm pc}} \sim \frac{fL}{\sqrt{2}\pi R}.
\end{align}
This is likely to give $A_{\rm coh} \gg 1$, so we can instead estimate the result of these fully coherent scatters by the sum of $A_{\rm coh}$ random 2-dimensional vectors of length $\theta_{\rm coh}$. This, in turn, is well approximated by a 2-dimensional Gaussian, which gives a probability for the fully coherent scatters to sum to an observable angle of
\begin{align}\begin{split}
    p(\theta_{\rm coh,total} \geq \theta_{\rm pc}) &\sim \exp\left(-\frac{3(nb\Delta r_\perp\lambda_0)^2}{4\sqrt{2}A_{\rm pc}^2}\right) \\
    &\sim \exp\left( -53 \left(\frac{n}{10\text{ nm}^{-3}}\right)^2 \left(\frac{b}{10\text{ fm}}\right)^2 \left(\frac{\Delta r_\perp}{10\text{ }\mu\text{m}}\right)^2 \left(\frac{\lambda_0}{10\text{ \AA}}\right)^2 \left(\frac{0.1}{A_{\rm pc}}\right)^2 \right) \label{eq:MultipleScatteringProbability}
\end{split}\end{align}
where $\lambda_0 = 2\pi/q_0$ is the incident neutron wavelength. This probability needs to be smaller than $10^{-6}$ in order to reach our desired control over systematic errors (see Section \ref{sub:SystematicErrors}). Assuming near-liquid densities, this is always satisfied for argon-36 (see Table \ref{tab:MaterialSLDs}), but may not be the case for xenon-136 if the neutron transverse size is less than approximately $7~\mu$m (for a wavelength of $0.6$ nm); previous experiments have measured values from roughly $1-100~\mu$m \cite{TransvereSizeMajkrzak,TransverseSizeBerk,TransverseSizeEbrahimi,TransverseSizeTreimer}.

Our discussion to this point has assumed scattering from a single material. However, as we discuss in the main text, it is likely that a realistic experiment would combine two materials: a solid lattice to form the granular structure, and a noble liquid or gas within it whose scattering distribution has simpler electromagnetic contributions. This leads to a suppression of the partially-coherent scattering peak by the fractional difference between the two materials' scattering length densities since partial coherence depends on the SLD contrast (see \eqref{eq:TwoMaterialStructureFactorIntegrated}) and an enhancement of the fully-coherent and incoherent scattering peaks since both materials contribute to them. While this does not affect the multiple scattering contribution of the intermediate angular range, $\theta_{\rm min}<\theta<\theta_{X}$, it does enhance the effect of small angles $\theta<\theta_{\rm min}$. This may prohibit certain material combinations and porosities, depending on the value of $\Delta r_\perp$. In particular, for two-material targets, the argument of the exponent in \eqref{eq:MultipleScatteringProbability} acquires a factor $f(\mathcal{S}_{\rm gas} - \mathcal{S}_{\rm solid})/(f\mathcal{S}_{\rm gas} + (1-f)\mathcal{S}_{\rm solid})$, accounting for the different scattering length densities relevant for fully- and partially-coherent scattering.

Finally, we note that the frequency of multiple scattering events can be further increased by short-range order, which tends to enhance scattering at momentum transfers comparable to the inverse length scale of that order. This is unlikely to be significant for short-range order within amorphous solids, as the expected length scales should be shorter than those of the target grains, leading to additional peaks at larger rather than smaller angles. Any correlations in grain positions, however, may be more consequential, as the necessarily large associated length scales would enhance scattering at angles below $\theta_{\rm pc}$. The magnitude of this effect will depend on the detailed structure of a particular target, but it appears unlikely that it would significantly affect the discussion above except in highly ordered targets. We will therefore not consider this effect further in this work.

\section{Thermal Effects}\label{app:ThermalEffects}

The observed neutron scattering distribution depends on the temperature of the target in several ways. At high temperatures, the velocity of atoms in the target leads to a significant difference between the center-of-mass velocity of the scattering event and the lab-frame velocity of the neutron, creating an apparent enhancement of the cross-section at low neutron energies \cite{NeutronScatteringIntroBerk, ENDFPaper}. Additionally, higher temperatures lead to larger populations of excited states of atoms, which can have different neutron scattering lengths due to electromagnetic effects. 

As discussed in Appendix \ref{app:NeutronInteractions}, nuclear scattering of neutrons is angle-independent in the center-of-mass frame of the scattering event. When considering scattering from a bulk target, however, it is more convenient to work in the laboratory frame, in which individual atoms within the target have a Maxwellian velocity distribution. Since the target atoms are no lighter (and typically much heavier) than the incident neutron, this leads to large differences in the center-of-mass velocity of the neutron. In particular, the apparent scattering cross-section of neutrons slower than the atoms in the target is enhanced because the scattering becomes the result of atoms striking an essentially stationary neutron. This leads to an enhancement of the low-energy cross section as $\sigma_{\rm lab} \propto 1/v$, with $v$ the neutron velocity, whenever this velocity is lower than the thermal velocity of atoms in the target, as well as a modification of the angular distribution of neutrons after scattering in the lab frame.

As we discuss in Section \ref{subapp:NeutronBeams}, typical neutron wavelengths at beamlines that may be appropriate for our purposes are around 0.4-0.8 nm. Targets with atomic weight $A$ then have thermal velocities equal to the neutron speed at temperature
\begin{align}
    T \sim A\left(\frac{\lambda}{0.4\text{ nm}}\right)^{-2}\times 10\text{ K}.
\end{align}

Most of the targets we consider in this work have $A \gg 1$, and can therefore likely be cooled below this temperature with little difficulty. In this case, frame differences will only lead to small corrections in the observed scattering distribution, which should not meaningfully affect final sensitivities. If this is not the case---for example if using helium, or for argon at larger neutron wavelengths--low velocity enhancement may become more significant. While this should not prevent measurements of the sort we describe in this work from being performed, it may have a more significant effect on the final sensitivity, which we do not attempt to estimate.

The other effect of high target temperatures on cold neutron scattering is the enhancement of populations of excited atomic states. In particular, our modeling of neutron scattering from noble elements throughout this work assumes their ground state electron configuration, with zero total spin or orbital angular momentum. These assumptions no longer hold for excited states of the atoms.

Fortunately, the excitation energies of noble elements are too high for this to be a significant effect: the lowest excited state of xenon is at an energy of $1.3$ eV \cite{NISTSpectroscopicTable, NIST_ASD, XenonSpectroscopicData}, corresponding to an excited state population fraction of order $10^{-22}$ even at room temperature; the excited states of other noble elements are even more suppressed. This is far below the $10^{-6}$ benchmark that we use throughout this work, so any effects of excited electronic states in the noble gas should be negligible. Any thermal effects in the solid component of a target should be accounted for when separating scattering contributions, so they do not affect our analysis further; see Appendix \ref{app:TwoMaterialSeparation}.

\section{Atomic Interaction Effects}\label{app:AtomicInteractions}

In this appendix, we discuss the effects of interactions between atoms on structured targets' scattering distributions. We consider both interactions between atoms within the noble gas (or liquid; while we have ignored the distinction in most of this work, the high density of the noble fluid will be relevant here) and between those noble atoms and the atoms in the solid. Interactions among atoms within the solid are not relevant for our purposes since we are generally impartial to the total scattering length of the solid.

\subsection{Interactions at the Grain Surface} \label{subapp:SurfaceInteractions}

There are two ways in which interactions at the surface between the solid component of the target and the liquid or gas within it can affect the observed neutron scattering distribution. First, they can modify the distribution of the noble atoms within their grain, modifying the structure factor. For example, an attractive potential near the surface could increase the density of the noble gas near the surface, whereas our estimates in Appendix \ref{app:StructuredScattering} assumed a uniform distribution of the gas within each grain. Second, they can modify the electron orbitals of the noble atoms, potentially changing the electromagnetic scattering length. In particular, we have assumed throughout this work that the noble atoms have zero net electron spin and angular momentum, but this may cease to be the case in the presence of electromagnetic fields induced by the solid.

The former effect does not significantly affect us, so long as it does not lead to an order-unity change in the structure factor, since it is independent of the scattered particle and thus will be accounted for when combining measurements, as described in Appendix \ref{app:TwoMaterialSeparation}. This condition should certainly be satisfied for typical materials, considering both the weakness of Van der Waals interactions and that optimal sensitivity is generally achieved for near-liquid densities of the noble element, in which case there is little room for atoms to deviate from uniform packing.

Significant modification of the noble atoms' electronic structure could be more difficult to handle, however. In the presence of an inhomogeneous magnetic field, the atomic Hamiltonian should acquire off-diagonal terms of order $|\boldsymbol{\mu}||\Delta\mathbf{B}|$, with $\mu \sim \mu_B$ the characteristic magnetic moment of atomic electrons and $\Delta\mathbf{B}$ the variation in the magnetic field over the extent of the atom; we ignore any numerical factors here, given the considerable uncertainties in the discussion below. This perturbation then leads to a mixing of the ground state by a fraction
\begin{align}
    \frac{\Delta H}{\Delta E} \sim \frac{\mu_B |\Delta\mathbf{B}|}{\Delta E}
\end{align}
of the excited state, with $\Delta E$ the energy of that state, and thus gives a correction to the neutron scattering length of order
\begin{align}
    |\Delta b| \sim \frac{|g_n| e^2}{8\pi m_e}\frac{\mu_B |\Delta\mathbf{B}|}{\Delta E};
\end{align}
see Appendix \ref{app:NeutronInteractions}.

The maximum magnetic field resulting from a single atom on the solid surface is of order
\begin{align}
    |\mathbf{B}|_{\rm max} \sim \frac{\mu_B}{R_{\rm atom}^3},
\end{align}
with $R_{\rm atom} \sim 0.1$ nm the characteristic size of the atoms. The neutron scattering length of such an atom is then modified by approximately
\begin{align}
    |\Delta b| \sim \frac{|g_n| e^2 \mu_B^2}{8\pi m_e R_{\rm atom}^3 \Delta E} \lesssim 3\times10^{-3} \text{ fm},
\end{align}
where we use the lowest excited state energy of a noble element ($1.3$ eV for xenon \cite{NISTSpectroscopicTable, NIST_ASD, XenonSpectroscopicData}) for this upper bound. This should be compared to nuclear scattering lengths of order 10 fm; see Table \ref{tab:MaterialSLDs}.

Such a correction to every atom would be well above our systematic error target of one part in $10^6$, if it were not further suppressed by two additional effects. First, the rapid decay of dipole magnetic fields with distance from the dipole means that only an order-unity number of atoms near the surface dipole will be affected; only a fraction of order $R_{\rm atom}/R_{\rm grain}$ of the atoms in a grain of radius $R_{\rm grain}$ are therefore affected. Second, since we are not considering ferromagnetic targets, we expect the magnetic field directions, and thus the signs of the neutron scattering length changes for a given momentum transfer, to vary within (as well as among) grains. Assuming that any correlations in the alignment of surface dipoles occur on length scales $\xi < R_{\rm grain}$, we expect a total of approximately $(R_{\rm grain}/\xi)^2$ independently chosen directions over the extent of the grain surface.

The average change in scattering length is then
\begin{align}
    \left|\overline{\Delta b}\right| \sim \frac{|g_n| e^2 \mu_B^2}{8\pi m_e R_{\rm atom}^3 \Delta E} \left(\frac{R_{\rm atom}}{R_{\rm grain}}\right) \frac{1}{\sqrt{(R_{\rm grain}/\xi)^2}} \lesssim \left(\frac{R_{\rm atom}\xi}{R_{\rm grain}^2}\right) 3\times10^{-3} \text{ fm},
\end{align}
which should be below our target of order $10^{-5}$ fm for all targets we consider.

Note that this average change in the scattering length is the correct quantity to consider (as opposed to, for example, its root mean square change) even for incoherent scattering: since the maximal change in the scattering length of one noble atom is much smaller than its nuclear scattering length, its non-negligible effect comes from the interference between nuclear and electromagnetic scattering (i.e. from the product of the nuclear and electromagnetic scattering lengths). This term is linear in the electromagnetic scattering length, and thus the change in the total scattering probability depends precisely on $\left|\overline{\Delta b}\right|$.

\subsection{Interactions Among Noble Atoms} \label{subapp:LiquidInteractions}

We now turn to the effects of interactions within the noble liquid or gas itself. Similarly to the surface interactions described above, the effects of interactions between the atoms in a fluid can affect the observed scattering distribution both by modifying the target's structure factor and by inducing changes in the electromagnetic scattering length of the atoms.

The high density of liquids leads to significant correlations between the positions of atoms and their neighbors, i.e. a non-trivial pair distribution function \cite{LiquidsBookAttard, LiquidsBookHansen, LiquidsBookSekerka, StructureOfLiquids, DynamicsOfLiquids, SlowNeutronsLiquids, BornCorrelations}. These correlations will then lead to a modification of the structure factor at inverse momentum transfers comparable to the atomic spacing in the liquid. Fortunately, these are not an issue for the measurements we propose: since this effect is independent of the scattered particle, it will automatically be accounted for when combining measurements (see Section \ref{sub:SingleMaterialSeparation}), and it should have little effect on achievable sensitivity since it occurs at much larger momentum transfers than the focus of this work.

Interactions between the atoms within the liquid may, in principle, also lead to changes in their electronic states, which would then modify the electromagnetic scattering lengths of the target atoms. Since we are considering noble elements, whose ground states have zero magnetic moment, there are no apparent mechanisms to induce significant changes in the atoms' electronic states. Thus this should not be a significant effect for liquids or gases. It is less clear whether spontaneous magnetization could appear in noble solids, but we leave consideration of this to future work.

\section{Instrument Parameters}\label{app:InstrumentParameters}

In this appendix, we summarize the key features of neutron and X-ray scattering instruments and describe the parameters of such instruments that we assume for our projections in the main text. We begin with an outline of small-angle neutron scattering instruments as a whole, before focusing on the properties of neutron sources and beams in particular, which are the limiting parameters for our proposal's reach. We then consider the analogous properties of X-ray beams, justifying our assumption in the main text that these contribute subdominant uncertainties.

\subsection{Target Geometry}\label{subapp:TargetGeometry}

A sketch of a simplified neutron scattering experiment layout is shown in Figure \ref{fig:ExperimentSummaryDiagram}. In this section, we consider a few aspects of this layout and their impact on the sensitivity of such an experiment to the new forces we consider.

The importance of the distance between the neutron source and the scattering target for collimation is discussed in the next section of this appendix. We begin instead with the distance beyond the target, between it and the neutron detectors, which is significant in comparison to three other length scales: the pixel size of the neutron detector, and the transverse and longitudinal target dimensions.

Modern neutron detectors can achieve resolutions approaching 1 $\mu$m \cite{PixelSizeFujiwara,PixelSizeHerrera,PixelSizeJakubekHoly,PixelSizeJakubekUher,PixelSizeKohler,PixelSizeLosko}, though this enormously exceeds our needs; as we discuss below, there is little benefit to resolutions significantly below the target's transverse size. The ratio of the pixel size to the distance between the target and the neutron detector sets the maximum angular resolution of the experiment. This angular resolution should certainly be no greater than the minimum scattering angle we wish to measure, and there is likely to be some benefit to an angular resolution a few times better, in order to better resolve the new force peak. Our projections below assume a minimum scattering angle of 3 mrad, requiring a detector distance of at least 300 times the pixel size.

The minimum useful pixel size, in turn, is set by the target's transverse length scale. Below this length scale, scattering events at the same angle can appear in different pixels if they occurred at different locations within the target. While this is not intrinsically problematic, there is little purpose to using much smaller pixels, since the angular distribution of the neutrons will be washed out by the target size at these scales. Note as well that, when the target transverse size is larger than the pixel size, it is the ratio of the detector distance to the former that sets the maximum resolution. Thus, the 10 cm$^2$ targets that we assume in our projections require a target to detector distance of approximately 10 m.

There is a similar effect from the target's longitudinal size (i.e. its depth), though it is likely to be sub-dominant for our purposes due to our focus on small-angle scattering. There are three other constraints on this depth, however. First, for a given set of target materials, the target depth sets the fraction of neutrons that are scattered into observable angles; as we discuss in Appendix \ref{app:MultipleScattering}, we assume that this is set to $0.1$, in order to maintain control over multiple scattering events. Second, as we noted in Section \ref{sub:TwoMaterialOptions}, excessively thin targets may be difficult to work with; combined with the maximum depth from multiple scattering, this may preclude certain material combinations.

The third effect of finite target depth is a suppression of fully coherent scattering at large angles. Our calculation of fully coherent scattering in Section \ref{app:StructuredScattering} assumed that the direction of momentum transfer was perpendicular to the neutron's incident direction, but this is only true in the limit of zero scattering angle. At larger angles, neutrons scattered from one end of the target become incoherent with those scattered from the other end, reducing the fully-coherent scattering contribution at these angles. This is completely insignificant for all parameters we consider, however, since full coherence is only significant at very small angles to begin with.

Finally, we consider the effects of the cell containing the target material(s), which we have heretofore ignored. The additional scattering contribution from the neutron beam passing through the walls of the cell must be separated in order to isolate the scattering distribution of the ideal gas within. However, the separation of contributions procedure discussed in Appendix \ref{app:TwoMaterialSeparation} is not immediately applicable to this issue because scattering from the ideal gas cannot be performed without the cell. Note, however, that this is only a concern if there is coherent scattering from the cell walls combined with the cell contents (i.e. the cross-term discussed in Appendix \ref{app:TwoMaterialSeparation}); otherwise the cell's scattering distribution can simply be measured separately and then subtracted out of the final distribution. Our projections assume that the cell walls lack any structure at length scales comparable to the inverse momentum transfer, such that this should be the case, leaving characterization of cell wall roughness to future work.

\subsection{Neutron Beam Parameters}\label{subapp:NeutronBeams}

Neutrons used in small-angle neutron scattering (SANS) experiments are typically produced in nuclear reactors (e.g. \cite{SANSInstrumentNIST, SANSInstrumentORNL, SANSInstrumentBERII, SANSInstrumentJRR3M, SANSInstrumentHANARO1, SANSInstrumentHANARO2}) or neutron generators (e.g. \cite{SANSInstrumentSINQ, SANSInstrumentESS}). Since such sources produce neutrons at much higher energies than desirable for SANS experiments, they are then cooled by passing through one or more cold moderators (e.g. water and liquid hydrogen), resulting in an uncollimated collection of neutrons with a broad (though not necessarily thermal) distribution of energies.

The simplest approach to forming a neutron beam is simply to reject all neutrons that do not pass through two small, widely separated apertures. Neutrons can also be transported via waveguides and focused with optics, though we will not review such devices here (see e.g. \cite{NeutronLensUseExample, NeutronOpticsSears}). The key feature of neutron collimation for our purposes is simply that it is statistically costly, with neutron count proportional to accepted phase space.

Neutron energy distributions are, similarly, narrowed primarily through rejection of velocities other than those desired. A typical design for a neutron velocity selector is described in \cite{NeutronVelocitySelector}: a rotating cylinder of absorbing material with helical channels, such that neutrons are absorbed unless they have the right velocity to pass through in a straight line without striking any surfaces. An alternative approach to velocity selection using slotted disks is described in \cite{NeutronVelocitySelectorDisks}. The statistical cost of these procedures depends somewhat on the particular neutron source used and the corresponding energy distribution after moderation.

A variety of existing neutron scattering instruments may be able to accommodate our proposal: the NIST Center for Neutron Research's VSANS instrument \cite{NIST_VSANS}, Oak Ridge National Laboratory's EQ-SANS diffractometer \cite{ORNL_EQSANS}, the Institut Laue-Langevin's D22 diffractometer \cite{ILL_D22}, and various instruments at J-PARC's Materials and Life Science Experimental Facility \cite{JPARC_MLF}, among others. Since this work is not specific to any particular source, our projections assume a set of parameters approximately representative of these optimal sources: a flux of $10^8$ cm$^{-2}$ s$^{-1}$ neutrons of $0.6$ nm wavelength over a target area of $10$ cm$^2$, with a minimum resolvable angle (including collimation, detector pixel size, and detector distance to target size ratio) of $3$ mrad, corresponding to a momentum transfer of approximately $(30\text{ nm})^{-1}$. 

Other parameters of the neutron beam should be less important for our sensitivity projections. In particular, energy spread acts only to ``wash out'' the low-angle peak that would indicate the presence of a new force; while this increases the uncertainty on a detected new force's mediator mass $\mu$, it has little effect on the total number of new force scattering events within that peak and therefore has minimal impact on our ability to detect a new force's presence (see Appendix \ref{app:Statistics}). Similarly, our proposal should not require particularly good angular resolution, so long as sufficiently small angles can be observed.

\subsection{X-Ray Beam Parameters}\label{subapp:XRayBeams}

Small-angle X-ray scattering (SAXS) sources are typically based on undulation of an electron beam by a series of magnets of alternating polarity \cite{SynchrotronFacilities}. SAXS instruments are available over a wide range of X-ray energies, including well above the keV momentum scales of the neutron sources we consider. Using such high-energy sources is likely to be necessary for most targets of interest, due to X-ray absorption: as discussed in Appendix \ref{app:XRayInteractions}, low-energy X-rays are rapidly attenuated within dense materials, especially at large atomic weights. For the xenon targets that are the focus of this work, energies of at least 40 keV are likely desirable.

A variety of SAXS facilities could potentially meet the requirements of our proposal; see, for example, \cite{XRay_Source_D2AM, XRay_Source_ID15A}. Just as for neutron sources, we do not choose a particular source in this work, and in fact we are generally agnostic to the  parameters of the source used so long as they are sufficient for the dominant uncertainties in the final measurement to arise from neutron scattering; we assume that this holds for all of our projections. The precise X-ray photon counts required for this to hold were discussed in Section \ref{sub:SingleMaterialSeparation}: an X-ray count of $10^5$ times the neutron count should always be sufficient, with as little as a few times the neutron count sufficient at the smaller mediator masses we focus on. Even the former condition can be satisfied by instruments such as the European Synchrotron Radiation Facility's ID15A \cite{XRay_Source_ID15A}, and an enormous variety of X-ray sources can meet the weaker flux requirement.

There is one other property of X-ray sources that complicates our proposal somewhat: their beamsize. Typical X-ray scattering instrument beams are far narrower than the centimeter-scale targets that are necessary to maximize neutron count. Since targets may not be spatially uniform over their transverse extent, it is necessary to scan the X-ray scattering measurement over the target in order to obtain a spatially-average structure factor applicable to the neutron scattering measurement. Note that the neutron beam may not be spatially uniform either, so the spatial distribution of neutrons must be measured as well. This has a negligible impact on the total neutron count needed for our experiment, however, as the absence of a scattering target for this measurement eliminates the usual factor of 10 loss in flux from most of the beam passing through a target without scattering (see Appendix \ref{app:MultipleScattering}).

It is similarly critical that the X-ray and neutron beamlines be precisely coaxial in order to see the same effective target thickness and structures. This may pose challenges for the use of recently developed techniques for simultaneous X-ray and neutron scattering measurements \cite{SimultaneousSAXSSANS}.

Other parameters of X-ray sources are generally not a concern for our proposal. As discussed in Appendix \ref{app:MultipleScattering}, the X-ray collimation requirement is a factor of a few stronger (in terms of transverse momentum) to the neutron requirement, due to the need to measure the structure factor at smaller momentum transfers in order to predict the impact of multiple scattering events; this is easily satisfied by many X-ray instruments. Similarly, the distribution of incident X-ray energies is generally far narrower than that of neutrons, so energy spread should not be a meaningful constraint for our purposes.

\section{Statistics}\label{app:Statistics}

In this appendix, we describe our approach to estimating the potential sensitivity of the various target materials considered in the main text. We first tabulate the various systematic errors faced by any implementation of our proposal. We then describe our approach to calculating the statistical reach of single-material targets, before explaining our approximation of the statistical error for two-material targets.

\subsection{Systematic Errors}\label{sub:SystematicErrors}

We begin by summarizing the systematic errors that limit the achievable sensitivity of new force searches implementing our proposed strategy. Most of the effects we consider in this section have been discussed elsewhere in this work; our goal here is to tabulate their respective magnitudes before we calculate realistically achievable experimental sensitivities.

A relatively fundamental limit on searches for new neutron-atom interactions is our limited ability to predict strong nuclear interactions. In most of this work, we have described nuclear scattering as entirely angle-independent, but this is an approximation: significant deviations from angle-independence occur at momentum transfers comparable to the inverse strong force range, i.e. inverse femtometers \cite{NeutronScatteringKoester, NeutronScatteringSearsAppendix, NeutronEffectiveRangeBlatt, NEScatteringKoester,  NeutronEffectiveRangeHackenburg}; see Appendix \ref{app:NeutronInteractions}. Corrections to the angle-independence of nuclear scattering are therefore suppressed by $\mathcal{O}((q_Tb_{\rm nuc})^2) \lesssim 10^{-8}$ even at the largest momentum transfers we consider, leaving them far below our systematic target. In fact, a more significant systematic error may arise from the contribution of the nuclear charge form factor to the neutron's electric polarizability scattering length, which is suppressed by $\mathcal{O}(q_T\sqrt{\left\langle r_n^2 \right\rangle}b_{P}/b_{\rm nuc}) \sim 10^{-9}$ (see Appendix \ref{subapp:EMScattering}). This error may be reducible using knowledge of this form factor, but we will not explore this here, given that we do not expect this to be a limiting error for our proposal.

Errors related to modeling of electromagnetic scattering may be more significant, though these depend considerably on the exact target used. As we discuss in Section \ref{sub:SingleMaterialProjections}, contributions from the new force and from electromagnetic scattering become difficult to distinguish once $q_o \sim \mu$; we avoid this issue in our projections by conservatively restricting to $\lambda > 10^{-1}$ nm. Moreover, even for ideal noble gases, our description of electromagnetic scattering (see Appendix \ref{app:NeutronInteractions}) ignored various terms suppressed by $m_e/m_n$, $r_N/r_A$, or $q_Tr_N$, with $r_N$ the nuclear radius and $r_A$ the atomic radius. All three of these factors are of order $10^{-6}$, but may be enhanced sufficiently to be relevant by the large atomic numbers of the target atoms we consider. We note, however, that these higher-order terms can likely be worked out more precisely if this is useful for future experiments, since, unlike the nuclear corrections described above, they are the result of well-understood electromagnetism.

Scattering from realistic targets leads to a number of other electromagnetic corrections, however. Non-noble elements generally have non-zero magnetic dipole moments, leading to additional neutron scattering (see Appendix \ref{app:NeutronInteractions}), and even noble atoms may have induced magnetic moments due to interatomic interactions (see Appendix \ref{app:AtomicInteractions}). The separation of scattering contributions explained in Appendix \ref{app:TwoMaterialSeparation} allows non-noble elements' contributions to be removed, so they do not lead to a systematic error (up to small caveats discussed below). Magnetic moments induced in the target noble atoms, however, cannot be separated out. We show in Appendix \ref{app:AtomicInteractions} that magnetic moments induced by surface interactions should lead to scattering length corrections of no more than roughly $3\times10^{-3} (R_{\rm atom}\xi/R_{\rm grain}^2) \text{ fm}$ when correlations in the magnetic dipole moments within the solid have length scale $\xi$; effects of interactions within noble liquids (or dense noble gases) should be negligible by comparison.

Separation of contributions may fail to entirely remove systematic electromagnetic backgrounds from non-noble elements if there are any changes to the target between neutron and X-ray scattering or between different measurements of one type will inhibit the accuracy of this separation, however. One potential cause of such changes is material degradation from the scattering processes themselves, though this is likely to be minimal for the low energies we consider. In the case of materials consisting of piles of grains, the structure may also change simply due to motion of the target between measurements (though this may be circumventable by, for example, rotating such targets continuously during measurements and using the resulting average structure, as one would need to do for aerosols or boiling liquids). Xenon snow may be so unstable that it degrades even if not moved. Since all of these effects are strongly dependent on the specific materials used, we will not include them in our sensitivity projections below, but we note that they may be significant in some circumstances.

A related systematic effect arises due to the differing energies of appropriate neutron and X-ray sources, which change their respective correspondences between momentum transfer and scattering angle. As a result, neutrons and X-rays at a fixed momentum transfer take different paths through the target and do not, in fact, see identical target structures, potentially changing their respective structure factors. While we expect this effect to be small, given the generally thin targets and small angles that we are most interested in, it will likely need to be simulated numerically or tested with additional measurements given our desired precision. We leave a more detailed treatment of this effect to future work.

Systematic errors may also arise if materials' compositions change between measurements. This could occur due to finite noble gas purity, or due to imperfect separation of materials, for example from adsorption of atoms by the solid components of a target. Again, these depend heavily on details beyond the general principles outlined in this work, so we will not include these effects below.

A final, more generic source of systematic error is low-angle multiple scattering, discussed in Appendix \ref{app:MultipleScattering}. As we note in that appendix, multiple scattering should not be an issue for scattering from argon, but may or may not be a significant constraint on xenon-based targets. The error introduced by multiple scattering grows exponentially as a function of various experimental parameters (see \eqref{eq:MultipleScatteringProbability}) so it is generally either enormous or irrelevant. In particular, this error depends on the transverse size of individual neutrons, which is not currently known sufficiently well for any apparatus to definitively determine which material combinations are compatible with a scattering fraction of $0.1$; reducing the scattering fraction suppressed this effect. We assume in most of this work that neutrons have sufficient transverse sizes to use the materials we consider with scattering fractions of $0.1$ without significant multiple scattering errors, but we note that some solids discussed in Section \ref{sub:TwoMaterialOptions} may in actuality require reduced statistics.

\subsection{Projecting Sensitivity for Single-Material Targets}

We can now estimate the sensitivity of a neutron scattering experiment implementing our proposal. Since we expect to have many neutrons observed in each angular bin, we can approximate the exact maximum-likelihood analysis of the data with an F-test \cite{DataReductionBook}, comparing the $\chi^2$ values obtained when fitting the observed data including ($\chi^2_{\rm with}$) and not including ($\chi^2_{\rm without}$) a new force. More precisely, let
\begin{align}
    F = \frac{\left(\chi^2_{\rm without}-\chi^2_{\rm with}\right)/2}{\chi^2_{\rm with}/N^{\rm dof}_{\rm with}}, \label{eq:FStatisticExpression}
\end{align}
for $N^{\rm dof}_{\rm with}$ degrees of freedom in the with-new force fit (and assuming two degrees of freedom for the new force: $\mu$ and $g$). Then the distribution of $F$ values will follow an $F$-distribution with $d_1 = 2$ and $d_2 = N^{\rm dof}_{\rm with}$ degrees of freedom. We can then constrain any new force for which the resulting $F$ is expected to exceed some threshold value.

In the case of single-material scattering, we can straightforwardly compute the expected values of $\chi^2$ both with and without the new force included in the fit as follows. We generically expect a $\chi^2$ contribution of $1$ for each degree of freedom in a fit, simply from Poisson statistics within each angular bin, independent of which fit is used. In the presence of a new force, we expect an additional number of scatterings into the bin at angle $\theta$ given by $2\kappa_{\rm new}N_{\rm expected}(\theta)/(1+q_T(\theta)^2)$, with $N_{\rm expected}$ the expected number of neutrons scattered into this bin by nuclear scattering. In the absence of any fit corrections---that is, using the fit parameters that would be optimal in the absence of a new force---this leads to an expected $\chi_{\rm without}^2$ contribution given by $4\kappa_{\rm new}^2(\theta)N_{\rm expected}(\theta)/(1+q_T(\theta)^2)^2$, but no contribution to $\chi_{\rm with}^2$. Note that this total expected contribution is independent of the number of bins in the limit that the nuclear scattering distribution (including coherence) is constant within each bin:
\begin{align}
    \Delta\chi^2_{\rm no-fit} = \sum\limits_{\rm bins} \frac{4\kappa_{\rm new}^2(\theta)N_{\rm expected}(\theta)}{(1+q_T(\theta)^2)^2} \to N_{\rm scattered}\int d\theta\frac{4\kappa_{\rm new}^2(\theta)}{(1+q_T(\theta)^2)^2}\frac{dp}{d\theta}
\end{align}
for $N_{\rm scattered}$ total scattered neutrons and (nuclear, coherence-enhanced) scattering distribution $dp/d\theta$, normalized to integrate to unity over angles above some minimum. This additional $\chi^2$ contribution is reduced after fitting empirically-determined Standard Model scattering parameters: fitting over the angle-independent nuclear scattering length, for example, eliminates the average number of additional scatters per bin (of constant solid angle), leaving only the variation in new force scattering over different angles.

In terms of this additional post-fit $\chi^2$ contribution, \eqref{eq:FStatisticExpression} predicts
\begin{align}
    F = 1 + \frac{1}{2}\Delta\chi^2_{\rm fit}.
\end{align}
For $N^{\rm dof}_{\rm with} \gg 1$, the cumulative distribution function is well approximated at $F$ of no more than a few by
\begin{align}
    {\rm CDF}_{2,N^{\rm dof}_{\rm with}}(F) \approx 1 - e^{-F},
\end{align}
so we expect to be able to detect a new force at 95\% confidence whenever $1 + \Delta\chi^2_{\rm fit}/2 > \ln(20) \approx 3.0$, i.e. when $\Delta\chi^2_{\rm fit} > 4.0$. Since the finite true number of angular bins weakens this somewhat, we conservatively require $\Delta\chi^2_{\rm fit} > 5.0$ for the projections in this work.

\subsection{Projecting Sensitivity for Two-Material Targets}\label{subapp:TwoMaterialStatistics}

In the previous subsection, we explained our method for estimating the sensitivity of a single-material neutron scattering experiment following the approach discussed in this work. An exact analysis of the two-material case is necessarily numerical. However, in this appendix, we describe an approximate, analytic approach to estimating the two-material sensitivity, which avoids the computational cost of the numerical analysis while offering additional insight into the parameter dependence of the sensitivity. 

Note that, throughout this appendix, we assume that the spin-dependent scattering cross-term discussed in Appendix \ref{app:TwoMaterialSeparation} is not significant, such that measurements from two noble gases are sufficient. Including the spin-dependent cross-term complicates the analysis further, but we will not consider it in detail in this work; in any case it is unlikely to have more than an order-unity effect on the final sensitivity.

In the single-material case, the effects of a new force can be resolved from those of similar-length scale structure simply by looking at ratios of neutron to X-ray scattering. Unfortunately, this fails in the two-material case due to interference between scattering from the solid and gas, which differs for neutrons and X-rays. In particular, the total neutron scattering distribution in this case can be written as (see \eqref{eq:TwoMaterialScatteringDistribution})
\begin{align}
    \left.\frac{dp}{d\Omega}\right|_{\rm 2-material} &= \left.\frac{dp}{d\Omega}\right|_{\rm s,inc} + \left.\frac{dp}{d\Omega}\right|_{\rm g,inc} + \left|\left\langle B_0(\mathbf{q}_T)\right\rangle + b_g(\mathbf{q}_T)\left\langle W(\mathbf{q}_T)\right\rangle\right|^2
\end{align}
where the first two terms on the right-hand side are the incoherent scattering distributions from the solid alone and the gas alone, while the last term is the total coherent scattering length, including both the unknown solid contribution $B_0(\mathbf{q}_T)$ and the gas contribution, which is given by the product of the gas's single-atom scattering length $b_g(\mathbf{q}_T)$ and the target phase sum $W(\mathbf{q}_T)$. Note that we have taken the expectation values of $B_0(\mathbf{q}_T)$ and $W(\mathbf{q}_T)$ above in order to separate the coherent and incoherent scattering contributions to the scattering distribution.

We can eliminate all of the terms not enhanced by the gas's structure by making three measurements---the two materials together, the solid alone, and the gas alone (which gives precisely the incoherent gas scattering distribution above)---and taking the following linear combination:
\begin{align}\begin{split}
    \left.\frac{dp}{d\Omega}\right|_{\rm difference} &= \left.\frac{dp}{d\Omega}\right|_{\rm 2-material} - \left.\frac{dp}{d\Omega}\right|_{\rm s\ only} - \left.\frac{dp}{d\Omega}\right|_{\rm g\ only} \\
    &= 2~\text{Re}\left(\left\langle B_0^*(\theta)\right\rangle b_g(\theta)\left\langle W(\theta)\right\rangle \right) + \left|b_g(\theta)\left\langle W(\theta)\right\rangle \right|^2.
\end{split}\end{align}
Now suppose that we have obtained $W(\mathbf{q}_T)$ from X-ray scattering measurements. This is not a precise description of the two-material analysis, since the X-ray scattering distribution similarly suffers from interference, but it should act as a reasonable approximation of the process, since the purpose of the X-ray measurements is precisely to distinguish any new force from the (shared) structure factor. 

Then, if we ignore the electromagnetic and new force contributions to the last term, we can predict it from a combination of the gas-only measurement of $b_g(\theta) \approx b_0$ and the X-ray measurement of $W(\theta)$; including an estimate of the electromagnetic contribution (from other measurements, or from theoretical calculations) can make this prediction even more precise. Similarly to the handling of the phase sum above, this is not necessary in the true numerical analysis, but is useful for our simple estimate here; note that this approximation is likely to be conservative, as we are discarding part of the effect of the new force. Then let
\begin{align}
    \left.\frac{dp}{d\Omega}\right|_{\rm cross} &= \left.\frac{dp}{d\Omega}\right|_{\rm 2-material} - \left.\frac{dp}{d\Omega}\right|_{\rm s\ only} - \left.\frac{dp}{d\Omega}\right|_{\rm g\ only} - \left|b_g(\theta)\left\langle W(\theta)\right\rangle\right|^2_{\rm predicted}.
\end{align}
Since the only complex component to the single-atom scattering length comes from the irrelevant Schwinger scattering length (see Appendix \ref{app:TwoMaterialSeparation}), we can also write this measurement combination as
\begin{align}
    \left.\frac{dp}{d\Omega}\right|_{\rm cross} &= 2~\text{Re}\left(\left\langle B_0^*(\theta)\right\rangle\left\langle W(\theta)\right\rangle\right)b_g(\theta).
\end{align}
Crucially, this is now fully factored into a gas-independent term $2~\text{Re}\left(\left\langle B_0^*(\theta)\right\rangle \left\langle W(\theta)\right\rangle\right)$ and a gas-specific scattering length $b_g(\theta)$. Thus, if we perform this process for two different noble elements, we can take the ratio
\begin{align}
    \frac{\left.dp/d\Omega\right|_{\rm cross,1}}{\left.dp/d\Omega\right|_{\rm cross,2}} &= \frac{b_{g,1}(\theta)}{b_{g,2}(\theta)},
\end{align}
which is now independent of both the solid and the structure factor.

We can now use this to detect a new force by detecting a difference in the angle-dependence of the two elements' single-atom scattering lengths. In particular, using \eqref{eq:CoherentScatteringLength}, we have
\begin{align}
    \frac{\left.dp/d\Omega\right|_{\rm cross,1}}{\left.dp/d\Omega\right|_{\rm cross,2}} &= ({\rm const.})\left(1 + \kappa_{\rm EM,1}f_1(q_T) - \kappa_{\rm EM,2}f_2(q_T) + \frac{\Delta\kappa_{\rm  new}}{1+(q_T/\mu)^2} + \mathcal{O}(\kappa^2)\right)
\end{align}
where $\Delta\kappa_{\rm new} = \kappa_{\rm new,1} - \kappa_{\rm new,2}$. From here, the fitting procedure is much the same as for the single-material case, with seven free parameters (the constant prefactor, two electromagnetic parameters for each noble element, and two new force parameters). The four electromagnetic parameters were not included in the fits for the rough projections of this work, as this significantly reduced computational expense. We do not expect this to have a meaningful impact on the final sensitivity, however, as electromagnetic scattering has a significantly different angular dependence from new force scattering; this was also confirmed numerically for the single-material case.

As we note in the main text, the dependence of this final ratio only on the difference $\Delta\kappa_{\rm new}$ makes it advantageous to use noble gases with very different atomic weights, in order to maximize this difference. However, this must be balanced against the weakness of the gas's scattering contribution relative to that of the solid, given the low value of helium's SLD in particular--see Table \ref{tab:MaterialSLDs}--leading to the approximate parity between argon- and helium-based scattering seen in Figure \ref{fig:TwoMaterialSensitivity}.

\bibliographystyle{apsrev4-2}
\bibliography{main}

\begin{thebibliography}{145}%
\makeatletter
\providecommand \@ifxundefined [1]{%
 \@ifx{#1\undefined}
}%
\providecommand \@ifnum [1]{%
 \ifnum #1\expandafter \@firstoftwo
 \else \expandafter \@secondoftwo
 \fi
}%
\providecommand \@ifx [1]{%
 \ifx #1\expandafter \@firstoftwo
 \else \expandafter \@secondoftwo
 \fi
}%
\providecommand \natexlab [1]{#1}%
\providecommand \enquote  [1]{``#1''}%
\providecommand \bibnamefont  [1]{#1}%
\providecommand \bibfnamefont [1]{#1}%
\providecommand \citenamefont [1]{#1}%
\providecommand \href@noop [0]{\@secondoftwo}%
\providecommand \href [0]{\begingroup \@sanitize@url \@href}%
\providecommand \@href[1]{\@@startlink{#1}\@@href}%
\providecommand \@@href[1]{\endgroup#1\@@endlink}%
\providecommand \@sanitize@url [0]{\catcode `\\12\catcode `\$12\catcode
  `\&12\catcode `\#12\catcode `\^12\catcode `\_12\catcode `\%12\relax}%
\providecommand \@@startlink[1]{}%
\providecommand \@@endlink[0]{}%
\providecommand \url  [0]{\begingroup\@sanitize@url \@url }%
\providecommand \@url [1]{\endgroup\@href {#1}{\urlprefix }}%
\providecommand \urlprefix  [0]{URL }%
\providecommand \Eprint [0]{\href }%
\providecommand \doibase [0]{https://doi.org/}%
\providecommand \selectlanguage [0]{\@gobble}%
\providecommand \bibinfo  [0]{\@secondoftwo}%
\providecommand \bibfield  [0]{\@secondoftwo}%
\providecommand \translation [1]{[#1]}%
\providecommand \BibitemOpen [0]{}%
\providecommand \bibitemStop [0]{}%
\providecommand \bibitemNoStop [0]{.\EOS\space}%
\providecommand \EOS [0]{\spacefactor3000\relax}%
\providecommand \BibitemShut  [1]{\csname bibitem#1\endcsname}%
\let\auto@bib@innerbib\@empty
\bibitem [{\citenamefont {Lee}\ and\ \citenamefont
  {Yang}(1955)}]{PhysRev.98.1501}%
  \BibitemOpen
  \bibfield  {author} {\bibinfo {author} {\bibfnamefont {T.~D.}\ \bibnamefont
  {Lee}}\ and\ \bibinfo {author} {\bibfnamefont {C.~N.}\ \bibnamefont {Yang}},\
  }\href {https://doi.org/10.1103/PhysRev.98.1501} {\bibfield  {journal}
  {\bibinfo  {journal} {Phys. Rev.}\ }\textbf {\bibinfo {volume} {98}},\
  \bibinfo {pages} {1501} (\bibinfo {year} {1955})}\BibitemShut {NoStop}%
\bibitem [{\citenamefont {Fayet}(1996)}]{Fayet_1996}%
  \BibitemOpen
  \bibfield  {author} {\bibinfo {author} {\bibfnamefont {P.}~\bibnamefont
  {Fayet}},\ }\href {https://doi.org/10.1088/0264-9381/13/11a/004} {\bibfield
  {journal} {\bibinfo  {journal} {Classical and Quantum Gravity}\ }\textbf
  {\bibinfo {volume} {13}},\ \bibinfo {pages} {A19} (\bibinfo {year}
  {1996})}\BibitemShut {NoStop}%
\bibitem [{\citenamefont {Fayet}(2001)}]{Fayet_2001}%
  \BibitemOpen
  \bibfield  {author} {\bibinfo {author} {\bibfnamefont {P.}~\bibnamefont
  {Fayet}},\ }\href
  {https://doi.org/https://doi.org/10.1016/S1296-2147(01)01265-3} {\bibfield
  {journal} {\bibinfo  {journal} {Comptes Rendus de l'Acad\'{e}mie des Sciences
  - Series IV - Physics}\ }\textbf {\bibinfo {volume} {2}},\ \bibinfo {pages}
  {1257} (\bibinfo {year} {2001})}\BibitemShut {NoStop}%
\bibitem [{\citenamefont {Long}\ \emph {et~al.}(1999)\citenamefont {Long},
  \citenamefont {Chan},\ and\ \citenamefont {Price}}]{Long_1999}%
  \BibitemOpen
  \bibfield  {author} {\bibinfo {author} {\bibfnamefont {J.}~\bibnamefont
  {Long}}, \bibinfo {author} {\bibfnamefont {H.}~\bibnamefont {Chan}},\ and\
  \bibinfo {author} {\bibfnamefont {J.}~\bibnamefont {Price}},\ }\href
  {https://doi.org/https://doi.org/10.1016/S0550-3213(98)00711-1} {\bibfield
  {journal} {\bibinfo  {journal} {Nuclear Physics B}\ }\textbf {\bibinfo
  {volume} {539}},\ \bibinfo {pages} {23} (\bibinfo {year} {1999})}\BibitemShut
  {NoStop}%
\bibitem [{\citenamefont {Arkani-Hamed}\ and\ \citenamefont
  {Dimopoulos}(2002)}]{PhysRevD.65.052003}%
  \BibitemOpen
  \bibfield  {author} {\bibinfo {author} {\bibfnamefont {N.}~\bibnamefont
  {Arkani-Hamed}}\ and\ \bibinfo {author} {\bibfnamefont {S.}~\bibnamefont
  {Dimopoulos}},\ }\href {https://doi.org/10.1103/PhysRevD.65.052003}
  {\bibfield  {journal} {\bibinfo  {journal} {Phys. Rev. D}\ }\textbf {\bibinfo
  {volume} {65}},\ \bibinfo {pages} {052003} (\bibinfo {year}
  {2002})}\BibitemShut {NoStop}%
\bibitem [{\citenamefont {Dimopoulos}\ and\ \citenamefont
  {Geraci}(2003)}]{PhysRevD.68.124021}%
  \BibitemOpen
  \bibfield  {author} {\bibinfo {author} {\bibfnamefont {S.}~\bibnamefont
  {Dimopoulos}}\ and\ \bibinfo {author} {\bibfnamefont {A.~A.}\ \bibnamefont
  {Geraci}},\ }\href {https://doi.org/10.1103/PhysRevD.68.124021} {\bibfield
  {journal} {\bibinfo  {journal} {Phys. Rev. D}\ }\textbf {\bibinfo {volume}
  {68}},\ \bibinfo {pages} {124021} (\bibinfo {year} {2003})}\BibitemShut
  {NoStop}%
\bibitem [{\citenamefont {Arkani-Hamed}\ \emph {et~al.}(1999)\citenamefont
  {Arkani-Hamed}, \citenamefont {Dimopoulos},\ and\ \citenamefont
  {Dvali}}]{PhysRevD.59.086004}%
  \BibitemOpen
  \bibfield  {author} {\bibinfo {author} {\bibfnamefont {N.}~\bibnamefont
  {Arkani-Hamed}}, \bibinfo {author} {\bibfnamefont {S.}~\bibnamefont
  {Dimopoulos}},\ and\ \bibinfo {author} {\bibfnamefont {G.}~\bibnamefont
  {Dvali}},\ }\href {https://doi.org/10.1103/PhysRevD.59.086004} {\bibfield
  {journal} {\bibinfo  {journal} {Phys. Rev. D}\ }\textbf {\bibinfo {volume}
  {59}},\ \bibinfo {pages} {086004} (\bibinfo {year} {1999})}\BibitemShut
  {NoStop}%
\bibitem [{\citenamefont {Sundrum}(2004)}]{MotivationCC}%
  \BibitemOpen
  \bibfield  {author} {\bibinfo {author} {\bibfnamefont {R.}~\bibnamefont
  {Sundrum}},\ }\href {https://doi.org/10.1103/PhysRevD.69.044014} {\bibfield
  {journal} {\bibinfo  {journal} {Phys. Rev. D}\ }\textbf {\bibinfo {volume}
  {69}},\ \bibinfo {pages} {044014} (\bibinfo {year} {2004})}\BibitemShut
  {NoStop}%
\bibitem [{\citenamefont {Choi}\ \emph {et~al.}(2020)\citenamefont {Choi},
  \citenamefont {Yanagida},\ and\ \citenamefont {Yokozaki}}]{CHOI2020135836}%
  \BibitemOpen
  \bibfield  {author} {\bibinfo {author} {\bibfnamefont {G.}~\bibnamefont
  {Choi}}, \bibinfo {author} {\bibfnamefont {T.~T.}\ \bibnamefont {Yanagida}},\
  and\ \bibinfo {author} {\bibfnamefont {N.}~\bibnamefont {Yokozaki}},\ }\href
  {https://doi.org/https://doi.org/10.1016/j.physletb.2020.135836} {\bibfield
  {journal} {\bibinfo  {journal} {Physics Letters B}\ }\textbf {\bibinfo
  {volume} {810}},\ \bibinfo {pages} {135836} (\bibinfo {year}
  {2020})}\BibitemShut {NoStop}%
\bibitem [{\citenamefont {Okada}\ \emph {et~al.}(2020)\citenamefont {Okada},
  \citenamefont {Okada}, \citenamefont {Raut},\ and\ \citenamefont
  {Shafi}}]{OKADA2020135785}%
  \BibitemOpen
  \bibfield  {author} {\bibinfo {author} {\bibfnamefont {N.}~\bibnamefont
  {Okada}}, \bibinfo {author} {\bibfnamefont {S.}~\bibnamefont {Okada}},
  \bibinfo {author} {\bibfnamefont {D.}~\bibnamefont {Raut}},\ and\ \bibinfo
  {author} {\bibfnamefont {Q.}~\bibnamefont {Shafi}},\ }\href
  {https://doi.org/https://doi.org/10.1016/j.physletb.2020.135785} {\bibfield
  {journal} {\bibinfo  {journal} {Physics Letters B}\ }\textbf {\bibinfo
  {volume} {810}},\ \bibinfo {pages} {135785} (\bibinfo {year}
  {2020})}\BibitemShut {NoStop}%
\bibitem [{\citenamefont {Fujii}(1991)}]{Fujii_1991}%
  \BibitemOpen
  \bibfield  {author} {\bibinfo {author} {\bibfnamefont {Y.}~\bibnamefont
  {Fujii}},\ }\href {https://doi.org/10.1142/S0217751X91001714} {\bibfield
  {journal} {\bibinfo  {journal} {International Journal of Modern Physics A}\
  }\textbf {\bibinfo {volume} {06}},\ \bibinfo {pages} {3505} (\bibinfo {year}
  {1991})}\BibitemShut {NoStop}%
\bibitem [{\citenamefont {Kim}(1987)}]{KimReview}%
  \BibitemOpen
  \bibfield  {author} {\bibinfo {author} {\bibfnamefont {J.~E.}\ \bibnamefont
  {Kim}},\ }\href
  {https://doi.org/https://doi.org/10.1016/0370-1573(87)90017-2} {\bibfield
  {journal} {\bibinfo  {journal} {Physics Reports}\ }\textbf {\bibinfo {volume}
  {150}},\ \bibinfo {pages} {1} (\bibinfo {year} {1987})}\BibitemShut {NoStop}%
\bibitem [{\citenamefont {Adelberger}\ \emph {et~al.}(2003)\citenamefont
  {Adelberger}, \citenamefont {Heckel},\ and\ \citenamefont
  {Nelson}}]{AdelbergerReview}%
  \BibitemOpen
  \bibfield  {author} {\bibinfo {author} {\bibfnamefont {E.}~\bibnamefont
  {Adelberger}}, \bibinfo {author} {\bibfnamefont {B.}~\bibnamefont {Heckel}},\
  and\ \bibinfo {author} {\bibfnamefont {A.}~\bibnamefont {Nelson}},\ }\href
  {https://doi.org/10.1146/annurev.nucl.53.041002.110503} {\bibfield  {journal}
  {\bibinfo  {journal} {Annual Review of Nuclear and Particle Science}\
  }\textbf {\bibinfo {volume} {53}},\ \bibinfo {pages} {77} (\bibinfo {year}
  {2003})}\BibitemShut {NoStop}%
\bibitem [{\citenamefont {Moody}\ and\ \citenamefont
  {Wilczek}(1984)}]{MoodyWilczek}%
  \BibitemOpen
  \bibfield  {author} {\bibinfo {author} {\bibfnamefont {J.~E.}\ \bibnamefont
  {Moody}}\ and\ \bibinfo {author} {\bibfnamefont {F.}~\bibnamefont
  {Wilczek}},\ }\href {https://doi.org/10.1103/PhysRevD.30.130} {\bibfield
  {journal} {\bibinfo  {journal} {Phys. Rev. D}\ }\textbf {\bibinfo {volume}
  {30}},\ \bibinfo {pages} {130} (\bibinfo {year} {1984})}\BibitemShut
  {NoStop}%
\bibitem [{\citenamefont {Bordag}\ \emph {et~al.}(2001)\citenamefont {Bordag},
  \citenamefont {Mohideen},\ and\ \citenamefont
  {Mostepanenko}}]{NewForceLimitsBordag}%
  \BibitemOpen
  \bibfield  {author} {\bibinfo {author} {\bibfnamefont {M.}~\bibnamefont
  {Bordag}}, \bibinfo {author} {\bibfnamefont {U.}~\bibnamefont {Mohideen}},\
  and\ \bibinfo {author} {\bibfnamefont {V.}~\bibnamefont {Mostepanenko}},\
  }\href {https://doi.org/https://doi.org/10.1016/S0370-1573(01)00015-1}
  {\bibfield  {journal} {\bibinfo  {journal} {Physics Reports}\ }\textbf
  {\bibinfo {volume} {353}},\ \bibinfo {pages} {1} (\bibinfo {year}
  {2001})}\BibitemShut {NoStop}%
\bibitem [{\citenamefont {Klimchitskaya}\ and\ \citenamefont
  {Mostepanenko}(2021)}]{NewForceLimitsKlimchitskaya}%
  \BibitemOpen
  \bibfield  {author} {\bibinfo {author} {\bibfnamefont {G.~L.}\ \bibnamefont
  {Klimchitskaya}}\ and\ \bibinfo {author} {\bibfnamefont {V.~M.}\ \bibnamefont
  {Mostepanenko}},\ }\href {https://www.mdpi.com/2218-1997/7/9/343} {\bibfield
  {journal} {\bibinfo  {journal} {Universe}\ }\textbf {\bibinfo {volume} {7}}
  (\bibinfo {year} {2021})}\BibitemShut {NoStop}%
\bibitem [{\citenamefont {Chen}\ \emph {et~al.}(2016)\citenamefont {Chen},
  \citenamefont {Tham}, \citenamefont {Krause}, \citenamefont {L\'opez},
  \citenamefont {Fischbach},\ and\ \citenamefont {Decca}}]{NewForceLimitsChen}%
  \BibitemOpen
  \bibfield  {author} {\bibinfo {author} {\bibfnamefont {Y.-J.}\ \bibnamefont
  {Chen}}, \bibinfo {author} {\bibfnamefont {W.~K.}\ \bibnamefont {Tham}},
  \bibinfo {author} {\bibfnamefont {D.~E.}\ \bibnamefont {Krause}}, \bibinfo
  {author} {\bibfnamefont {D.}~\bibnamefont {L\'opez}}, \bibinfo {author}
  {\bibfnamefont {E.}~\bibnamefont {Fischbach}},\ and\ \bibinfo {author}
  {\bibfnamefont {R.~S.}\ \bibnamefont {Decca}},\ }\href
  {https://doi.org/10.1103/PhysRevLett.116.221102} {\bibfield  {journal}
  {\bibinfo  {journal} {Phys. Rev. Lett.}\ }\textbf {\bibinfo {volume} {116}},\
  \bibinfo {pages} {221102} (\bibinfo {year} {2016})}\BibitemShut {NoStop}%
\bibitem [{\citenamefont {Lee}\ \emph {et~al.}(2020)\citenamefont {Lee},
  \citenamefont {Adelberger}, \citenamefont {Cook}, \citenamefont {Fleischer},\
  and\ \citenamefont {Heckel}}]{NewForceLimitsLee}%
  \BibitemOpen
  \bibfield  {author} {\bibinfo {author} {\bibfnamefont {J.~G.}\ \bibnamefont
  {Lee}}, \bibinfo {author} {\bibfnamefont {E.~G.}\ \bibnamefont {Adelberger}},
  \bibinfo {author} {\bibfnamefont {T.~S.}\ \bibnamefont {Cook}}, \bibinfo
  {author} {\bibfnamefont {S.~M.}\ \bibnamefont {Fleischer}},\ and\ \bibinfo
  {author} {\bibfnamefont {B.~R.}\ \bibnamefont {Heckel}},\ }\href
  {https://doi.org/10.1103/PhysRevLett.124.101101} {\bibfield  {journal}
  {\bibinfo  {journal} {Phys. Rev. Lett.}\ }\textbf {\bibinfo {volume} {124}},\
  \bibinfo {pages} {101101} (\bibinfo {year} {2020})}\BibitemShut {NoStop}%
\bibitem [{\citenamefont {Adelberger}\ \emph {et~al.}(2022)\citenamefont
  {Adelberger}, \citenamefont {Budker}, \citenamefont {Folman}, \citenamefont
  {Geraci}, \citenamefont {Harke}, \citenamefont {Kaplan}, \citenamefont
  {Kimball}, \citenamefont {Lehnert}, \citenamefont {Moore}, \citenamefont
  {Morley}, \citenamefont {Palladino}, \citenamefont {Phillips}, \citenamefont
  {Piacentino}, \citenamefont {Snow},\ and\ \citenamefont
  {Sudhir}}]{SNOWMASSAdelberger}%
  \BibitemOpen
  \bibfield  {author} {\bibinfo {author} {\bibfnamefont {E.}~\bibnamefont
  {Adelberger}}, \bibinfo {author} {\bibfnamefont {D.}~\bibnamefont {Budker}},
  \bibinfo {author} {\bibfnamefont {R.}~\bibnamefont {Folman}}, \bibinfo
  {author} {\bibfnamefont {A.~A.}\ \bibnamefont {Geraci}}, \bibinfo {author}
  {\bibfnamefont {J.~T.}\ \bibnamefont {Harke}}, \bibinfo {author}
  {\bibfnamefont {D.~M.}\ \bibnamefont {Kaplan}}, \bibinfo {author}
  {\bibfnamefont {D.~F.~J.}\ \bibnamefont {Kimball}}, \bibinfo {author}
  {\bibfnamefont {R.}~\bibnamefont {Lehnert}}, \bibinfo {author} {\bibfnamefont
  {D.}~\bibnamefont {Moore}}, \bibinfo {author} {\bibfnamefont {G.~W.}\
  \bibnamefont {Morley}}, \bibinfo {author} {\bibfnamefont {A.}~\bibnamefont
  {Palladino}}, \bibinfo {author} {\bibfnamefont {T.~J.}\ \bibnamefont
  {Phillips}}, \bibinfo {author} {\bibfnamefont {G.~M.}\ \bibnamefont
  {Piacentino}}, \bibinfo {author} {\bibfnamefont {W.~M.}\ \bibnamefont
  {Snow}},\ and\ \bibinfo {author} {\bibfnamefont {V.}~\bibnamefont {Sudhir}},\
  }\href {https://doi.org/10.48550/ARXIV.2203.09691} {\bibinfo {title}
  {Snowmass white paper: Precision studies of spacetime symmetries and
  gravitational physics}} (\bibinfo {year} {2022})\BibitemShut {NoStop}%
\bibitem [{\citenamefont {Blakemore}\ \emph {et~al.}(2021)\citenamefont
  {Blakemore}, \citenamefont {Fieguth}, \citenamefont {Kawasaki}, \citenamefont
  {Priel}, \citenamefont {Martin}, \citenamefont {Rider}, \citenamefont
  {Wang},\ and\ \citenamefont {Gratta}}]{LevitatedBeadSearch}%
  \BibitemOpen
  \bibfield  {author} {\bibinfo {author} {\bibfnamefont {C.~P.}\ \bibnamefont
  {Blakemore}}, \bibinfo {author} {\bibfnamefont {A.}~\bibnamefont {Fieguth}},
  \bibinfo {author} {\bibfnamefont {A.}~\bibnamefont {Kawasaki}}, \bibinfo
  {author} {\bibfnamefont {N.}~\bibnamefont {Priel}}, \bibinfo {author}
  {\bibfnamefont {D.}~\bibnamefont {Martin}}, \bibinfo {author} {\bibfnamefont
  {A.~D.}\ \bibnamefont {Rider}}, \bibinfo {author} {\bibfnamefont
  {Q.}~\bibnamefont {Wang}},\ and\ \bibinfo {author} {\bibfnamefont
  {G.}~\bibnamefont {Gratta}},\ }\href
  {https://doi.org/10.1103/PhysRevD.104.L061101} {\bibfield  {journal}
  {\bibinfo  {journal} {Phys. Rev. D}\ }\textbf {\bibinfo {volume} {104}},\
  \bibinfo {pages} {L061101} (\bibinfo {year} {2021})}\BibitemShut {NoStop}%
\bibitem [{\citenamefont {Decca}\ \emph {et~al.}(2005)\citenamefont {Decca},
  \citenamefont {L\'opez}, \citenamefont {Chan}, \citenamefont {Fischbach},
  \citenamefont {Krause},\ and\ \citenamefont {Jamell}}]{NewForceLimitsDecca}%
  \BibitemOpen
  \bibfield  {author} {\bibinfo {author} {\bibfnamefont {R.~S.}\ \bibnamefont
  {Decca}}, \bibinfo {author} {\bibfnamefont {D.}~\bibnamefont {L\'opez}},
  \bibinfo {author} {\bibfnamefont {H.~B.}\ \bibnamefont {Chan}}, \bibinfo
  {author} {\bibfnamefont {E.}~\bibnamefont {Fischbach}}, \bibinfo {author}
  {\bibfnamefont {D.~E.}\ \bibnamefont {Krause}},\ and\ \bibinfo {author}
  {\bibfnamefont {C.~R.}\ \bibnamefont {Jamell}},\ }\href
  {https://doi.org/10.1103/PhysRevLett.94.240401} {\bibfield  {journal}
  {\bibinfo  {journal} {Phys. Rev. Lett.}\ }\textbf {\bibinfo {volume} {94}},\
  \bibinfo {pages} {240401} (\bibinfo {year} {2005})}\BibitemShut {NoStop}%
\bibitem [{\citenamefont {Kamiya}\ \emph {et~al.}(2015)\citenamefont {Kamiya},
  \citenamefont {Itagaki}, \citenamefont {Tani}, \citenamefont {Kim},\ and\
  \citenamefont {Komamiya}}]{XenonPaper1}%
  \BibitemOpen
  \bibfield  {author} {\bibinfo {author} {\bibfnamefont {Y.}~\bibnamefont
  {Kamiya}}, \bibinfo {author} {\bibfnamefont {K.}~\bibnamefont {Itagaki}},
  \bibinfo {author} {\bibfnamefont {M.}~\bibnamefont {Tani}}, \bibinfo {author}
  {\bibfnamefont {G.~N.}\ \bibnamefont {Kim}},\ and\ \bibinfo {author}
  {\bibfnamefont {S.}~\bibnamefont {Komamiya}},\ }\href
  {https://doi.org/10.1103/PhysRevLett.114.161101} {\bibfield  {journal}
  {\bibinfo  {journal} {Phys. Rev. Lett.}\ }\textbf {\bibinfo {volume} {114}},\
  \bibinfo {pages} {161101} (\bibinfo {year} {2015})}\BibitemShut {NoStop}%
\bibitem [{\citenamefont {Kamiya}\ \emph {et~al.}(2021)\citenamefont {Kamiya},
  \citenamefont {Cubitt}, \citenamefont {Porcar}, \citenamefont {Zimmer},
  \citenamefont {Kim},\ and\ \citenamefont {Komamiya}}]{XenonPaper2}%
  \BibitemOpen
  \bibfield  {author} {\bibinfo {author} {\bibfnamefont {Y.}~\bibnamefont
  {Kamiya}}, \bibinfo {author} {\bibfnamefont {R.}~\bibnamefont {Cubitt}},
  \bibinfo {author} {\bibfnamefont {L.}~\bibnamefont {Porcar}}, \bibinfo
  {author} {\bibfnamefont {O.}~\bibnamefont {Zimmer}}, \bibinfo {author}
  {\bibfnamefont {G.~N.}\ \bibnamefont {Kim}},\ and\ \bibinfo {author}
  {\bibfnamefont {S.}~\bibnamefont {Komamiya}},\ }\href
  {https://doi.org/10.1063/5.0036985} {\bibfield  {journal} {\bibinfo
  {journal} {AIP Conference Proceedings}\ }\textbf {\bibinfo {volume} {2319}},\
  \bibinfo {pages} {040017} (\bibinfo {year} {2021})}\BibitemShut {NoStop}%
\bibitem [{\citenamefont {Haddock}\ \emph {et~al.}(2018)\citenamefont
  {Haddock}, \citenamefont {Oi}, \citenamefont {Hirota}, \citenamefont {Ino},
  \citenamefont {Kitaguchi}, \citenamefont {Matsumoto}, \citenamefont
  {Mishima}, \citenamefont {Shima}, \citenamefont {Shimizu}, \citenamefont
  {Snow},\ and\ \citenamefont {Yoshioka}}]{NewForceLimitsHaddock}%
  \BibitemOpen
  \bibfield  {author} {\bibinfo {author} {\bibfnamefont {C.~C.}\ \bibnamefont
  {Haddock}}, \bibinfo {author} {\bibfnamefont {N.}~\bibnamefont {Oi}},
  \bibinfo {author} {\bibfnamefont {K.}~\bibnamefont {Hirota}}, \bibinfo
  {author} {\bibfnamefont {T.}~\bibnamefont {Ino}}, \bibinfo {author}
  {\bibfnamefont {M.}~\bibnamefont {Kitaguchi}}, \bibinfo {author}
  {\bibfnamefont {S.}~\bibnamefont {Matsumoto}}, \bibinfo {author}
  {\bibfnamefont {K.}~\bibnamefont {Mishima}}, \bibinfo {author} {\bibfnamefont
  {T.}~\bibnamefont {Shima}}, \bibinfo {author} {\bibfnamefont {H.~M.}\
  \bibnamefont {Shimizu}}, \bibinfo {author} {\bibfnamefont {W.~M.}\
  \bibnamefont {Snow}},\ and\ \bibinfo {author} {\bibfnamefont
  {T.}~\bibnamefont {Yoshioka}},\ }\href
  {https://doi.org/10.1103/PhysRevD.97.062002} {\bibfield  {journal} {\bibinfo
  {journal} {Phys. Rev. D}\ }\textbf {\bibinfo {volume} {97}},\ \bibinfo
  {pages} {062002} (\bibinfo {year} {2018})}\BibitemShut {NoStop}%
\bibitem [{\citenamefont {Pokotilovski}(2006)}]{NewForceLimitsPokotilovski}%
  \BibitemOpen
  \bibfield  {author} {\bibinfo {author} {\bibfnamefont {Y.~N.}\ \bibnamefont
  {Pokotilovski}},\ }\href {https://doi.org/10.1134/S1063778806060020}
  {\bibfield  {journal} {\bibinfo  {journal} {Physics of Atomic Nuclei}\
  }\textbf {\bibinfo {volume} {69}},\ \bibinfo {pages} {924} (\bibinfo {year}
  {2006})}\BibitemShut {NoStop}%
\bibitem [{\citenamefont {Casimir}\ and\ \citenamefont
  {Polder}(1948)}]{CasimirForces}%
  \BibitemOpen
  \bibfield  {author} {\bibinfo {author} {\bibfnamefont {H.~B.~G.}\
  \bibnamefont {Casimir}}\ and\ \bibinfo {author} {\bibfnamefont
  {D.}~\bibnamefont {Polder}},\ }\href {https://doi.org/10.1103/PhysRev.73.360}
  {\bibfield  {journal} {\bibinfo  {journal} {Phys. Rev.}\ }\textbf {\bibinfo
  {volume} {73}},\ \bibinfo {pages} {360} (\bibinfo {year} {1948})}\BibitemShut
  {NoStop}%
\bibitem [{\citenamefont {Heacock}\ \emph {et~al.}(2021)\citenamefont
  {Heacock}, \citenamefont {Fujiie}, \citenamefont {Haun}, \citenamefont
  {Henins}, \citenamefont {Hirota}, \citenamefont {Hosobata}, \citenamefont
  {Huber}, \citenamefont {Kitaguchi}, \citenamefont {Pushin}, \citenamefont
  {Shimizu}, \citenamefont {Takeda}, \citenamefont {Valdillez}, \citenamefont
  {Yamagata},\ and\ \citenamefont {Young}}]{PendellosungResult}%
  \BibitemOpen
  \bibfield  {author} {\bibinfo {author} {\bibfnamefont {B.}~\bibnamefont
  {Heacock}}, \bibinfo {author} {\bibfnamefont {T.}~\bibnamefont {Fujiie}},
  \bibinfo {author} {\bibfnamefont {R.~W.}\ \bibnamefont {Haun}}, \bibinfo
  {author} {\bibfnamefont {A.}~\bibnamefont {Henins}}, \bibinfo {author}
  {\bibfnamefont {K.}~\bibnamefont {Hirota}}, \bibinfo {author} {\bibfnamefont
  {T.}~\bibnamefont {Hosobata}}, \bibinfo {author} {\bibfnamefont {M.~G.}\
  \bibnamefont {Huber}}, \bibinfo {author} {\bibfnamefont {M.}~\bibnamefont
  {Kitaguchi}}, \bibinfo {author} {\bibfnamefont {D.~A.}\ \bibnamefont
  {Pushin}}, \bibinfo {author} {\bibfnamefont {H.}~\bibnamefont {Shimizu}},
  \bibinfo {author} {\bibfnamefont {M.}~\bibnamefont {Takeda}}, \bibinfo
  {author} {\bibfnamefont {R.}~\bibnamefont {Valdillez}}, \bibinfo {author}
  {\bibfnamefont {Y.}~\bibnamefont {Yamagata}},\ and\ \bibinfo {author}
  {\bibfnamefont {A.~R.}\ \bibnamefont {Young}},\ }\href
  {https://doi.org/10.1126/science.abc2794} {\bibfield  {journal} {\bibinfo
  {journal} {Science}\ }\textbf {\bibinfo {volume} {373}},\ \bibinfo {pages}
  {1239} (\bibinfo {year} {2021})},\ \Eprint
  {https://arxiv.org/abs/https://www.science.org/doi/pdf/10.1126/science.abc2794}
  {https://www.science.org/doi/pdf/10.1126/science.abc2794} \BibitemShut
  {NoStop}%
\bibitem [{\citenamefont {Gratta}\ \emph {et~al.}(2020)\citenamefont {Gratta},
  \citenamefont {Kaplan},\ and\ \citenamefont {Rajendran}}]{MossbauerProposal}%
  \BibitemOpen
  \bibfield  {author} {\bibinfo {author} {\bibfnamefont {G.}~\bibnamefont
  {Gratta}}, \bibinfo {author} {\bibfnamefont {D.~E.}\ \bibnamefont {Kaplan}},\
  and\ \bibinfo {author} {\bibfnamefont {S.}~\bibnamefont {Rajendran}},\ }\href
  {https://doi.org/10.1103/PhysRevD.102.115031} {\bibfield  {journal} {\bibinfo
   {journal} {Phys. Rev. D}\ }\textbf {\bibinfo {volume} {102}},\ \bibinfo
  {pages} {115031} (\bibinfo {year} {2020})}\BibitemShut {NoStop}%
\bibitem [{\citenamefont {Glinka}\ \emph {et~al.}(1998)\citenamefont {Glinka},
  \citenamefont {Barker}, \citenamefont {Hammouda}, \citenamefont {Krueger},
  \citenamefont {Moyer},\ and\ \citenamefont {Orts}}]{SANSInstrumentNIST}%
  \BibitemOpen
  \bibfield  {author} {\bibinfo {author} {\bibfnamefont {C.~J.}\ \bibnamefont
  {Glinka}}, \bibinfo {author} {\bibfnamefont {J.~G.}\ \bibnamefont {Barker}},
  \bibinfo {author} {\bibfnamefont {B.}~\bibnamefont {Hammouda}}, \bibinfo
  {author} {\bibfnamefont {S.}~\bibnamefont {Krueger}}, \bibinfo {author}
  {\bibfnamefont {J.~J.}\ \bibnamefont {Moyer}},\ and\ \bibinfo {author}
  {\bibfnamefont {W.~J.}\ \bibnamefont {Orts}},\ }\href
  {https://doi.org/10.1107/S0021889897017020} {\bibfield  {journal} {\bibinfo
  {journal} {Journal of Applied Crystallography}\ }\textbf {\bibinfo {volume}
  {31}},\ \bibinfo {pages} {430} (\bibinfo {year} {1998})}\BibitemShut
  {NoStop}%
\bibitem [{\citenamefont {Lynn}\ \emph {et~al.}(2003)\citenamefont {Lynn},
  \citenamefont {Buchanan}, \citenamefont {Butler}, \citenamefont {Magid},\
  and\ \citenamefont {Wignall}}]{SANSInstrumentORNL}%
  \BibitemOpen
  \bibfield  {author} {\bibinfo {author} {\bibfnamefont {G.}~\bibnamefont
  {Lynn}}, \bibinfo {author} {\bibfnamefont {M.}~\bibnamefont {Buchanan}},
  \bibinfo {author} {\bibfnamefont {P.}~\bibnamefont {Butler}}, \bibinfo
  {author} {\bibfnamefont {L.~J.}\ \bibnamefont {Magid}},\ and\ \bibinfo
  {author} {\bibfnamefont {G.}~\bibnamefont {Wignall}},\ }\href
  {https://doi.org/10.1107/S0021889803000670} {\bibfield  {journal} {\bibinfo
  {journal} {Journal of Applied Crystallography}\ }\textbf {\bibinfo {volume}
  {36}},\ \bibinfo {pages} {829} (\bibinfo {year} {2003})}\BibitemShut
  {NoStop}%
\bibitem [{\citenamefont {Keiderling}\ and\ \citenamefont
  {Wiedenmann}(1995)}]{SANSInstrumentBERII}%
  \BibitemOpen
  \bibfield  {author} {\bibinfo {author} {\bibfnamefont {U.}~\bibnamefont
  {Keiderling}}\ and\ \bibinfo {author} {\bibfnamefont {A.}~\bibnamefont
  {Wiedenmann}},\ }\href
  {https://doi.org/https://doi.org/10.1016/0921-4526(95)00316-2} {\bibfield
  {journal} {\bibinfo  {journal} {Physica B: Condensed Matter}\ }\textbf
  {\bibinfo {volume} {213-214}},\ \bibinfo {pages} {895} (\bibinfo {year}
  {1995})}\BibitemShut {NoStop}%
\bibitem [{\citenamefont {Ito}\ \emph {et~al.}(1995)\citenamefont {Ito},
  \citenamefont {Imai},\ and\ \citenamefont {Takahashi}}]{SANSInstrumentJRR3M}%
  \BibitemOpen
  \bibfield  {author} {\bibinfo {author} {\bibfnamefont {Y.}~\bibnamefont
  {Ito}}, \bibinfo {author} {\bibfnamefont {M.}~\bibnamefont {Imai}},\ and\
  \bibinfo {author} {\bibfnamefont {S.}~\bibnamefont {Takahashi}},\ }\href
  {https://doi.org/https://doi.org/10.1016/0921-4526(95)00314-Y} {\bibfield
  {journal} {\bibinfo  {journal} {Physica B: Condensed Matter}\ }\textbf
  {\bibinfo {volume} {213-214}},\ \bibinfo {pages} {889} (\bibinfo {year}
  {1995})}\BibitemShut {NoStop}%
\bibitem [{\citenamefont {Seong}\ \emph {et~al.}(2002)\citenamefont {Seong},
  \citenamefont {Han}, \citenamefont {Lee}, \citenamefont {Lee}, \citenamefont
  {Hong}, \citenamefont {Park},\ and\ \citenamefont
  {Kim}}]{SANSInstrumentHANARO1}%
  \BibitemOpen
  \bibfield  {author} {\bibinfo {author} {\bibfnamefont {B.-S.}\ \bibnamefont
  {Seong}}, \bibinfo {author} {\bibfnamefont {Y.-S.}\ \bibnamefont {Han}},
  \bibinfo {author} {\bibfnamefont {C.-H.}\ \bibnamefont {Lee}}, \bibinfo
  {author} {\bibfnamefont {J.-S.}\ \bibnamefont {Lee}}, \bibinfo {author}
  {\bibfnamefont {K.-P.}\ \bibnamefont {Hong}}, \bibinfo {author}
  {\bibfnamefont {K.-N.}\ \bibnamefont {Park}},\ and\ \bibinfo {author}
  {\bibfnamefont {H.-J.}\ \bibnamefont {Kim}},\ }\href
  {https://doi.org/10.1007/s003390101091} {\bibfield  {journal} {\bibinfo
  {journal} {Applied Physics A}\ }\textbf {\bibinfo {volume} {74}},\ \bibinfo
  {pages} {s201} (\bibinfo {year} {2002})}\BibitemShut {NoStop}%
\bibitem [{\citenamefont {Han}\ \emph {et~al.}(2013)\citenamefont {Han},
  \citenamefont {Choi}, \citenamefont {Kim}, \citenamefont {Lee}, \citenamefont
  {Cho},\ and\ \citenamefont {Seong}}]{SANSInstrumentHANARO2}%
  \BibitemOpen
  \bibfield  {author} {\bibinfo {author} {\bibfnamefont {Y.-S.}\ \bibnamefont
  {Han}}, \bibinfo {author} {\bibfnamefont {S.-M.}\ \bibnamefont {Choi}},
  \bibinfo {author} {\bibfnamefont {T.-H.}\ \bibnamefont {Kim}}, \bibinfo
  {author} {\bibfnamefont {C.-H.}\ \bibnamefont {Lee}}, \bibinfo {author}
  {\bibfnamefont {S.-J.}\ \bibnamefont {Cho}},\ and\ \bibinfo {author}
  {\bibfnamefont {B.-S.}\ \bibnamefont {Seong}},\ }\href
  {https://doi.org/https://doi.org/10.1016/j.nima.2013.04.052} {\bibfield
  {journal} {\bibinfo  {journal} {Nuclear Instruments and Methods in Physics
  Research Section A: Accelerators, Spectrometers, Detectors and Associated
  Equipment}\ }\textbf {\bibinfo {volume} {721}},\ \bibinfo {pages} {17}
  (\bibinfo {year} {2013})}\BibitemShut {NoStop}%
\bibitem [{\citenamefont {Kohlbrecher}\ and\ \citenamefont
  {Wagner}(2000)}]{SANSInstrumentSINQ}%
  \BibitemOpen
  \bibfield  {author} {\bibinfo {author} {\bibfnamefont {J.}~\bibnamefont
  {Kohlbrecher}}\ and\ \bibinfo {author} {\bibfnamefont {W.}~\bibnamefont
  {Wagner}},\ }\href {https://doi.org/10.1107/S0021889800099775} {\bibfield
  {journal} {\bibinfo  {journal} {Journal of Applied Crystallography}\ }\textbf
  {\bibinfo {volume} {33}},\ \bibinfo {pages} {804} (\bibinfo {year}
  {2000})}\BibitemShut {NoStop}%
\bibitem [{\citenamefont {Andersen}\ \emph {et~al.}(2020)\citenamefont
  {Andersen} \emph {et~al.}}]{SANSInstrumentESS}%
  \BibitemOpen
  \bibfield  {author} {\bibinfo {author} {\bibfnamefont {K.}~\bibnamefont
  {Andersen}} \emph {et~al.},\ }\href
  {https://doi.org/https://doi.org/10.1016/j.nima.2020.163402} {\bibfield
  {journal} {\bibinfo  {journal} {Nuclear Instruments and Methods in Physics
  Research Section A: Accelerators, Spectrometers, Detectors and Associated
  Equipment}\ }\textbf {\bibinfo {volume} {957}},\ \bibinfo {pages} {163402}
  (\bibinfo {year} {2020})}\BibitemShut {NoStop}%
\bibitem [{\citenamefont {Friedrich}\ \emph {et~al.}(1989)\citenamefont
  {Friedrich}, \citenamefont {Wagner},\ and\ \citenamefont
  {Wille}}]{NeutronVelocitySelector}%
  \BibitemOpen
  \bibfield  {author} {\bibinfo {author} {\bibfnamefont {H.}~\bibnamefont
  {Friedrich}}, \bibinfo {author} {\bibfnamefont {V.}~\bibnamefont {Wagner}},\
  and\ \bibinfo {author} {\bibfnamefont {P.}~\bibnamefont {Wille}},\
  }\href@noop {} {\bibfield  {journal} {\bibinfo  {journal} {Physica
  B-condensed Matter}\ }\textbf {\bibinfo {volume} {156}},\ \bibinfo {pages}
  {547} (\bibinfo {year} {1989})}\BibitemShut {NoStop}%
\bibitem [{\citenamefont {Clark}\ \emph {et~al.}(1966)\citenamefont {Clark},
  \citenamefont {Mitchell}, \citenamefont {Palmer},\ and\ \citenamefont
  {Wilson}}]{NeutronVelocitySelectorDisks}%
  \BibitemOpen
  \bibfield  {author} {\bibinfo {author} {\bibfnamefont {C.~D.}\ \bibnamefont
  {Clark}}, \bibinfo {author} {\bibfnamefont {E.~W.~J.}\ \bibnamefont
  {Mitchell}}, \bibinfo {author} {\bibfnamefont {D.~W.}\ \bibnamefont
  {Palmer}},\ and\ \bibinfo {author} {\bibfnamefont {I.~H.}\ \bibnamefont
  {Wilson}},\ }\href@noop {} {\bibfield  {journal} {\bibinfo  {journal}
  {Journal of Scientific Instruments}\ }\textbf {\bibinfo {volume} {43}},\
  \bibinfo {pages} {1} (\bibinfo {year} {1966})}\BibitemShut {NoStop}%
\bibitem [{\citenamefont {Kopetka}\ \emph {et~al.}(2006)\citenamefont
  {Kopetka}, \citenamefont {Williams},\ and\ \citenamefont
  {Rowe}}]{NISTColdSource}%
  \BibitemOpen
  \bibfield  {author} {\bibinfo {author} {\bibfnamefont {P.}~\bibnamefont
  {Kopetka}}, \bibinfo {author} {\bibfnamefont {R.~E.}\ \bibnamefont
  {Williams}},\ and\ \bibinfo {author} {\bibfnamefont {J.~M.}\ \bibnamefont
  {Rowe}},\ }\href@noop {} {\emph {\bibinfo {title} {NIST Liquid Hydrogen Cold
  Source}}},\ \bibinfo {type} {Tech. Rep.}\ \bibinfo {number} {NISTIR 7352}\
  (\bibinfo  {institution} {National Institute of Standards and Technology},\
  \bibinfo {address} {Gaithersburg, MD},\ \bibinfo {year} {2006})\BibitemShut
  {NoStop}%
\bibitem [{\citenamefont {Taylor}(1972)}]{ScatteringTheoryBook}%
  \BibitemOpen
  \bibfield  {author} {\bibinfo {author} {\bibfnamefont {J.~R.}\ \bibnamefont
  {Taylor}},\ }\href@noop {} {\emph {\bibinfo {title} {Scattering theory : the
  quantum theory on nonrelativistic collisions}}}\ (\bibinfo  {publisher}
  {Wiley},\ \bibinfo {address} {New York, NY},\ \bibinfo {year}
  {1972})\BibitemShut {NoStop}%
\bibitem [{\citenamefont {Hammouda}(1995)}]{NeutronScatteringPolymersHammouda}%
  \BibitemOpen
  \bibfield  {author} {\bibinfo {author} {\bibfnamefont {B.}~\bibnamefont
  {Hammouda}},\ }in\ \href@noop {} {\emph {\bibinfo {booktitle} {A tutorial on
  small-angle neutron scattering from polymers}}}\ (\bibinfo {organization}
  {National Institute of Standards and Technology Materials Science and
  Engineering Laboratory},\ \bibinfo {address} {Gaithersburg, Maryland},\
  \bibinfo {year} {1995})\BibitemShut {NoStop}%
\bibitem [{\citenamefont {Windsor}(1988)}]{NeutronScatteringWindsor}%
  \BibitemOpen
  \bibfield  {author} {\bibinfo {author} {\bibfnamefont {C.~G.}\ \bibnamefont
  {Windsor}},\ }\href
  {https://doi.org/https://doi.org/10.1107/S0021889888008404} {\bibfield
  {journal} {\bibinfo  {journal} {Journal of Applied Crystallography}\ }\textbf
  {\bibinfo {volume} {21}},\ \bibinfo {pages} {582} (\bibinfo {year}
  {1988})}\BibitemShut {NoStop}%
\bibitem [{\citenamefont {Fermi}\ and\ \citenamefont
  {Marshall}(1947)}]{NobleGasChoiceFermi}%
  \BibitemOpen
  \bibfield  {author} {\bibinfo {author} {\bibfnamefont {E.}~\bibnamefont
  {Fermi}}\ and\ \bibinfo {author} {\bibfnamefont {L.}~\bibnamefont
  {Marshall}},\ }\href {https://doi.org/10.1103/PhysRev.72.1139} {\bibfield
  {journal} {\bibinfo  {journal} {Phys. Rev.}\ }\textbf {\bibinfo {volume}
  {72}},\ \bibinfo {pages} {1139} (\bibinfo {year} {1947})}\BibitemShut
  {NoStop}%
\bibitem [{\citenamefont {Chadwick}\ \emph {et~al.}(2011)\citenamefont
  {Chadwick} \emph {et~al.}}]{NeutronScatteringDataENDF}%
  \BibitemOpen
  \bibfield  {author} {\bibinfo {author} {\bibfnamefont {M.}~\bibnamefont
  {Chadwick}} \emph {et~al.},\ }\href
  {https://doi.org/https://doi.org/10.1016/j.nds.2011.11.002} {\bibfield
  {journal} {\bibinfo  {journal} {Nuclear Data Sheets}\ }\textbf {\bibinfo
  {volume} {112}},\ \bibinfo {pages} {2887} (\bibinfo {year} {2011})},\
  \bibinfo {note} {special Issue on ENDF/B-VII.1 Library}\BibitemShut {NoStop}%
\bibitem [{\citenamefont {Sears}(1992)}]{NeutronScatteringDataSears}%
  \BibitemOpen
  \bibfield  {author} {\bibinfo {author} {\bibfnamefont {V.~F.}\ \bibnamefont
  {Sears}},\ }\href {https://doi.org/10.1080/10448639208218770} {\bibfield
  {journal} {\bibinfo  {journal} {Neutron News}\ }\textbf {\bibinfo {volume}
  {3}},\ \bibinfo {pages} {26} (\bibinfo {year} {1992})}\BibitemShut {NoStop}%
\bibitem [{\citenamefont {Young}(1991)}]{PhaseDiagramBook}%
  \BibitemOpen
  \bibfield  {author} {\bibinfo {author} {\bibfnamefont {D.~A.}\ \bibnamefont
  {Young}},\ }\href@noop {} {\emph {\bibinfo {title} {Phase diagrams of the
  elements}}}\ (\bibinfo  {publisher} {University of California Press},\
  \bibinfo {address} {Berkeley and Los Angeles},\ \bibinfo {year}
  {1991})\BibitemShut {NoStop}%
\bibitem [{\citenamefont {Balakishiyeva}\ \emph {et~al.}(2010)\citenamefont
  {Balakishiyeva}, \citenamefont {Mahapatra}, \citenamefont {Saab},\ and\
  \citenamefont {Yoo}}]{XenonSnow}%
  \BibitemOpen
  \bibfield  {author} {\bibinfo {author} {\bibfnamefont {D.~N.}\ \bibnamefont
  {Balakishiyeva}}, \bibinfo {author} {\bibfnamefont {R.}~\bibnamefont
  {Mahapatra}}, \bibinfo {author} {\bibfnamefont {T.}~\bibnamefont {Saab}},\
  and\ \bibinfo {author} {\bibfnamefont {J.}~\bibnamefont {Yoo}},\ }\href
  {https://doi.org/10.1063/1.3489544} {\bibfield  {journal} {\bibinfo
  {journal} {AIP Conference Proceedings}\ }\textbf {\bibinfo {volume} {1274}},\
  \bibinfo {pages} {138} (\bibinfo {year} {2010})}\BibitemShut {NoStop}%
\bibitem [{\citenamefont {Schulze}\ and\ \citenamefont
  {Kolb}(1974)}]{NobleSolids}%
  \BibitemOpen
  \bibfield  {author} {\bibinfo {author} {\bibfnamefont {W.}~\bibnamefont
  {Schulze}}\ and\ \bibinfo {author} {\bibfnamefont {D.~M.}\ \bibnamefont
  {Kolb}},\ }\href {https://doi.org/10.1039/F29747001098} {\bibfield  {journal}
  {\bibinfo  {journal} {J. Chem. Soc.{,} Faraday Trans. 2}\ }\textbf {\bibinfo
  {volume} {70}},\ \bibinfo {pages} {1098} (\bibinfo {year}
  {1974})}\BibitemShut {NoStop}%
\bibitem [{\citenamefont {Friedlander}(2000)}]{AerosolDynamics}%
  \BibitemOpen
  \bibfield  {author} {\bibinfo {author} {\bibfnamefont {S.~K.}\ \bibnamefont
  {Friedlander}},\ }\href
  {https://app.knovel.com/hotlink/toc/id:kpSDHFADE1/smoke-dust-haze-fundamentals/smoke-dust-haze-fundamentals}
  {\emph {\bibinfo {title} {Smoke, Dust, and Haze - Fundamentals of Aerosol
  Dynamics (2nd Edition)}}}\ (\bibinfo  {publisher} {Oxford University Press},\
  \bibinfo {year} {2000})\BibitemShut {NoStop}%
\bibitem [{\citenamefont {Hidy}(1984)}]{AerosolScience}%
  \BibitemOpen
  \bibfield  {author} {\bibinfo {author} {\bibfnamefont {G.~M.}\ \bibnamefont
  {Hidy}},\ }\href@noop {} {\emph {\bibinfo {title} {Aerosols: An Industrial
  and environmental science}}}\ (\bibinfo  {publisher} {Academic Press},\
  \bibinfo {year} {1984})\BibitemShut {NoStop}%
\bibitem [{\citenamefont {Metwalli}\ \emph {et~al.}(2020)\citenamefont
  {Metwalli}, \citenamefont {G{\"{o}}tz}, \citenamefont {Lages}, \citenamefont
  {B{\"{a}}r}, \citenamefont {Zech}, \citenamefont {Noll}, \citenamefont
  {Schuldes}, \citenamefont {Schindler}, \citenamefont {Prihoda}, \citenamefont
  {Lang}, \citenamefont {Grasser}, \citenamefont {Jacques}, \citenamefont
  {Didier}, \citenamefont {Cyril}, \citenamefont {Martel}, \citenamefont
  {Porcar},\ and\ \citenamefont {Unruh}}]{SimultaneousSAXSSANS}%
  \BibitemOpen
  \bibfield  {author} {\bibinfo {author} {\bibfnamefont {E.}~\bibnamefont
  {Metwalli}}, \bibinfo {author} {\bibfnamefont {K.}~\bibnamefont
  {G{\"{o}}tz}}, \bibinfo {author} {\bibfnamefont {S.}~\bibnamefont {Lages}},
  \bibinfo {author} {\bibfnamefont {C.}~\bibnamefont {B{\"{a}}r}}, \bibinfo
  {author} {\bibfnamefont {T.}~\bibnamefont {Zech}}, \bibinfo {author}
  {\bibfnamefont {D.~M.}\ \bibnamefont {Noll}}, \bibinfo {author}
  {\bibfnamefont {I.}~\bibnamefont {Schuldes}}, \bibinfo {author}
  {\bibfnamefont {T.}~\bibnamefont {Schindler}}, \bibinfo {author}
  {\bibfnamefont {A.}~\bibnamefont {Prihoda}}, \bibinfo {author} {\bibfnamefont
  {H.}~\bibnamefont {Lang}}, \bibinfo {author} {\bibfnamefont {J.}~\bibnamefont
  {Grasser}}, \bibinfo {author} {\bibfnamefont {M.}~\bibnamefont {Jacques}},
  \bibinfo {author} {\bibfnamefont {L.}~\bibnamefont {Didier}}, \bibinfo
  {author} {\bibfnamefont {A.}~\bibnamefont {Cyril}}, \bibinfo {author}
  {\bibfnamefont {A.}~\bibnamefont {Martel}}, \bibinfo {author} {\bibfnamefont
  {L.}~\bibnamefont {Porcar}},\ and\ \bibinfo {author} {\bibfnamefont
  {T.}~\bibnamefont {Unruh}},\ }\href
  {https://doi.org/10.1107/S1600576720005208} {\bibfield  {journal} {\bibinfo
  {journal} {Journal of Applied Crystallography}\ }\textbf {\bibinfo {volume}
  {53}},\ \bibinfo {pages} {722} (\bibinfo {year} {2020})}\BibitemShut
  {NoStop}%
\bibitem [{\citenamefont {Tong}\ and\ \citenamefont
  {Tang}(1997)}]{BoilingTong}%
  \BibitemOpen
  \bibfield  {author} {\bibinfo {author} {\bibfnamefont {L.}~\bibnamefont
  {Tong}}\ and\ \bibinfo {author} {\bibfnamefont {Y.}~\bibnamefont {Tang}},\
  }\href@noop {} {\emph {\bibinfo {title} {Boiling Heat Transfer and Two-Phase
  Flow (2nd Edition)}}}\ (\bibinfo  {publisher} {Taylor and Francis},\ \bibinfo
  {year} {1997})\BibitemShut {NoStop}%
\bibitem [{\citenamefont {Stephan}(1992)}]{BoilingStephan}%
  \BibitemOpen
  \bibfield  {author} {\bibinfo {author} {\bibfnamefont {K.}~\bibnamefont
  {Stephan}},\ }\href@noop {} {\emph {\bibinfo {title} {Heat Transfer in
  Condensation and Boiling}}}\ (\bibinfo  {publisher} {Springer-Verlan Berlin
  Heidenberg},\ \bibinfo {year} {1992})\BibitemShut {NoStop}%
\bibitem [{\citenamefont {Leadbetter}\ and\ \citenamefont
  {Thomas}(1965)}]{XenonMeasurementsLeadbetter}%
  \BibitemOpen
  \bibfield  {author} {\bibinfo {author} {\bibfnamefont {A.~J.}\ \bibnamefont
  {Leadbetter}}\ and\ \bibinfo {author} {\bibfnamefont {H.~E.}\ \bibnamefont
  {Thomas}},\ }\href {https://doi.org/10.1039/TF9656100010} {\bibfield
  {journal} {\bibinfo  {journal} {Trans. Faraday Soc.}\ }\textbf {\bibinfo
  {volume} {61}},\ \bibinfo {pages} {10} (\bibinfo {year} {1965})}\BibitemShut
  {NoStop}%
\bibitem [{\citenamefont {Smith}\ \emph {et~al.}(1967)\citenamefont {Smith},
  \citenamefont {Gardner},\ and\ \citenamefont
  {Parker}}]{XenonMeasurementsSmith}%
  \BibitemOpen
  \bibfield  {author} {\bibinfo {author} {\bibfnamefont {B.~L.}\ \bibnamefont
  {Smith}}, \bibinfo {author} {\bibfnamefont {P.~R.}\ \bibnamefont {Gardner}},\
  and\ \bibinfo {author} {\bibfnamefont {E.~H.~C.}\ \bibnamefont {Parker}},\
  }\href {https://doi.org/10.1063/1.1712000} {\bibfield  {journal} {\bibinfo
  {journal} {The Journal of Chemical Physics}\ }\textbf {\bibinfo {volume}
  {47}},\ \bibinfo {pages} {1148} (\bibinfo {year} {1967})}\BibitemShut
  {NoStop}%
\bibitem [{\citenamefont {Zollweg}\ \emph {et~al.}(1971)\citenamefont
  {Zollweg}, \citenamefont {Hawkins},\ and\ \citenamefont
  {Benedek}}]{XenonMeasurementZollweg}%
  \BibitemOpen
  \bibfield  {author} {\bibinfo {author} {\bibfnamefont {J.}~\bibnamefont
  {Zollweg}}, \bibinfo {author} {\bibfnamefont {G.}~\bibnamefont {Hawkins}},\
  and\ \bibinfo {author} {\bibfnamefont {G.~B.}\ \bibnamefont {Benedek}},\
  }\href {https://doi.org/10.1103/PhysRevLett.27.1182} {\bibfield  {journal}
  {\bibinfo  {journal} {Phys. Rev. Lett.}\ }\textbf {\bibinfo {volume} {27}},\
  \bibinfo {pages} {1182} (\bibinfo {year} {1971})}\BibitemShut {NoStop}%
\bibitem [{\citenamefont {Safari}\ \emph {et~al.}(2015)\citenamefont {Safari},
  \citenamefont {Santos}, \citenamefont {Amaro}, \citenamefont {Jänkälä},\
  and\ \citenamefont {Fratini}}]{FormFactorsSafari}%
  \BibitemOpen
  \bibfield  {author} {\bibinfo {author} {\bibfnamefont {L.}~\bibnamefont
  {Safari}}, \bibinfo {author} {\bibfnamefont {J.~P.}\ \bibnamefont {Santos}},
  \bibinfo {author} {\bibfnamefont {P.}~\bibnamefont {Amaro}}, \bibinfo
  {author} {\bibfnamefont {K.}~\bibnamefont {Jänkälä}},\ and\ \bibinfo
  {author} {\bibfnamefont {F.}~\bibnamefont {Fratini}},\ }\href
  {https://doi.org/10.1063/1.4921227} {\bibfield  {journal} {\bibinfo
  {journal} {Journal of Mathematical Physics}\ }\textbf {\bibinfo {volume}
  {56}},\ \bibinfo {pages} {052105} (\bibinfo {year} {2015})}\BibitemShut
  {NoStop}%
\bibitem [{\citenamefont {Chantler}(1993)}]{FormFactorsChantler}%
  \BibitemOpen
  \bibfield  {author} {\bibinfo {author} {\bibfnamefont {C.}~\bibnamefont
  {Chantler}},\ }\href
  {https://doi.org/https://doi.org/10.1016/0969-806X(93)90323-M} {\bibfield
  {journal} {\bibinfo  {journal} {Radiation Physics and Chemistry}\ }\textbf
  {\bibinfo {volume} {41}},\ \bibinfo {pages} {759} (\bibinfo {year}
  {1993})}\BibitemShut {NoStop}%
\bibitem [{\citenamefont {Hubbell}\ \emph {et~al.}(1975)\citenamefont
  {Hubbell}, \citenamefont {Veigele}, \citenamefont {Briggs}, \citenamefont
  {Brown}, \citenamefont {Cromer},\ and\ \citenamefont
  {Howerton}}]{FormFactorsHubbell1}%
  \BibitemOpen
  \bibfield  {author} {\bibinfo {author} {\bibfnamefont {J.~H.}\ \bibnamefont
  {Hubbell}}, \bibinfo {author} {\bibfnamefont {W.~J.}\ \bibnamefont
  {Veigele}}, \bibinfo {author} {\bibfnamefont {E.~A.}\ \bibnamefont {Briggs}},
  \bibinfo {author} {\bibfnamefont {R.~T.}\ \bibnamefont {Brown}}, \bibinfo
  {author} {\bibfnamefont {D.~T.}\ \bibnamefont {Cromer}},\ and\ \bibinfo
  {author} {\bibfnamefont {R.~J.}\ \bibnamefont {Howerton}},\ }\href
  {https://doi.org/10.1063/1.555523} {\bibfield  {journal} {\bibinfo  {journal}
  {Journal of Physical and Chemical Reference Data}\ }\textbf {\bibinfo
  {volume} {4}},\ \bibinfo {pages} {471} (\bibinfo {year} {1975})}\BibitemShut
  {NoStop}%
\bibitem [{\citenamefont {Hubbell}\ and\ \citenamefont
  {{\O}verb{\o}}(1979)}]{FormFactorsHubbell2}%
  \BibitemOpen
  \bibfield  {author} {\bibinfo {author} {\bibfnamefont {J.~H.}\ \bibnamefont
  {Hubbell}}\ and\ \bibinfo {author} {\bibfnamefont {I.}~\bibnamefont
  {{\O}verb{\o}}},\ }\href {https://doi.org/10.1063/1.555593} {\bibfield
  {journal} {\bibinfo  {journal} {Journal of Physical and Chemical Reference
  Data}\ }\textbf {\bibinfo {volume} {8}},\ \bibinfo {pages} {69} (\bibinfo
  {year} {1979})}\BibitemShut {NoStop}%
\bibitem [{\citenamefont {Lehman}\ \emph {et~al.}(2011)\citenamefont {Lehman},
  \citenamefont {Terrones}, \citenamefont {Mansfield}, \citenamefont {Hurst},\
  and\ \citenamefont {Meunier}}]{CNTReview}%
  \BibitemOpen
  \bibfield  {author} {\bibinfo {author} {\bibfnamefont {J.~H.}\ \bibnamefont
  {Lehman}}, \bibinfo {author} {\bibfnamefont {M.}~\bibnamefont {Terrones}},
  \bibinfo {author} {\bibfnamefont {E.}~\bibnamefont {Mansfield}}, \bibinfo
  {author} {\bibfnamefont {K.~E.}\ \bibnamefont {Hurst}},\ and\ \bibinfo
  {author} {\bibfnamefont {V.}~\bibnamefont {Meunier}},\ }\href
  {https://doi.org/https://doi.org/10.1016/j.carbon.2011.03.028} {\bibfield
  {journal} {\bibinfo  {journal} {Carbon}\ }\textbf {\bibinfo {volume} {49}},\
  \bibinfo {pages} {2581} (\bibinfo {year} {2011})}\BibitemShut {NoStop}%
\bibitem [{\citenamefont {Kim}\ \emph {et~al.}(2009)\citenamefont {Kim},
  \citenamefont {Mulholland},\ and\ \citenamefont
  {Zachariah}}]{CNTDensityMeasurement}%
  \BibitemOpen
  \bibfield  {author} {\bibinfo {author} {\bibfnamefont {S.}~\bibnamefont
  {Kim}}, \bibinfo {author} {\bibfnamefont {G.}~\bibnamefont {Mulholland}},\
  and\ \bibinfo {author} {\bibfnamefont {M.}~\bibnamefont {Zachariah}},\ }\href
  {https://doi.org/https://doi.org/10.1016/j.carbon.2009.01.011} {\bibfield
  {journal} {\bibinfo  {journal} {Carbon}\ }\textbf {\bibinfo {volume} {47}},\
  \bibinfo {pages} {1297} (\bibinfo {year} {2009})}\BibitemShut {NoStop}%
\bibitem [{\citenamefont {Esconjauregui}\ \emph {et~al.}(2013)\citenamefont
  {Esconjauregui}, \citenamefont {Xie}, \citenamefont {Fouquet}, \citenamefont
  {Cartwright}, \citenamefont {Hardeman}, \citenamefont {Yang},\ and\
  \citenamefont {Robertson}}]{MWCNTDensityDiscussionEsconjauregui}%
  \BibitemOpen
  \bibfield  {author} {\bibinfo {author} {\bibfnamefont {S.}~\bibnamefont
  {Esconjauregui}}, \bibinfo {author} {\bibfnamefont {R.}~\bibnamefont {Xie}},
  \bibinfo {author} {\bibfnamefont {M.}~\bibnamefont {Fouquet}}, \bibinfo
  {author} {\bibfnamefont {R.}~\bibnamefont {Cartwright}}, \bibinfo {author}
  {\bibfnamefont {D.}~\bibnamefont {Hardeman}}, \bibinfo {author}
  {\bibfnamefont {J.}~\bibnamefont {Yang}},\ and\ \bibinfo {author}
  {\bibfnamefont {J.}~\bibnamefont {Robertson}},\ }\href
  {https://doi.org/10.1063/1.4799417} {\bibfield  {journal} {\bibinfo
  {journal} {Journal of Applied Physics}\ }\textbf {\bibinfo {volume} {113}},\
  \bibinfo {pages} {144309} (\bibinfo {year} {2013})}\BibitemShut {NoStop}%
\bibitem [{\citenamefont {Laurent}\ \emph {et~al.}(2010)\citenamefont
  {Laurent}, \citenamefont {Flahaut},\ and\ \citenamefont
  {Peigney}}]{MWCNTDensityDiscussionLaurent}%
  \BibitemOpen
  \bibfield  {author} {\bibinfo {author} {\bibfnamefont {C.}~\bibnamefont
  {Laurent}}, \bibinfo {author} {\bibfnamefont {E.}~\bibnamefont {Flahaut}},\
  and\ \bibinfo {author} {\bibfnamefont {A.}~\bibnamefont {Peigney}},\ }\href
  {https://doi.org/https://doi.org/10.1016/j.carbon.2010.04.010} {\bibfield
  {journal} {\bibinfo  {journal} {Carbon}\ }\textbf {\bibinfo {volume} {48}},\
  \bibinfo {pages} {2994} (\bibinfo {year} {2010})}\BibitemShut {NoStop}%
\bibitem [{\citenamefont {Lide}(2021)}]{CRCHandbook}%
  \BibitemOpen
  \bibfield  {author} {\bibinfo {author} {\bibfnamefont {D.~R.~E.}\
  \bibnamefont {Lide}},\ }\href@noop {} {\emph {\bibinfo {title} {CRC Handbook
  of Chemistry and Physics (102nd Edition)}}}\ (\bibinfo  {publisher} {CRC
  Press},\ \bibinfo {year} {2021})\BibitemShut {NoStop}%
\bibitem [{\citenamefont {{National Center for Biotechnology
  Information}}(2022)}]{PubChemAl2Ti3O9}%
  \BibitemOpen
  \bibfield  {author} {\bibinfo {author} {\bibnamefont {{National Center for
  Biotechnology Information}}},\ }\href@noop {} {\bibinfo {title} {{PubChem
  Compound Summary for CID 129627815, Aluminium titanate}}},\ \bibinfo
  {howpublished}
  {\url{https://pubchem.ncbi.nlm.nih.gov/compound/Aluminium-titanate}}
  (\bibinfo {year} {2022}),\ \bibinfo {note} {accessed: 2022-11-09}\BibitemShut
  {NoStop}%
\bibitem [{\citenamefont {Alothman}(2012)}]{SilicaReview}%
  \BibitemOpen
  \bibfield  {author} {\bibinfo {author} {\bibfnamefont {Z.~A.}\ \bibnamefont
  {Alothman}},\ }\href {https://doi.org/10.3390/ma5122874} {\bibfield
  {journal} {\bibinfo  {journal} {Materials}\ }\textbf {\bibinfo {volume}
  {5}},\ \bibinfo {pages} {2874} (\bibinfo {year} {2012})}\BibitemShut
  {NoStop}%
\bibitem [{\citenamefont {Bogush}\ \emph {et~al.}(1988)\citenamefont {Bogush},
  \citenamefont {Tracy},\ and\ \citenamefont {Zukoski}}]{StoberProcess}%
  \BibitemOpen
  \bibfield  {author} {\bibinfo {author} {\bibfnamefont {G.}~\bibnamefont
  {Bogush}}, \bibinfo {author} {\bibfnamefont {M.}~\bibnamefont {Tracy}},\ and\
  \bibinfo {author} {\bibfnamefont {C.}~\bibnamefont {Zukoski}},\ }\href
  {https://doi.org/https://doi.org/10.1016/0022-3093(88)90187-1} {\bibfield
  {journal} {\bibinfo  {journal} {Journal of Non-Crystalline Solids}\ }\textbf
  {\bibinfo {volume} {104}},\ \bibinfo {pages} {95} (\bibinfo {year}
  {1988})}\BibitemShut {NoStop}%
\bibitem [{Sil(2017)}]{SilicaMicrospheres}%
  \BibitemOpen
  \href
  {https://www.polysciences.com/default/awfile/index/attach/file/TDS_635.pdf/}
  {\emph {\bibinfo {title} {Uniform Silica Microspheres}}},\ \bibinfo
  {organization} {Polysciences, Inc.},\ \bibinfo {address} {Warrington, PA,
  USA} (\bibinfo {year} {2017})\BibitemShut {NoStop}%
\bibitem [{\citenamefont {Reinhardt}\ and\ \citenamefont
  {Winkler}(2000)}]{CeO2Review}%
  \BibitemOpen
  \bibfield  {author} {\bibinfo {author} {\bibfnamefont {K.}~\bibnamefont
  {Reinhardt}}\ and\ \bibinfo {author} {\bibfnamefont {H.}~\bibnamefont
  {Winkler}},\ }\bibinfo {title} {Cerium mischmetal, cerium alloys, and cerium
  compounds},\ in\ \href
  {https://onlinelibrary.wiley.com/doi/abs/10.1002/14356007.a06_139} {\emph
  {\bibinfo {booktitle} {Ullmann's Encyclopedia of Industrial Chemistry}}}\
  (\bibinfo  {publisher} {John Wiley \& Sons, Ltd},\ \bibinfo {year}
  {2000})\BibitemShut {NoStop}%
\bibitem [{\citenamefont {{Nishimura Advanced
  Ceramics}}(2020)}]{AluminaCompany}%
  \BibitemOpen
  \bibfield  {author} {\bibinfo {author} {\bibnamefont {{Nishimura Advanced
  Ceramics}}},\ }\href@noop {} {}\bibinfo {howpublished} {personal
  communication} (\bibinfo {year} {2020})\BibitemShut {NoStop}%
\bibitem [{\citenamefont {Mamalis}\ \emph {et~al.}(2004)\citenamefont
  {Mamalis}, \citenamefont {Vogtländer},\ and\ \citenamefont
  {Markopoulos}}]{CNTDiametersMamalis}%
  \BibitemOpen
  \bibfield  {author} {\bibinfo {author} {\bibfnamefont {A.}~\bibnamefont
  {Mamalis}}, \bibinfo {author} {\bibfnamefont {L.}~\bibnamefont
  {Vogtländer}},\ and\ \bibinfo {author} {\bibfnamefont {A.}~\bibnamefont
  {Markopoulos}},\ }\href
  {https://doi.org/https://doi.org/10.1016/j.precisioneng.2002.11.002}
  {\bibfield  {journal} {\bibinfo  {journal} {Precision Engineering}\ }\textbf
  {\bibinfo {volume} {28}},\ \bibinfo {pages} {16} (\bibinfo {year}
  {2004})}\BibitemShut {NoStop}%
\bibitem [{\citenamefont {Hinds}\ \emph {et~al.}(2004)\citenamefont {Hinds},
  \citenamefont {Chopra}, \citenamefont {Rantell}, \citenamefont {Andrews},
  \citenamefont {Gavalas},\ and\ \citenamefont {Bachas}}]{CNTDiametersHinds}%
  \BibitemOpen
  \bibfield  {author} {\bibinfo {author} {\bibfnamefont {B.~J.}\ \bibnamefont
  {Hinds}}, \bibinfo {author} {\bibfnamefont {N.}~\bibnamefont {Chopra}},
  \bibinfo {author} {\bibfnamefont {T.}~\bibnamefont {Rantell}}, \bibinfo
  {author} {\bibfnamefont {R.}~\bibnamefont {Andrews}}, \bibinfo {author}
  {\bibfnamefont {V.}~\bibnamefont {Gavalas}},\ and\ \bibinfo {author}
  {\bibfnamefont {L.~G.}\ \bibnamefont {Bachas}},\ }\href
  {https://doi.org/10.1126/science.1092048} {\bibfield  {journal} {\bibinfo
  {journal} {Science}\ }\textbf {\bibinfo {volume} {303}},\ \bibinfo {pages}
  {62} (\bibinfo {year} {2004})}\BibitemShut {NoStop}%
\bibitem [{\citenamefont {Ibrahim}(2013)}]{CNTDiametersIbrahim}%
  \BibitemOpen
  \bibfield  {author} {\bibinfo {author} {\bibfnamefont {K.}~\bibnamefont
  {Ibrahim}},\ }\href@noop {} {\bibfield  {journal} {\bibinfo  {journal}
  {Carbon letters}\ }\textbf {\bibinfo {volume} {14}},\ \bibinfo {pages} {131}
  (\bibinfo {year} {2013})}\BibitemShut {NoStop}%
\bibitem [{\citenamefont {Truong}\ \emph {et~al.}(2010)\citenamefont {Truong},
  \citenamefont {McMahon}, \citenamefont {Olsson-Jacques}, \citenamefont
  {Wilson},\ and\ \citenamefont {Mathys}}]{CNTDiametersTruong}%
  \BibitemOpen
  \bibfield  {author} {\bibinfo {author} {\bibfnamefont {V.-T.}\ \bibnamefont
  {Truong}}, \bibinfo {author} {\bibfnamefont {P.~J.}\ \bibnamefont {McMahon}},
  \bibinfo {author} {\bibfnamefont {C.~L.}\ \bibnamefont {Olsson-Jacques}},
  \bibinfo {author} {\bibfnamefont {A.~R.}\ \bibnamefont {Wilson}},\ and\
  \bibinfo {author} {\bibfnamefont {G.~I.}\ \bibnamefont {Mathys}},\ }in\ \href
  {https://doi.org/10.1109/ICONN.2010.6045178} {\emph {\bibinfo {booktitle}
  {2010 International Conference on Nanoscience and Nanotechnology}}}\
  (\bibinfo {year} {2010})\ pp.\ \bibinfo {pages} {66--69}\BibitemShut
  {NoStop}%
\bibitem [{\citenamefont {Nesvizhevsky}\ \emph {et~al.}(2008)\citenamefont
  {Nesvizhevsky}, \citenamefont {Pignol},\ and\ \citenamefont
  {Protasov}}]{NeutronScatteringNesvizhevsky}%
  \BibitemOpen
  \bibfield  {author} {\bibinfo {author} {\bibfnamefont {V.~V.}\ \bibnamefont
  {Nesvizhevsky}}, \bibinfo {author} {\bibfnamefont {G.}~\bibnamefont
  {Pignol}},\ and\ \bibinfo {author} {\bibfnamefont {K.~V.}\ \bibnamefont
  {Protasov}},\ }\href {https://doi.org/10.1103/PhysRevD.77.034020} {\bibfield
  {journal} {\bibinfo  {journal} {Phys. Rev. D}\ }\textbf {\bibinfo {volume}
  {77}},\ \bibinfo {pages} {034020} (\bibinfo {year} {2008})}\BibitemShut
  {NoStop}%
\bibitem [{\citenamefont {Kamyshkov}\ \emph {et~al.}(2008)\citenamefont
  {Kamyshkov}, \citenamefont {Tithof},\ and\ \citenamefont
  {Vysotsky}}]{NeutronLimitsKamyshkov}%
  \BibitemOpen
  \bibfield  {author} {\bibinfo {author} {\bibfnamefont {Y.}~\bibnamefont
  {Kamyshkov}}, \bibinfo {author} {\bibfnamefont {J.}~\bibnamefont {Tithof}},\
  and\ \bibinfo {author} {\bibfnamefont {M.}~\bibnamefont {Vysotsky}},\ }\href
  {https://doi.org/10.1103/PhysRevD.78.114029} {\bibfield  {journal} {\bibinfo
  {journal} {Phys. Rev. D}\ }\textbf {\bibinfo {volume} {78}},\ \bibinfo
  {pages} {114029} (\bibinfo {year} {2008})}\BibitemShut {NoStop}%
\bibitem [{\citenamefont {Leeb}\ and\ \citenamefont
  {Schmiedmayer}(1992)}]{NeutronOpticsLimits}%
  \BibitemOpen
  \bibfield  {author} {\bibinfo {author} {\bibfnamefont {H.}~\bibnamefont
  {Leeb}}\ and\ \bibinfo {author} {\bibfnamefont {J.}~\bibnamefont
  {Schmiedmayer}},\ }\href {https://doi.org/10.1103/PhysRevLett.68.1472}
  {\bibfield  {journal} {\bibinfo  {journal} {Phys. Rev. Lett.}\ }\textbf
  {\bibinfo {volume} {68}},\ \bibinfo {pages} {1472} (\bibinfo {year}
  {1992})}\BibitemShut {NoStop}%
\bibitem [{\citenamefont {Murata}\ and\ \citenamefont
  {Tanaka}(2015)}]{AtomicTestsMurata}%
  \BibitemOpen
  \bibfield  {author} {\bibinfo {author} {\bibfnamefont {J.}~\bibnamefont
  {Murata}}\ and\ \bibinfo {author} {\bibfnamefont {S.}~\bibnamefont
  {Tanaka}},\ }\href {https://doi.org/10.1088/0264-9381/32/3/033001} {\bibfield
   {journal} {\bibinfo  {journal} {Classical and Quantum Gravity}\ }\textbf
  {\bibinfo {volume} {32}},\ \bibinfo {pages} {033001} (\bibinfo {year}
  {2015})}\BibitemShut {NoStop}%
\bibitem [{\citenamefont {Tanaka}\ \emph {et~al.}(2014)\citenamefont {Tanaka},
  \citenamefont {Nakaya}, \citenamefont {Narikawa}, \citenamefont {Ninomiya},
  \citenamefont {Onishi}, \citenamefont {Pearson}, \citenamefont {Openshaw},
  \citenamefont {Saiba}, \citenamefont {Tanuma}, \citenamefont {Totsuka},\ and\
  \citenamefont {Murata}}]{AtomicTestsTanaka}%
  \BibitemOpen
  \bibfield  {author} {\bibinfo {author} {\bibfnamefont {S.}~\bibnamefont
  {Tanaka}}, \bibinfo {author} {\bibfnamefont {Y.}~\bibnamefont {Nakaya}},
  \bibinfo {author} {\bibfnamefont {R.}~\bibnamefont {Narikawa}}, \bibinfo
  {author} {\bibfnamefont {K.}~\bibnamefont {Ninomiya}}, \bibinfo {author}
  {\bibfnamefont {J.}~\bibnamefont {Onishi}}, \bibinfo {author} {\bibfnamefont
  {M.}~\bibnamefont {Pearson}}, \bibinfo {author} {\bibfnamefont
  {R.}~\bibnamefont {Openshaw}}, \bibinfo {author} {\bibfnamefont
  {S.}~\bibnamefont {Saiba}}, \bibinfo {author} {\bibfnamefont
  {R.}~\bibnamefont {Tanuma}}, \bibinfo {author} {\bibfnamefont
  {Y.}~\bibnamefont {Totsuka}},\ and\ \bibinfo {author} {\bibfnamefont
  {J.}~\bibnamefont {Murata}},\ }\href
  {https://doi.org/10.1051/epjconf/20146605021} {\bibfield  {journal} {\bibinfo
   {journal} {EPJ Web of Conferences}\ }\textbf {\bibinfo {volume} {66}},\
  \bibinfo {pages} {05021} (\bibinfo {year} {2014})}\BibitemShut {NoStop}%
\bibitem [{\citenamefont {Klimchitskaya}\ \emph {et~al.}(2019)\citenamefont
  {Klimchitskaya}, \citenamefont {Mostepanenko}, \citenamefont {Sedmik},\ and\
  \citenamefont {Abele}}]{CannexProposal}%
  \BibitemOpen
  \bibfield  {author} {\bibinfo {author} {\bibfnamefont {G.~L.}\ \bibnamefont
  {Klimchitskaya}}, \bibinfo {author} {\bibfnamefont {V.~M.}\ \bibnamefont
  {Mostepanenko}}, \bibinfo {author} {\bibfnamefont {R.~I.~P.}\ \bibnamefont
  {Sedmik}},\ and\ \bibinfo {author} {\bibfnamefont {H.}~\bibnamefont
  {Abele}},\ }\bibfield  {journal} {\bibinfo  {journal} {Symmetry}\ }\textbf
  {\bibinfo {volume} {11}},\ \href {https://doi.org/10.3390/sym11030407}
  {10.3390/sym11030407} (\bibinfo {year} {2019})\BibitemShut {NoStop}%
\bibitem [{\citenamefont {Hardy}\ and\ \citenamefont
  {Lasenby}(2017)}]{AstroConstraintsLasenby}%
  \BibitemOpen
  \bibfield  {author} {\bibinfo {author} {\bibfnamefont {E.}~\bibnamefont
  {Hardy}}\ and\ \bibinfo {author} {\bibfnamefont {R.}~\bibnamefont
  {Lasenby}},\ }\href {https://doi.org/10.1007/JHEP02(2017)033} {\bibfield
  {journal} {\bibinfo  {journal} {Journal of High Energy Physics}\ }\textbf
  {\bibinfo {volume} {2017}},\ \bibinfo {pages} {33} (\bibinfo {year}
  {2017})}\BibitemShut {NoStop}%
\bibitem [{\citenamefont {Bottaro}\ \emph {et~al.}(2023)\citenamefont
  {Bottaro}, \citenamefont {Caputo}, \citenamefont {Raffelt},\ and\
  \citenamefont {Vitagliano}}]{WDLuminosity}%
  \BibitemOpen
  \bibfield  {author} {\bibinfo {author} {\bibfnamefont {S.}~\bibnamefont
  {Bottaro}}, \bibinfo {author} {\bibfnamefont {A.}~\bibnamefont {Caputo}},
  \bibinfo {author} {\bibfnamefont {G.}~\bibnamefont {Raffelt}},\ and\ \bibinfo
  {author} {\bibfnamefont {E.}~\bibnamefont {Vitagliano}},\ }\href
  {https://doi.org/10.1088/1475-7516/2023/07/071} {\bibfield  {journal}
  {\bibinfo  {journal} {Journal of Cosmology and Astroparticle Physics}\
  }\textbf {\bibinfo {volume} {2023}}\bibinfo  {number} { (07)},\ \bibinfo
  {pages} {071}}\BibitemShut {NoStop}%
\bibitem [{\citenamefont {Brax}\ \emph {et~al.}(2007)\citenamefont {Brax},
  \citenamefont {van~de Bruck},\ and\ \citenamefont {Davis}}]{EvasionBrax}%
  \BibitemOpen
\bibfield  {number} {  }\bibfield  {author} {\bibinfo {author} {\bibfnamefont
  {P.}~\bibnamefont {Brax}}, \bibinfo {author} {\bibfnamefont {C.}~\bibnamefont
  {van~de Bruck}},\ and\ \bibinfo {author} {\bibfnamefont {A.-C.}\ \bibnamefont
  {Davis}},\ }\href {https://doi.org/10.1103/PhysRevLett.99.121103} {\bibfield
  {journal} {\bibinfo  {journal} {Phys. Rev. Lett.}\ }\textbf {\bibinfo
  {volume} {99}},\ \bibinfo {pages} {121103} (\bibinfo {year}
  {2007})}\BibitemShut {NoStop}%
\bibitem [{\citenamefont {DeRocco}\ \emph {et~al.}(2020)\citenamefont
  {DeRocco}, \citenamefont {Graham},\ and\ \citenamefont
  {Rajendran}}]{EvasionDeRocco}%
  \BibitemOpen
  \bibfield  {author} {\bibinfo {author} {\bibfnamefont {W.}~\bibnamefont
  {DeRocco}}, \bibinfo {author} {\bibfnamefont {P.~W.}\ \bibnamefont
  {Graham}},\ and\ \bibinfo {author} {\bibfnamefont {S.}~\bibnamefont
  {Rajendran}},\ }\href {https://doi.org/10.1103/PhysRevD.102.075015}
  {\bibfield  {journal} {\bibinfo  {journal} {Phys. Rev. D}\ }\textbf {\bibinfo
  {volume} {102}},\ \bibinfo {pages} {075015} (\bibinfo {year}
  {2020})}\BibitemShut {NoStop}%
\bibitem [{\citenamefont {Jaeckel}\ \emph {et~al.}(2007)\citenamefont
  {Jaeckel}, \citenamefont {Mass\'o}, \citenamefont {Redondo}, \citenamefont
  {Ringwald},\ and\ \citenamefont {Takahashi}}]{EvasionJaeckel}%
  \BibitemOpen
  \bibfield  {author} {\bibinfo {author} {\bibfnamefont {J.}~\bibnamefont
  {Jaeckel}}, \bibinfo {author} {\bibfnamefont {E.}~\bibnamefont {Mass\'o}},
  \bibinfo {author} {\bibfnamefont {J.}~\bibnamefont {Redondo}}, \bibinfo
  {author} {\bibfnamefont {A.}~\bibnamefont {Ringwald}},\ and\ \bibinfo
  {author} {\bibfnamefont {F.}~\bibnamefont {Takahashi}},\ }\href
  {https://doi.org/10.1103/PhysRevD.75.013004} {\bibfield  {journal} {\bibinfo
  {journal} {Phys. Rev. D}\ }\textbf {\bibinfo {volume} {75}},\ \bibinfo
  {pages} {013004} (\bibinfo {year} {2007})}\BibitemShut {NoStop}%
\bibitem [{\citenamefont {Jain}\ and\ \citenamefont
  {Mandal}(2006)}]{EvasionJain}%
  \BibitemOpen
  \bibfield  {author} {\bibinfo {author} {\bibfnamefont {P.}~\bibnamefont
  {Jain}}\ and\ \bibinfo {author} {\bibfnamefont {S.}~\bibnamefont {Mandal}},\
  }\href {https://doi.org/10.1142/s0218271806009558} {\bibfield  {journal}
  {\bibinfo  {journal} {International Journal of Modern Physics D}\ }\textbf
  {\bibinfo {volume} {15}},\ \bibinfo {pages} {2095} (\bibinfo {year}
  {2006})}\BibitemShut {NoStop}%
\bibitem [{\citenamefont {Mass\'{o}}\ and\ \citenamefont
  {Redondo}(2005)}]{EvasionMasso1}%
  \BibitemOpen
  \bibfield  {author} {\bibinfo {author} {\bibfnamefont {E.}~\bibnamefont
  {Mass\'{o}}}\ and\ \bibinfo {author} {\bibfnamefont {J.}~\bibnamefont
  {Redondo}},\ }\href {https://doi.org/10.1088/1475-7516/2005/09/015}
  {\bibfield  {journal} {\bibinfo  {journal} {Journal of Cosmology and
  Astroparticle Physics}\ }\textbf {\bibinfo {volume} {2005}}\bibinfo  {number}
  { (09)},\ \bibinfo {pages} {015}}\BibitemShut {NoStop}%
\bibitem [{\citenamefont {Mass\'{o}}\ and\ \citenamefont
  {Redondo}(2006)}]{EvasionMasso2}%
  \BibitemOpen
\bibfield  {number} {  }\bibfield  {author} {\bibinfo {author} {\bibfnamefont
  {E.}~\bibnamefont {Mass\'{o}}}\ and\ \bibinfo {author} {\bibfnamefont
  {J.}~\bibnamefont {Redondo}},\ }\href
  {https://doi.org/10.1103/PhysRevLett.97.151802} {\bibfield  {journal}
  {\bibinfo  {journal} {Phys. Rev. Lett.}\ }\textbf {\bibinfo {volume} {97}},\
  \bibinfo {pages} {151802} (\bibinfo {year} {2006})}\BibitemShut {NoStop}%
\bibitem [{\citenamefont {Mohapatra}\ and\ \citenamefont
  {Nasri}(2007)}]{EvasionMohapatra}%
  \BibitemOpen
  \bibfield  {author} {\bibinfo {author} {\bibfnamefont {R.~N.}\ \bibnamefont
  {Mohapatra}}\ and\ \bibinfo {author} {\bibfnamefont {S.}~\bibnamefont
  {Nasri}},\ }\href {https://doi.org/10.1103/PhysRevLett.98.050402} {\bibfield
  {journal} {\bibinfo  {journal} {Phys. Rev. Lett.}\ }\textbf {\bibinfo
  {volume} {98}},\ \bibinfo {pages} {050402} (\bibinfo {year}
  {2007})}\BibitemShut {NoStop}%
\bibitem [{\citenamefont {Raffelt}(2012)}]{AstroConstraintsRaffelt}%
  \BibitemOpen
  \bibfield  {author} {\bibinfo {author} {\bibfnamefont {G.}~\bibnamefont
  {Raffelt}},\ }\href {https://doi.org/10.1103/PhysRevD.86.015001} {\bibfield
  {journal} {\bibinfo  {journal} {Phys. Rev. D}\ }\textbf {\bibinfo {volume}
  {86}},\ \bibinfo {pages} {015001} (\bibinfo {year} {2012})}\BibitemShut
  {NoStop}%
\bibitem [{\citenamefont {O'Hare}\ and\ \citenamefont
  {Vitagliano}(2020)}]{AxionForceBounds}%
  \BibitemOpen
  \bibfield  {author} {\bibinfo {author} {\bibfnamefont {C.~A.~J.}\
  \bibnamefont {O'Hare}}\ and\ \bibinfo {author} {\bibfnamefont
  {E.}~\bibnamefont {Vitagliano}},\ }\href
  {https://doi.org/10.1103/PhysRevD.102.115026} {\bibfield  {journal} {\bibinfo
   {journal} {Phys. Rev. D}\ }\textbf {\bibinfo {volume} {102}},\ \bibinfo
  {pages} {115026} (\bibinfo {year} {2020})}\BibitemShut {NoStop}%
\bibitem [{\citenamefont
  {Sears}(1986{\natexlab{a}})}]{SearsNeutronAtomInteractions}%
  \BibitemOpen
  \bibfield  {author} {\bibinfo {author} {\bibfnamefont {V.}~\bibnamefont
  {Sears}},\ }\href
  {https://doi.org/https://doi.org/10.1016/0370-1573(86)90129-8} {\bibfield
  {journal} {\bibinfo  {journal} {Physics Reports}\ }\textbf {\bibinfo {volume}
  {141}},\ \bibinfo {pages} {281} (\bibinfo {year}
  {1986}{\natexlab{a}})}\BibitemShut {NoStop}%
\bibitem [{\citenamefont {Abele}(2008)}]{NeutronReviewAbele}%
  \BibitemOpen
  \bibfield  {author} {\bibinfo {author} {\bibfnamefont {H.}~\bibnamefont
  {Abele}},\ }\href
  {https://doi.org/https://doi.org/10.1016/j.ppnp.2007.05.002} {\bibfield
  {journal} {\bibinfo  {journal} {Progress in Particle and Nuclear Physics}\
  }\textbf {\bibinfo {volume} {60}},\ \bibinfo {pages} {1} (\bibinfo {year}
  {2008})}\BibitemShut {NoStop}%
\bibitem [{\citenamefont {Fermi}(1936)}]{FermiPseudopotentialOriginal}%
  \BibitemOpen
  \bibfield  {author} {\bibinfo {author} {\bibfnamefont {E.}~\bibnamefont
  {Fermi}},\ }\href@noop {} {\bibfield  {journal} {\bibinfo  {journal} {Ricerca
  Scientifica}\ }\textbf {\bibinfo {volume} {7}},\ \bibinfo {pages} {13}
  (\bibinfo {year} {1936})}\BibitemShut {NoStop}%
\bibitem [{\citenamefont {Koester}\ \emph {et~al.}(1991)\citenamefont
  {Koester}, \citenamefont {Rauch},\ and\ \citenamefont
  {Seymann}}]{NeutronScatteringKoester}%
  \BibitemOpen
  \bibfield  {author} {\bibinfo {author} {\bibfnamefont {L.}~\bibnamefont
  {Koester}}, \bibinfo {author} {\bibfnamefont {H.}~\bibnamefont {Rauch}},\
  and\ \bibinfo {author} {\bibfnamefont {E.}~\bibnamefont {Seymann}},\ }\href
  {https://doi.org/https://doi.org/10.1016/0092-640X(91)90012-S} {\bibfield
  {journal} {\bibinfo  {journal} {Atomic Data and Nuclear Data Tables}\
  }\textbf {\bibinfo {volume} {49}},\ \bibinfo {pages} {65} (\bibinfo {year}
  {1991})}\BibitemShut {NoStop}%
\bibitem [{\citenamefont
  {Sears}(1986{\natexlab{b}})}]{NeutronScatteringSearsAppendix}%
  \BibitemOpen
  \bibfield  {author} {\bibinfo {author} {\bibfnamefont {V.~F.}\ \bibnamefont
  {Sears}},\ }in\ \href
  {https://doi.org/https://doi.org/10.1016/S0076-695X(08)60561-X} {\emph
  {\bibinfo {booktitle} {Neutron Scattering}}},\ \bibinfo {series} {Methods in
  Experimental Physics}, Vol.~\bibinfo {volume} {23},\ \bibinfo {editor}
  {edited by\ \bibinfo {editor} {\bibfnamefont {K.}~\bibnamefont {Sk\"{o}ld}}\
  and\ \bibinfo {editor} {\bibfnamefont {D.~L.}\ \bibnamefont {Price}}}\
  (\bibinfo  {publisher} {Academic Press},\ \bibinfo {year} {1986})\ pp.\
  \bibinfo {pages} {521--550}\BibitemShut {NoStop}%
\bibitem [{\citenamefont {Blatt}\ and\ \citenamefont
  {Jackson}(1949)}]{NeutronEffectiveRangeBlatt}%
  \BibitemOpen
  \bibfield  {author} {\bibinfo {author} {\bibfnamefont {J.~M.}\ \bibnamefont
  {Blatt}}\ and\ \bibinfo {author} {\bibfnamefont {J.~D.}\ \bibnamefont
  {Jackson}},\ }\href {https://doi.org/10.1103/PhysRev.76.18} {\bibfield
  {journal} {\bibinfo  {journal} {Phys. Rev.}\ }\textbf {\bibinfo {volume}
  {76}},\ \bibinfo {pages} {18} (\bibinfo {year} {1949})}\BibitemShut {NoStop}%
\bibitem [{\citenamefont {Koester}\ \emph {et~al.}(1995)\citenamefont
  {Koester}, \citenamefont {Waschkowski}, \citenamefont {Mitsyna},
  \citenamefont {Samosvat}, \citenamefont {Prokofjevs},\ and\ \citenamefont
  {Tambergs}}]{NEScatteringKoester}%
  \BibitemOpen
  \bibfield  {author} {\bibinfo {author} {\bibfnamefont {L.}~\bibnamefont
  {Koester}}, \bibinfo {author} {\bibfnamefont {W.}~\bibnamefont
  {Waschkowski}}, \bibinfo {author} {\bibfnamefont {L.~V.}\ \bibnamefont
  {Mitsyna}}, \bibinfo {author} {\bibfnamefont {G.~S.}\ \bibnamefont
  {Samosvat}}, \bibinfo {author} {\bibfnamefont {P.}~\bibnamefont
  {Prokofjevs}},\ and\ \bibinfo {author} {\bibfnamefont {J.}~\bibnamefont
  {Tambergs}},\ }\href {https://doi.org/10.1103/PhysRevC.51.3363} {\bibfield
  {journal} {\bibinfo  {journal} {Phys. Rev. C}\ }\textbf {\bibinfo {volume}
  {51}},\ \bibinfo {pages} {3363} (\bibinfo {year} {1995})}\BibitemShut
  {NoStop}%
\bibitem [{\citenamefont {Hackenburg}(2006)}]{NeutronEffectiveRangeHackenburg}%
  \BibitemOpen
  \bibfield  {author} {\bibinfo {author} {\bibfnamefont {R.~W.}\ \bibnamefont
  {Hackenburg}},\ }\href {https://doi.org/10.1103/PhysRevC.73.044002}
  {\bibfield  {journal} {\bibinfo  {journal} {Phys. Rev. C}\ }\textbf {\bibinfo
  {volume} {73}},\ \bibinfo {pages} {044002} (\bibinfo {year}
  {2006})}\BibitemShut {NoStop}%
\bibitem [{\citenamefont {Kopecky}\ \emph {et~al.}(1997)\citenamefont
  {Kopecky}, \citenamefont {Krenn}, \citenamefont {Riehs}, \citenamefont
  {Steiner}, \citenamefont {Harvey}, \citenamefont {Hill},\ and\ \citenamefont
  {Pernicka}}]{NEScatteringKopecky}%
  \BibitemOpen
  \bibfield  {author} {\bibinfo {author} {\bibfnamefont {S.}~\bibnamefont
  {Kopecky}}, \bibinfo {author} {\bibfnamefont {M.}~\bibnamefont {Krenn}},
  \bibinfo {author} {\bibfnamefont {P.}~\bibnamefont {Riehs}}, \bibinfo
  {author} {\bibfnamefont {S.}~\bibnamefont {Steiner}}, \bibinfo {author}
  {\bibfnamefont {J.~A.}\ \bibnamefont {Harvey}}, \bibinfo {author}
  {\bibfnamefont {N.~W.}\ \bibnamefont {Hill}},\ and\ \bibinfo {author}
  {\bibfnamefont {M.}~\bibnamefont {Pernicka}},\ }\href
  {https://doi.org/10.1103/PhysRevC.56.2229} {\bibfield  {journal} {\bibinfo
  {journal} {Phys. Rev. C}\ }\textbf {\bibinfo {volume} {56}},\ \bibinfo
  {pages} {2229} (\bibinfo {year} {1997})}\BibitemShut {NoStop}%
\bibitem [{\citenamefont {Workman}\ and\ \citenamefont
  {Others}(2022)}]{PDG2022}%
  \BibitemOpen
  \bibfield  {author} {\bibinfo {author} {\bibfnamefont {R.~L.}\ \bibnamefont
  {Workman}}\ and\ \bibinfo {author} {\bibnamefont {Others}} (\bibinfo
  {collaboration} {Particle Data Group}),\ }\href
  {https://doi.org/10.1093/ptep/ptac097} {\bibfield  {journal} {\bibinfo
  {journal} {PTEP}\ }\textbf {\bibinfo {volume} {2022}},\ \bibinfo {pages}
  {083C01} (\bibinfo {year} {2022})}\BibitemShut {NoStop}%
\bibitem [{\citenamefont {Marshall}\ and\ \citenamefont
  {Lovesey}(1971)}]{AngularMomentumFormFactorBook}%
  \BibitemOpen
  \bibfield  {author} {\bibinfo {author} {\bibfnamefont {W.}~\bibnamefont
  {Marshall}}\ and\ \bibinfo {author} {\bibfnamefont {S.~W.}\ \bibnamefont
  {Lovesey}},\ }\href@noop {} {\emph {\bibinfo {title} {Theory of thermal
  neutron scattering: the use of neutrons for the investigation of condensed
  matter}}}\ (\bibinfo  {publisher} {Clarendon Press},\ \bibinfo {address}
  {Oxford},\ \bibinfo {year} {1971})\BibitemShut {NoStop}%
\bibitem [{\citenamefont {Schwinger}(1948)}]{SchwingerTerm}%
  \BibitemOpen
  \bibfield  {author} {\bibinfo {author} {\bibfnamefont {J.}~\bibnamefont
  {Schwinger}},\ }\href {https://doi.org/10.1103/PhysRev.73.407} {\bibfield
  {journal} {\bibinfo  {journal} {Phys. Rev.}\ }\textbf {\bibinfo {volume}
  {73}},\ \bibinfo {pages} {407} (\bibinfo {year} {1948})}\BibitemShut
  {NoStop}%
\bibitem [{\citenamefont {Foldy}(1952)}]{FoldyTermOriginal}%
  \BibitemOpen
  \bibfield  {author} {\bibinfo {author} {\bibfnamefont {L.~L.}\ \bibnamefont
  {Foldy}},\ }\href {https://doi.org/10.1103/PhysRev.87.688} {\bibfield
  {journal} {\bibinfo  {journal} {Phys. Rev.}\ }\textbf {\bibinfo {volume}
  {87}},\ \bibinfo {pages} {688} (\bibinfo {year} {1952})}\BibitemShut
  {NoStop}%
\bibitem [{\citenamefont {Bawin}\ and\ \citenamefont
  {Coon}(1999)}]{FoldyTermDiracEquation}%
  \BibitemOpen
  \bibfield  {author} {\bibinfo {author} {\bibfnamefont {M.}~\bibnamefont
  {Bawin}}\ and\ \bibinfo {author} {\bibfnamefont {S.~A.}\ \bibnamefont
  {Coon}},\ }\href {https://doi.org/10.1103/PhysRevC.60.025207} {\bibfield
  {journal} {\bibinfo  {journal} {Phys. Rev. C}\ }\textbf {\bibinfo {volume}
  {60}},\ \bibinfo {pages} {025207} (\bibinfo {year} {1999})}\BibitemShut
  {NoStop}%
\bibitem [{\citenamefont {Isgur}(1999)}]{FoldyTermFormFactor}%
  \BibitemOpen
  \bibfield  {author} {\bibinfo {author} {\bibfnamefont {N.}~\bibnamefont
  {Isgur}},\ }\href {https://doi.org/10.1103/PhysRevLett.83.272} {\bibfield
  {journal} {\bibinfo  {journal} {Phys. Rev. Lett.}\ }\textbf {\bibinfo
  {volume} {83}},\ \bibinfo {pages} {272} (\bibinfo {year} {1999})}\BibitemShut
  {NoStop}%
\bibitem [{\citenamefont {Blume}(1985)}]{XRayScatteringBlume}%
  \BibitemOpen
  \bibfield  {author} {\bibinfo {author} {\bibfnamefont {M.}~\bibnamefont
  {Blume}},\ }\href {https://doi.org/10.1063/1.335023} {\bibfield  {journal}
  {\bibinfo  {journal} {Journal of Applied Physics}\ }\textbf {\bibinfo
  {volume} {57}},\ \bibinfo {pages} {3615} (\bibinfo {year}
  {1985})}\BibitemShut {NoStop}%
\bibitem [{\citenamefont {Blume}\ and\ \citenamefont
  {Gibbs}(1988)}]{XRayScatteringGibbs}%
  \BibitemOpen
  \bibfield  {author} {\bibinfo {author} {\bibfnamefont {M.}~\bibnamefont
  {Blume}}\ and\ \bibinfo {author} {\bibfnamefont {D.}~\bibnamefont {Gibbs}},\
  }\href {https://doi.org/10.1103/PhysRevB.37.1779} {\bibfield  {journal}
  {\bibinfo  {journal} {Phys. Rev. B}\ }\textbf {\bibinfo {volume} {37}},\
  \bibinfo {pages} {1779} (\bibinfo {year} {1988})}\BibitemShut {NoStop}%
\bibitem [{\citenamefont {Stirling}\ and\ \citenamefont
  {Cooper}(1999)}]{XRayScatteringStirling}%
  \BibitemOpen
  \bibfield  {author} {\bibinfo {author} {\bibfnamefont {W.}~\bibnamefont
  {Stirling}}\ and\ \bibinfo {author} {\bibfnamefont {M.}~\bibnamefont
  {Cooper}},\ }\href
  {https://doi.org/https://doi.org/10.1016/S0304-8853(99)00307-8} {\bibfield
  {journal} {\bibinfo  {journal} {Journal of Magnetism and Magnetic Materials}\
  }\textbf {\bibinfo {volume} {200}},\ \bibinfo {pages} {755} (\bibinfo {year}
  {1999})}\BibitemShut {NoStop}%
\bibitem [{\citenamefont {Levinger}(1951)}]{PhysRev.84.523}%
  \BibitemOpen
  \bibfield  {author} {\bibinfo {author} {\bibfnamefont {J.~S.}\ \bibnamefont
  {Levinger}},\ }\href {https://doi.org/10.1103/PhysRev.84.523} {\bibfield
  {journal} {\bibinfo  {journal} {Phys. Rev.}\ }\textbf {\bibinfo {volume}
  {84}},\ \bibinfo {pages} {523} (\bibinfo {year} {1951})}\BibitemShut
  {NoStop}%
\bibitem [{\citenamefont {Henke}\ \emph {et~al.}(1993)\citenamefont {Henke},
  \citenamefont {Gullikson},\ and\ \citenamefont {Davis}}]{XRayAttenuation}%
  \BibitemOpen
  \bibfield  {author} {\bibinfo {author} {\bibfnamefont {B.}~\bibnamefont
  {Henke}}, \bibinfo {author} {\bibfnamefont {E.}~\bibnamefont {Gullikson}},\
  and\ \bibinfo {author} {\bibfnamefont {J.}~\bibnamefont {Davis}},\ }\href
  {https://doi.org/https://doi.org/10.1006/adnd.1993.1013} {\bibfield
  {journal} {\bibinfo  {journal} {Atomic Data and Nuclear Data Tables}\
  }\textbf {\bibinfo {volume} {54}},\ \bibinfo {pages} {181} (\bibinfo {year}
  {1993})}\BibitemShut {NoStop}%
\bibitem [{\citenamefont {Mitzenmacher}\ and\ \citenamefont
  {Upfal}(2017)}]{PoissonTailBound}%
  \BibitemOpen
  \bibfield  {author} {\bibinfo {author} {\bibfnamefont {M.}~\bibnamefont
  {Mitzenmacher}}\ and\ \bibinfo {author} {\bibfnamefont {E.}~\bibnamefont
  {Upfal}},\ }\href@noop {} {\emph {\bibinfo {title} {Probability and
  computing: randomization and probabilistic techniques in algorithms and data
  analysis}}}\ (\bibinfo  {publisher} {Cambridge University Press},\ \bibinfo
  {address} {Cambridge, United Kingdom; New York, NY},\ \bibinfo {year}
  {2017})\BibitemShut {NoStop}%
\bibitem [{\citenamefont {Majkrzak}\ \emph {et~al.}(2014)\citenamefont
  {Majkrzak}, \citenamefont {Metting}, \citenamefont {Maranville},
  \citenamefont {Dura}, \citenamefont {Satija}, \citenamefont {Udovic},\ and\
  \citenamefont {Berk}}]{TransvereSizeMajkrzak}%
  \BibitemOpen
  \bibfield  {author} {\bibinfo {author} {\bibfnamefont {C.~F.}\ \bibnamefont
  {Majkrzak}}, \bibinfo {author} {\bibfnamefont {C.}~\bibnamefont {Metting}},
  \bibinfo {author} {\bibfnamefont {B.~B.}\ \bibnamefont {Maranville}},
  \bibinfo {author} {\bibfnamefont {J.~A.}\ \bibnamefont {Dura}}, \bibinfo
  {author} {\bibfnamefont {S.}~\bibnamefont {Satija}}, \bibinfo {author}
  {\bibfnamefont {T.}~\bibnamefont {Udovic}},\ and\ \bibinfo {author}
  {\bibfnamefont {N.~F.}\ \bibnamefont {Berk}},\ }\href
  {https://doi.org/10.1103/PhysRevA.89.033851} {\bibfield  {journal} {\bibinfo
  {journal} {Phys. Rev. A}\ }\textbf {\bibinfo {volume} {89}},\ \bibinfo
  {pages} {033851} (\bibinfo {year} {2014})}\BibitemShut {NoStop}%
\bibitem [{\citenamefont {Berk}(2014)}]{TransverseSizeBerk}%
  \BibitemOpen
  \bibfield  {author} {\bibinfo {author} {\bibfnamefont {N.~F.}\ \bibnamefont
  {Berk}},\ }\href {https://doi.org/10.1103/PhysRevA.89.033852} {\bibfield
  {journal} {\bibinfo  {journal} {Phys. Rev. A}\ }\textbf {\bibinfo {volume}
  {89}},\ \bibinfo {pages} {033852} (\bibinfo {year} {2014})}\BibitemShut
  {NoStop}%
\bibitem [{\citenamefont {Ebrahimi}\ \emph {et~al.}(2010)\citenamefont
  {Ebrahimi}, \citenamefont {Treimer}, \citenamefont {Strobl}, \citenamefont
  {Feye-Treimer}, \citenamefont {Beul}, \citenamefont {Jericha},\ and\
  \citenamefont {Seidel}}]{TransverseSizeEbrahimi}%
  \BibitemOpen
  \bibfield  {author} {\bibinfo {author} {\bibfnamefont {O.}~\bibnamefont
  {Ebrahimi}}, \bibinfo {author} {\bibfnamefont {W.}~\bibnamefont {Treimer}},
  \bibinfo {author} {\bibfnamefont {M.}~\bibnamefont {Strobl}}, \bibinfo
  {author} {\bibfnamefont {U.}~\bibnamefont {Feye-Treimer}}, \bibinfo {author}
  {\bibfnamefont {N.}~\bibnamefont {Beul}}, \bibinfo {author} {\bibfnamefont
  {E.}~\bibnamefont {Jericha}},\ and\ \bibinfo {author} {\bibfnamefont {S.-O.}\
  \bibnamefont {Seidel}},\ }\href
  {https://doi.org/10.1088/1742-6596/251/1/012072} {\bibfield  {journal}
  {\bibinfo  {journal} {Journal of Physics: Conference Series}\ }\textbf
  {\bibinfo {volume} {251}} (\bibinfo {year} {2010})}\BibitemShut {NoStop}%
\bibitem [{\citenamefont {Treimer}\ \emph {et~al.}(2006)\citenamefont
  {Treimer}, \citenamefont {Hilger},\ and\ \citenamefont
  {Strobl}}]{TransverseSizeTreimer}%
  \BibitemOpen
  \bibfield  {author} {\bibinfo {author} {\bibfnamefont {W.}~\bibnamefont
  {Treimer}}, \bibinfo {author} {\bibfnamefont {A.}~\bibnamefont {Hilger}},\
  and\ \bibinfo {author} {\bibfnamefont {M.}~\bibnamefont {Strobl}},\ }\href
  {https://doi.org/https://doi.org/10.1016/j.physb.2006.05.205} {\bibfield
  {journal} {\bibinfo  {journal} {Physica B: Condensed Matter}\ }\textbf
  {\bibinfo {volume} {385-386}},\ \bibinfo {pages} {1388} (\bibinfo {year}
  {2006})}\BibitemShut {NoStop}%
\bibitem [{\citenamefont {Berk}(1993)}]{NeutronScatteringIntroBerk}%
  \BibitemOpen
  \bibfield  {author} {\bibinfo {author} {\bibfnamefont {N.~F.}\ \bibnamefont
  {Berk}},\ }\href {https://doi.org/10.6028/jres.098.002} {\bibfield  {journal}
  {\bibinfo  {journal} {Journal of research of the National Institute of
  Standards and Technology}\ }\textbf {\bibinfo {volume} {98}},\ \bibinfo
  {pages} {15} (\bibinfo {year} {1993})}\BibitemShut {NoStop}%
\bibitem [{\citenamefont {Pritychenko}\ and\ \citenamefont
  {Mughabghab}(2012)}]{ENDFPaper}%
  \BibitemOpen
  \bibfield  {author} {\bibinfo {author} {\bibfnamefont {B.}~\bibnamefont
  {Pritychenko}}\ and\ \bibinfo {author} {\bibfnamefont {S.}~\bibnamefont
  {Mughabghab}},\ }\href
  {https://doi.org/https://doi.org/10.1016/j.nds.2012.11.007} {\bibfield
  {journal} {\bibinfo  {journal} {Nuclear Data Sheets}\ }\textbf {\bibinfo
  {volume} {113}},\ \bibinfo {pages} {3120} (\bibinfo {year}
  {2012})}\BibitemShut {NoStop}%
\bibitem [{\citenamefont {Sansonetti}\ and\ \citenamefont
  {Martin}(2005)}]{NISTSpectroscopicTable}%
  \BibitemOpen
  \bibfield  {author} {\bibinfo {author} {\bibfnamefont {J.~E.}\ \bibnamefont
  {Sansonetti}}\ and\ \bibinfo {author} {\bibfnamefont {W.~C.}\ \bibnamefont
  {Martin}},\ }\href {https://doi.org/10.1063/1.1800011} {\bibfield  {journal}
  {\bibinfo  {journal} {Journal of Physical and Chemical Reference Data}\
  }\textbf {\bibinfo {volume} {34}},\ \bibinfo {pages} {1559} (\bibinfo {year}
  {2005})},\ \Eprint
  {https://arxiv.org/abs/https://pubs.aip.org/aip/jpr/article-pdf/34/4/1559/8183749/1559\_1\_online.pdf}
  {https://pubs.aip.org/aip/jpr/article-pdf/34/4/1559/8183749/1559\_1\_online.pdf}
  \BibitemShut {NoStop}%
\bibitem [{\citenamefont {Kramida}\ \emph {et~al.}(2021)\citenamefont
  {Kramida}, \citenamefont {{Yu. Ralchenko}}, \citenamefont {Reader},\ and\
  \citenamefont {{and NIST ASD Team}}}]{NIST_ASD}%
  \BibitemOpen
  \bibfield  {author} {\bibinfo {author} {\bibfnamefont {A.}~\bibnamefont
  {Kramida}}, \bibinfo {author} {\bibnamefont {{Yu. Ralchenko}}}, \bibinfo
  {author} {\bibfnamefont {J.}~\bibnamefont {Reader}},\ and\ \bibinfo {author}
  {\bibnamefont {{and NIST ASD Team}}},\ }\href@noop {} {}\bibinfo
  {howpublished} {{NIST Atomic Spectra Database (ver. 5.9), [Online].
  Available: {\tt{https://physics.nist.gov/asd}} [2022, April 4]. National
  Institute of Standards and Technology, Gaithersburg, MD.}} (\bibinfo {year}
  {2021})\BibitemShut {NoStop}%
\bibitem [{\citenamefont {Humphreys}\ and\ \citenamefont
  {Paul}(1970)}]{XenonSpectroscopicData}%
  \BibitemOpen
  \bibfield  {author} {\bibinfo {author} {\bibfnamefont {C.~J.}\ \bibnamefont
  {Humphreys}}\ and\ \bibinfo {author} {\bibfnamefont {E.}~\bibnamefont
  {Paul}},\ }\href {https://doi.org/10.1364/JOSA.60.001302} {\bibfield
  {journal} {\bibinfo  {journal} {J. Opt. Soc. Am.}\ }\textbf {\bibinfo
  {volume} {60}},\ \bibinfo {pages} {1302} (\bibinfo {year}
  {1970})}\BibitemShut {NoStop}%
\bibitem [{\citenamefont {Attard}(2002)}]{LiquidsBookAttard}%
  \BibitemOpen
  \bibfield  {author} {\bibinfo {author} {\bibfnamefont {P.}~\bibnamefont
  {Attard}},\ }\href@noop {} {\emph {\bibinfo {title} {Thermodynamics and
  Statistical Mechanics}}}\ (\bibinfo  {publisher} {Academic Press},\ \bibinfo
  {address} {London},\ \bibinfo {year} {2002})\BibitemShut {NoStop}%
\bibitem [{\citenamefont {Hansen}\ and\ \citenamefont
  {McDonald}(2006)}]{LiquidsBookHansen}%
  \BibitemOpen
  \bibfield  {author} {\bibinfo {author} {\bibfnamefont {J.-P.}\ \bibnamefont
  {Hansen}}\ and\ \bibinfo {author} {\bibfnamefont {I.}~\bibnamefont
  {McDonald}},\ }\href@noop {} {\emph {\bibinfo {title} {Theory of Simple
  Liquids}}}\ (\bibinfo  {publisher} {Academic Press},\ \bibinfo {address}
  {London},\ \bibinfo {year} {2006})\BibitemShut {NoStop}%
\bibitem [{\citenamefont {Sekerka}(2015)}]{LiquidsBookSekerka}%
  \BibitemOpen
  \bibfield  {author} {\bibinfo {author} {\bibfnamefont {R.}~\bibnamefont
  {Sekerka}},\ }\href@noop {} {\emph {\bibinfo {title} {Thermal Physics:
  Thermodynamics and Statistical Mechanics for Scientists and Engineers}}}\
  (\bibinfo  {publisher} {Elsevier},\ \bibinfo {year} {2015})\BibitemShut
  {NoStop}%
\bibitem [{\citenamefont {Powles}(1973)}]{StructureOfLiquids}%
  \BibitemOpen
  \bibfield  {author} {\bibinfo {author} {\bibfnamefont {J.}~\bibnamefont
  {Powles}},\ }\href {https://doi.org/10.1080/00018737300101259} {\bibfield
  {journal} {\bibinfo  {journal} {Advances in Physics}\ }\textbf {\bibinfo
  {volume} {22}},\ \bibinfo {pages} {1} (\bibinfo {year} {1973})}\BibitemShut
  {NoStop}%
\bibitem [{\citenamefont {Gläser}(1990)}]{DynamicsOfLiquids}%
  \BibitemOpen
  \bibfield  {author} {\bibinfo {author} {\bibfnamefont {W.}~\bibnamefont
  {Gläser}},\ }\href
  {https://doi.org/https://doi.org/10.1002/bbpc.19900940323} {\bibfield
  {journal} {\bibinfo  {journal} {Berichte der Bunsengesellschaft für
  physikalische Chemie}\ }\textbf {\bibinfo {volume} {94}},\ \bibinfo {pages}
  {315} (\bibinfo {year} {1990})}\BibitemShut {NoStop}%
\bibitem [{\citenamefont {Rahman}\ \emph {et~al.}(1962)\citenamefont {Rahman},
  \citenamefont {Singwi},\ and\ \citenamefont
  {Sj\"olander}}]{SlowNeutronsLiquids}%
  \BibitemOpen
  \bibfield  {author} {\bibinfo {author} {\bibfnamefont {A.}~\bibnamefont
  {Rahman}}, \bibinfo {author} {\bibfnamefont {K.~S.}\ \bibnamefont {Singwi}},\
  and\ \bibinfo {author} {\bibfnamefont {A.}~\bibnamefont {Sj\"olander}},\
  }\href {https://doi.org/10.1103/PhysRev.126.986} {\bibfield  {journal}
  {\bibinfo  {journal} {Phys. Rev.}\ }\textbf {\bibinfo {volume} {126}},\
  \bibinfo {pages} {986} (\bibinfo {year} {1962})}\BibitemShut {NoStop}%
\bibitem [{\citenamefont {Van~Hove}(1954)}]{BornCorrelations}%
  \BibitemOpen
  \bibfield  {author} {\bibinfo {author} {\bibfnamefont {L.}~\bibnamefont
  {Van~Hove}},\ }\href {https://doi.org/10.1103/PhysRev.95.249} {\bibfield
  {journal} {\bibinfo  {journal} {Phys. Rev.}\ }\textbf {\bibinfo {volume}
  {95}},\ \bibinfo {pages} {249} (\bibinfo {year} {1954})}\BibitemShut
  {NoStop}%
\bibitem [{\citenamefont {Fujiwara}\ \emph {et~al.}(2022)\citenamefont
  {Fujiwara}, \citenamefont {Miyoshi}, \citenamefont {Mitsuya}, \citenamefont
  {Yamada}, \citenamefont {Wakabayashi}, \citenamefont {Otake}, \citenamefont
  {Hino}, \citenamefont {Kino}, \citenamefont {Tanaka}, \citenamefont
  {Oshima},\ and\ \citenamefont {Takahashi}}]{PixelSizeFujiwara}%
  \BibitemOpen
  \bibfield  {author} {\bibinfo {author} {\bibfnamefont {T.}~\bibnamefont
  {Fujiwara}}, \bibinfo {author} {\bibfnamefont {H.}~\bibnamefont {Miyoshi}},
  \bibinfo {author} {\bibfnamefont {Y.}~\bibnamefont {Mitsuya}}, \bibinfo
  {author} {\bibfnamefont {N.~L.}\ \bibnamefont {Yamada}}, \bibinfo {author}
  {\bibfnamefont {Y.}~\bibnamefont {Wakabayashi}}, \bibinfo {author}
  {\bibfnamefont {Y.}~\bibnamefont {Otake}}, \bibinfo {author} {\bibfnamefont
  {M.}~\bibnamefont {Hino}}, \bibinfo {author} {\bibfnamefont {K.}~\bibnamefont
  {Kino}}, \bibinfo {author} {\bibfnamefont {M.}~\bibnamefont {Tanaka}},
  \bibinfo {author} {\bibfnamefont {N.}~\bibnamefont {Oshima}},\ and\ \bibinfo
  {author} {\bibfnamefont {H.}~\bibnamefont {Takahashi}},\ }\href
  {https://doi.org/10.1063/5.0066557} {\bibfield  {journal} {\bibinfo
  {journal} {Review of Scientific Instruments}\ }\textbf {\bibinfo {volume}
  {93}},\ \bibinfo {pages} {013304} (\bibinfo {year} {2022})}\BibitemShut
  {NoStop}%
\bibitem [{\citenamefont {Herrera}\ \emph {et~al.}(2016)\citenamefont
  {Herrera}, \citenamefont {Hamm}, \citenamefont {Wiggins}, \citenamefont
  {Milburn}, \citenamefont {Burger}, \citenamefont {Bilheux}, \citenamefont
  {Santodonato}, \citenamefont {Chvala}, \citenamefont {Stowe},\ and\
  \citenamefont {Lukosi}}]{PixelSizeHerrera}%
  \BibitemOpen
  \bibfield  {author} {\bibinfo {author} {\bibfnamefont {E.}~\bibnamefont
  {Herrera}}, \bibinfo {author} {\bibfnamefont {D.}~\bibnamefont {Hamm}},
  \bibinfo {author} {\bibfnamefont {B.}~\bibnamefont {Wiggins}}, \bibinfo
  {author} {\bibfnamefont {R.}~\bibnamefont {Milburn}}, \bibinfo {author}
  {\bibfnamefont {A.}~\bibnamefont {Burger}}, \bibinfo {author} {\bibfnamefont
  {H.}~\bibnamefont {Bilheux}}, \bibinfo {author} {\bibfnamefont
  {L.}~\bibnamefont {Santodonato}}, \bibinfo {author} {\bibfnamefont
  {O.}~\bibnamefont {Chvala}}, \bibinfo {author} {\bibfnamefont
  {A.}~\bibnamefont {Stowe}},\ and\ \bibinfo {author} {\bibfnamefont
  {E.}~\bibnamefont {Lukosi}},\ }\href
  {https://doi.org/https://doi.org/10.1016/j.nima.2016.07.035} {\bibfield
  {journal} {\bibinfo  {journal} {Nuclear Instruments and Methods in Physics
  Research Section A: Accelerators, Spectrometers, Detectors and Associated
  Equipment}\ }\textbf {\bibinfo {volume} {833}},\ \bibinfo {pages} {142}
  (\bibinfo {year} {2016})}\BibitemShut {NoStop}%
\bibitem [{\citenamefont {Jakubek}\ \emph {et~al.}(2004)\citenamefont
  {Jakubek}, \citenamefont {Holy}, \citenamefont {Lehmann}, \citenamefont
  {Pospisil}, \citenamefont {Uher}, \citenamefont {Vacik}, \citenamefont
  {Vavrik},\ and\ \citenamefont {Author}}]{PixelSizeJakubekHoly}%
  \BibitemOpen
  \bibfield  {author} {\bibinfo {author} {\bibfnamefont {J.}~\bibnamefont
  {Jakubek}}, \bibinfo {author} {\bibfnamefont {T.}~\bibnamefont {Holy}},
  \bibinfo {author} {\bibfnamefont {E.}~\bibnamefont {Lehmann}}, \bibinfo
  {author} {\bibfnamefont {S.}~\bibnamefont {Pospisil}}, \bibinfo {author}
  {\bibfnamefont {J.}~\bibnamefont {Uher}}, \bibinfo {author} {\bibfnamefont
  {J.}~\bibnamefont {Vacik}}, \bibinfo {author} {\bibfnamefont
  {D.}~\bibnamefont {Vavrik}},\ and\ \bibinfo {author} {\bibfnamefont
  {S.}~\bibnamefont {Author}},\ }in\ \href
  {https://doi.org/10.1109/NSSMIC.2004.1462363} {\emph {\bibinfo {booktitle}
  {IEEE Symposium Conference Record Nuclear Science 2004.}}},\ Vol.~\bibinfo
  {volume} {2}\ (\bibinfo {year} {2004})\ pp.\ \bibinfo {pages} {945--949 Vol.
  2}\BibitemShut {NoStop}%
\bibitem [{\citenamefont {Jakubek}\ and\ \citenamefont
  {Uher}(2009)}]{PixelSizeJakubekUher}%
  \BibitemOpen
  \bibfield  {author} {\bibinfo {author} {\bibfnamefont {J.}~\bibnamefont
  {Jakubek}}\ and\ \bibinfo {author} {\bibfnamefont {J.}~\bibnamefont {Uher}},\
  }in\ \href {https://doi.org/10.1109/NSSMIC.2009.5402420} {\emph {\bibinfo
  {booktitle} {2009 IEEE Nuclear Science Symposium Conference Record
  (NSS/MIC)}}}\ (\bibinfo {year} {2009})\ pp.\ \bibinfo {pages}
  {1113--1116}\BibitemShut {NoStop}%
\bibitem [{\citenamefont {Treimer}\ and\ \citenamefont
  {K\"{o}hler}(2021)}]{PixelSizeKohler}%
  \BibitemOpen
  \bibfield  {author} {\bibinfo {author} {\bibfnamefont {W.}~\bibnamefont
  {Treimer}}\ and\ \bibinfo {author} {\bibfnamefont {R.}~\bibnamefont
  {K\"{o}hler}},\ }\bibfield  {journal} {\bibinfo  {journal} {Applied
  Sciences}\ }\textbf {\bibinfo {volume} {11}},\ \href
  {https://doi.org/10.3390/app11156973} {10.3390/app11156973} (\bibinfo {year}
  {2021})\BibitemShut {NoStop}%
\bibitem [{\citenamefont {Losko}\ \emph {et~al.}(2021)\citenamefont {Losko},
  \citenamefont {Han}, \citenamefont {Schillinger}, \citenamefont
  {Tartaglione}, \citenamefont {Morgano}, \citenamefont {Strobl}, \citenamefont
  {Long}, \citenamefont {Tremsin},\ and\ \citenamefont
  {Schulz}}]{PixelSizeLosko}%
  \BibitemOpen
  \bibfield  {author} {\bibinfo {author} {\bibfnamefont {A.~S.}\ \bibnamefont
  {Losko}}, \bibinfo {author} {\bibfnamefont {Y.}~\bibnamefont {Han}}, \bibinfo
  {author} {\bibfnamefont {B.}~\bibnamefont {Schillinger}}, \bibinfo {author}
  {\bibfnamefont {A.}~\bibnamefont {Tartaglione}}, \bibinfo {author}
  {\bibfnamefont {M.}~\bibnamefont {Morgano}}, \bibinfo {author} {\bibfnamefont
  {M.}~\bibnamefont {Strobl}}, \bibinfo {author} {\bibfnamefont
  {J.}~\bibnamefont {Long}}, \bibinfo {author} {\bibfnamefont {A.~S.}\
  \bibnamefont {Tremsin}},\ and\ \bibinfo {author} {\bibfnamefont
  {M.}~\bibnamefont {Schulz}},\ }\href
  {https://doi.org/10.1038/s41598-021-00822-5} {\bibfield  {journal} {\bibinfo
  {journal} {Scientific Reports}\ }\textbf {\bibinfo {volume} {11}},\ \bibinfo
  {pages} {21360} (\bibinfo {year} {2021})}\BibitemShut {NoStop}%
\bibitem [{\citenamefont {Radulescu}\ \emph {et~al.}(2016)\citenamefont
  {Radulescu}, \citenamefont {Szekely}, \citenamefont {Appavou}, \citenamefont
  {Pipich}, \citenamefont {Kohnke}, \citenamefont {Ossovyi}, \citenamefont
  {Staringer}, \citenamefont {Schneider}, \citenamefont {Amann}, \citenamefont
  {Zhang-Haagen}, \citenamefont {Brandl}, \citenamefont {Drochner},
  \citenamefont {Engels}, \citenamefont {Hanslik},\ and\ \citenamefont
  {Kemmerling}}]{NeutronLensUseExample}%
  \BibitemOpen
  \bibfield  {author} {\bibinfo {author} {\bibfnamefont {A.}~\bibnamefont
  {Radulescu}}, \bibinfo {author} {\bibfnamefont {N.~K.}\ \bibnamefont
  {Szekely}}, \bibinfo {author} {\bibfnamefont {M.-S.}\ \bibnamefont
  {Appavou}}, \bibinfo {author} {\bibfnamefont {V.}~\bibnamefont {Pipich}},
  \bibinfo {author} {\bibfnamefont {T.}~\bibnamefont {Kohnke}}, \bibinfo
  {author} {\bibfnamefont {V.}~\bibnamefont {Ossovyi}}, \bibinfo {author}
  {\bibfnamefont {S.}~\bibnamefont {Staringer}}, \bibinfo {author}
  {\bibfnamefont {G.~J.}\ \bibnamefont {Schneider}}, \bibinfo {author}
  {\bibfnamefont {M.}~\bibnamefont {Amann}}, \bibinfo {author} {\bibfnamefont
  {B.}~\bibnamefont {Zhang-Haagen}}, \bibinfo {author} {\bibfnamefont
  {G.}~\bibnamefont {Brandl}}, \bibinfo {author} {\bibfnamefont
  {M.}~\bibnamefont {Drochner}}, \bibinfo {author} {\bibfnamefont
  {R.}~\bibnamefont {Engels}}, \bibinfo {author} {\bibfnamefont
  {R.}~\bibnamefont {Hanslik}},\ and\ \bibinfo {author} {\bibfnamefont
  {G.}~\bibnamefont {Kemmerling}},\ }\href {https://doi.org/10.3791/54639}
  {\bibfield  {journal} {\bibinfo  {journal} {JoVE}\ ,\ \bibinfo {pages}
  {e54639}} (\bibinfo {year} {2016})}\BibitemShut {NoStop}%
\bibitem [{\citenamefont {Sears}(1989)}]{NeutronOpticsSears}%
  \BibitemOpen
  \bibfield  {author} {\bibinfo {author} {\bibfnamefont {V.~F.}\ \bibnamefont
  {Sears}},\ }\href@noop {} {\emph {\bibinfo {title} {Neutron optics : an
  introduction to the theory of neutron optical phenomena and their
  applications}}}\ (\bibinfo  {publisher} {Oxford University Press},\ \bibinfo
  {year} {1989})\BibitemShut {NoStop}%
\bibitem [{\citenamefont {Barker}\ \emph {et~al.}(2022)\citenamefont {Barker},
  \citenamefont {Moyer}, \citenamefont {Kline}, \citenamefont {Jensen},
  \citenamefont {Cook}, \citenamefont {Gagnon}, \citenamefont {Kelley},
  \citenamefont {Chabot}, \citenamefont {Maliszewskyj}, \citenamefont {Parikh},
  \citenamefont {Chen}, \citenamefont {Murphy},\ and\ \citenamefont
  {Glinka}}]{NIST_VSANS}%
  \BibitemOpen
  \bibfield  {author} {\bibinfo {author} {\bibfnamefont {J.}~\bibnamefont
  {Barker}}, \bibinfo {author} {\bibfnamefont {J.}~\bibnamefont {Moyer}},
  \bibinfo {author} {\bibfnamefont {S.}~\bibnamefont {Kline}}, \bibinfo
  {author} {\bibfnamefont {G.}~\bibnamefont {Jensen}}, \bibinfo {author}
  {\bibfnamefont {J.}~\bibnamefont {Cook}}, \bibinfo {author} {\bibfnamefont
  {C.}~\bibnamefont {Gagnon}}, \bibinfo {author} {\bibfnamefont
  {E.}~\bibnamefont {Kelley}}, \bibinfo {author} {\bibfnamefont {J.~P.}\
  \bibnamefont {Chabot}}, \bibinfo {author} {\bibfnamefont {N.}~\bibnamefont
  {Maliszewskyj}}, \bibinfo {author} {\bibfnamefont {C.}~\bibnamefont
  {Parikh}}, \bibinfo {author} {\bibfnamefont {W.}~\bibnamefont {Chen}},
  \bibinfo {author} {\bibfnamefont {R.~P.}\ \bibnamefont {Murphy}},\ and\
  \bibinfo {author} {\bibfnamefont {C.}~\bibnamefont {Glinka}},\ }\href
  {https://doi.org/10.1107/S1600576722000826} {\bibfield  {journal} {\bibinfo
  {journal} {Journal of Applied Crystallography}\ }\textbf {\bibinfo {volume}
  {55}},\ \bibinfo {pages} {271} (\bibinfo {year} {2022})}\BibitemShut
  {NoStop}%
\bibitem [{\citenamefont {Zhao}\ \emph {et~al.}(2010)\citenamefont {Zhao},
  \citenamefont {Gao},\ and\ \citenamefont {Liu}}]{ORNL_EQSANS}%
  \BibitemOpen
  \bibfield  {author} {\bibinfo {author} {\bibfnamefont {J.~K.}\ \bibnamefont
  {Zhao}}, \bibinfo {author} {\bibfnamefont {C.~Y.}\ \bibnamefont {Gao}},\ and\
  \bibinfo {author} {\bibfnamefont {D.}~\bibnamefont {Liu}},\ }\href
  {https://doi.org/10.1107/S002188981002217X} {\bibfield  {journal} {\bibinfo
  {journal} {Journal of Applied Crystallography}\ }\textbf {\bibinfo {volume}
  {43}},\ \bibinfo {pages} {1068} (\bibinfo {year} {2010})}\BibitemShut
  {NoStop}%
\bibitem [{ILL()}]{ILL_D22}%
  \BibitemOpen
  \href
  {https://www.ill.eu/users/instruments/instruments-list/d22/more/d22-manual}
  {\emph {\bibinfo {title} {D22 Manual}}},\ \bibinfo {organization} {Institut
  Laue-Langevin},\ \bibinfo {address} {Grenoble, France}\BibitemShut {NoStop}%
\bibitem [{\citenamefont {Akutsu}\ \emph {et~al.}(2020)\citenamefont {Akutsu},
  \citenamefont {Su}, \citenamefont {Iida}, \citenamefont {Naoe}, \citenamefont
  {Igarashi}, \citenamefont {Nagatani}, \citenamefont {Ino}, \citenamefont
  {Matsukawa}, \citenamefont {Oikawa},\ and\ \citenamefont
  {Yamazaki}}]{JPARC_MLF}%
  \BibitemOpen
  \bibfield  {author} {\bibinfo {author} {\bibfnamefont {K.}~\bibnamefont
  {Akutsu}}, \bibinfo {author} {\bibfnamefont {Y.}~\bibnamefont {Su}}, \bibinfo
  {author} {\bibfnamefont {K.}~\bibnamefont {Iida}}, \bibinfo {author}
  {\bibfnamefont {T.}~\bibnamefont {Naoe}}, \bibinfo {author} {\bibfnamefont
  {M.}~\bibnamefont {Igarashi}}, \bibinfo {author} {\bibfnamefont
  {Y.}~\bibnamefont {Nagatani}}, \bibinfo {author} {\bibfnamefont
  {T.}~\bibnamefont {Ino}}, \bibinfo {author} {\bibfnamefont {T.}~\bibnamefont
  {Matsukawa}}, \bibinfo {author} {\bibfnamefont {K.}~\bibnamefont {Oikawa}},\
  and\ \bibinfo {author} {\bibfnamefont {D.}~\bibnamefont {Yamazaki}},\ }\href
  {http://j-parc.jp/researcher/MatLife/ja/publication/files/MLF-AR_2020.pdf}
  {\emph {\bibinfo {title} {J-PARC Annual Report, Vol. 2: Materials and Life
  Science Experimental Facility}}},\ \bibinfo {type} {Tech. Rep.}\ \bibinfo
  {number} {J-PARC 21-02}\ (\bibinfo  {institution} {Japan Proton Accelerator
  Research Complex},\ \bibinfo {address} {T\={o}kai, Ibaraki, Japan},\ \bibinfo
  {year} {2020})\BibitemShut {NoStop}%
\bibitem [{\citenamefont {Vylet}\ and\ \citenamefont
  {Liu}(2007)}]{SynchrotronFacilities}%
  \BibitemOpen
  \bibfield  {author} {\bibinfo {author} {\bibfnamefont {V.}~\bibnamefont
  {Vylet}}\ and\ \bibinfo {author} {\bibfnamefont {J.}~\bibnamefont {Liu}},\
  }\href@noop {} {\emph {\bibinfo {title} {Synchrotron Facilities And Free
  Electron Lasers}}},\ \bibinfo {type} {Tech. Rep.}\ \bibinfo {number}
  {SLAC-PUB-13049}\ (\bibinfo  {institution} {Stanford Linear Accelerator
  Center},\ \bibinfo {address} {Stanford, CA},\ \bibinfo {year}
  {2007})\BibitemShut {NoStop}%
\bibitem [{\citenamefont {Ferrer}\ \emph {et~al.}(1998)\citenamefont {Ferrer},
  \citenamefont {Simon}, \citenamefont {B{\'{e}}rar}, \citenamefont {Caillot},
  \citenamefont {Fanchon}, \citenamefont {Ka{\"\i}kati}, \citenamefont
  {Arnaud}, \citenamefont {Guidotti}, \citenamefont {Pirocchi},\ and\
  \citenamefont {Roth}}]{XRay_Source_D2AM}%
  \BibitemOpen
  \bibfield  {author} {\bibinfo {author} {\bibfnamefont {J.-L.}\ \bibnamefont
  {Ferrer}}, \bibinfo {author} {\bibfnamefont {J.-P.}\ \bibnamefont {Simon}},
  \bibinfo {author} {\bibfnamefont {J.-F.}\ \bibnamefont {B{\'{e}}rar}},
  \bibinfo {author} {\bibfnamefont {B.}~\bibnamefont {Caillot}}, \bibinfo
  {author} {\bibfnamefont {E.}~\bibnamefont {Fanchon}}, \bibinfo {author}
  {\bibfnamefont {O.}~\bibnamefont {Ka{\"\i}kati}}, \bibinfo {author}
  {\bibfnamefont {S.}~\bibnamefont {Arnaud}}, \bibinfo {author} {\bibfnamefont
  {M.}~\bibnamefont {Guidotti}}, \bibinfo {author} {\bibfnamefont
  {M.}~\bibnamefont {Pirocchi}},\ and\ \bibinfo {author} {\bibfnamefont
  {M.}~\bibnamefont {Roth}},\ }\href
  {https://doi.org/10.1107/S0909049598004257} {\bibfield  {journal} {\bibinfo
  {journal} {Journal of Synchrotron Radiation}\ }\textbf {\bibinfo {volume}
  {5}},\ \bibinfo {pages} {1346} (\bibinfo {year} {1998})}\BibitemShut
  {NoStop}%
\bibitem [{\citenamefont {Vaughan}\ \emph {et~al.}(2020)\citenamefont
  {Vaughan}, \citenamefont {Baker}, \citenamefont {Barret}, \citenamefont
  {Bonnefoy}, \citenamefont {Buslaps}, \citenamefont {Checchia}, \citenamefont
  {Duran}, \citenamefont {Fihman}, \citenamefont {Got}, \citenamefont
  {Kieffer}, \citenamefont {Kimber}, \citenamefont {Martel}, \citenamefont
  {Morawe}, \citenamefont {Mottin}, \citenamefont {Papillon}, \citenamefont
  {Petitdemange}, \citenamefont {Vamvakeros}, \citenamefont {Vieux},\ and\
  \citenamefont {Di~Michiel}}]{XRay_Source_ID15A}%
  \BibitemOpen
  \bibfield  {author} {\bibinfo {author} {\bibfnamefont {G.~B.~M.}\
  \bibnamefont {Vaughan}}, \bibinfo {author} {\bibfnamefont {R.}~\bibnamefont
  {Baker}}, \bibinfo {author} {\bibfnamefont {R.}~\bibnamefont {Barret}},
  \bibinfo {author} {\bibfnamefont {J.}~\bibnamefont {Bonnefoy}}, \bibinfo
  {author} {\bibfnamefont {T.}~\bibnamefont {Buslaps}}, \bibinfo {author}
  {\bibfnamefont {S.}~\bibnamefont {Checchia}}, \bibinfo {author}
  {\bibfnamefont {D.}~\bibnamefont {Duran}}, \bibinfo {author} {\bibfnamefont
  {F.}~\bibnamefont {Fihman}}, \bibinfo {author} {\bibfnamefont
  {P.}~\bibnamefont {Got}}, \bibinfo {author} {\bibfnamefont {J.}~\bibnamefont
  {Kieffer}}, \bibinfo {author} {\bibfnamefont {S.~A.~J.}\ \bibnamefont
  {Kimber}}, \bibinfo {author} {\bibfnamefont {K.}~\bibnamefont {Martel}},
  \bibinfo {author} {\bibfnamefont {C.}~\bibnamefont {Morawe}}, \bibinfo
  {author} {\bibfnamefont {D.}~\bibnamefont {Mottin}}, \bibinfo {author}
  {\bibfnamefont {E.}~\bibnamefont {Papillon}}, \bibinfo {author}
  {\bibfnamefont {S.}~\bibnamefont {Petitdemange}}, \bibinfo {author}
  {\bibfnamefont {A.}~\bibnamefont {Vamvakeros}}, \bibinfo {author}
  {\bibfnamefont {J.-P.}\ \bibnamefont {Vieux}},\ and\ \bibinfo {author}
  {\bibfnamefont {M.}~\bibnamefont {Di~Michiel}},\ }\href
  {https://doi.org/10.1107/S1600577519016813} {\bibfield  {journal} {\bibinfo
  {journal} {Journal of Synchrotron Radiation}\ }\textbf {\bibinfo {volume}
  {27}},\ \bibinfo {pages} {515} (\bibinfo {year} {2020})}\BibitemShut
  {NoStop}%
\bibitem [{\citenamefont {Bevington}\ and\ \citenamefont
  {Robinson}(2015)}]{DataReductionBook}%
  \BibitemOpen
  \bibfield  {author} {\bibinfo {author} {\bibfnamefont {P.~R.}\ \bibnamefont
  {Bevington}}\ and\ \bibinfo {author} {\bibfnamefont {D.~K.}\ \bibnamefont
  {Robinson}},\ }\href@noop {} {\emph {\bibinfo {title} {Data Reduction and
  Error Analysis for the Physical Sciences, 3rd Ed.}}}\ (\bibinfo  {publisher}
  {McGraw-Hill Education},\ \bibinfo {address} {New York},\ \bibinfo {year}
  {2015})\BibitemShut {NoStop}%
\end{thebibliography}%

\end{document}